%% file: CG-EcoSta-01.tex
\documentclass[a4paper,12pt]{article} 
\pdfoutput=1

\usepackage[paper = a4paper, left = 2.5cm, right = 2.5cm, top = 2.5cm, bottom = 2.5cm]{geometry}
\usepackage[english]{babel}                         
\usepackage{makeidx}                                
\usepackage{amssymb,amsmath,dsfont}                 
\usepackage{graphicx}
\usepackage{natbib}
\usepackage{rotating}                               
\usepackage{setspace}                               
\usepackage{url}                                    
\usepackage{bm}                                     
\usepackage{bbm}
\usepackage{boxedminipage}                          
\usepackage{fancybox}                               
\usepackage{verbatim}                               
\usepackage{times}
\usepackage{wasysym}
\usepackage{longtable}
\usepackage{lscape}
\usepackage{lineno}
\usepackage{multirow}
\usepackage{hyperref}
\usepackage{listings}                               
\usepackage[usenames,dvipsnames]{xcolor}
\usepackage{textcomp}
\usepackage{times}
\usepackage{amsthm}   
\usepackage{booktabs} 
\usepackage{epstopdf} 
\usepackage{scalefnt} 
\usepackage{dcolumn}
\usepackage{subcaption}

\newcommand{\info}[2]{\mathcal{#1}_{#2}}
\newcommand{\Ind}{I}

\DeclareMathOperator{\avar}{Avar}
\DeclareMathOperator{\iid}{i.i.d.}
\DeclareMathOperator{\diag}{diag}

\newcommand{\MEM}{$\mathsf{MEM}$}
\newcommand{\HAR}{$\mathsf{HAR}$}

\newcommand{\MSAMEM}{$\mathsf{MS-AMEM}$}
\newcommand{\STAMEM}{$\mathsf{ST-AMEM}$}

\newcommand{\ACD}{$\mathsf{ACD}$}
\newcommand{\SpMEM}{$\mathsf{SpMEM}$}
\newcommand{\vMEM}{$\mathsf{vMEM}$}
\newcommand{\SpvMEM}{$\mathsf{SpvMEM}$}

\title{Multiplicative Error Models: 20 years on}

\author{Fabrizio Cipollini, Giampiero M. Gallo}

\date{}

\begin{document}

  \maketitle

\begin{abstract}
	Several phenomena are available representing market activity: volumes, number of trades, durations between trades or quotes, volatility -- however measured -- all share the feature to be represented as positive valued time series. When modeled, persistence in their behavior and reaction to new information suggested to adopt an autoregressive--type framework. The Multiplicative Error Model (\MEM) is borne of an extension of the popular GARCH approach for modeling and forecasting conditional volatility of asset returns. It is obtained by multiplicatively combining the conditional expectation of a process (deterministically dependent upon an information set at a previous time period) with a random disturbance representing unpredictable news: \MEM s have proved to parsimoniously achieve their task of producing good performing forecasts. In this paper we discuss various aspects of model specification and inference both for the univariate and the multivariate case. The applications are illustrative examples of how the presence of a slow moving low--frequency component can improve the properties of the estimated models.
	
\end{abstract}
  
  \section{Introduction}

Twenty years have gone by since the \citet{Engle:2002} paper in which Rob Engle provided a framework of generalization to the ARCH \citep{Engle:1982} and GARCH \citep{Bollerslev:1986} modeling of the conditional variance of financial returns. Following the work in \citet{Engle:Russell:1998}, and the overall remarks in \citet{Engle:2000}, it was recognized that the availability of ultra--high frequency data had opened the venue to widening the range of the analysis of market activity. 
New financial time series became readily available: \citet{Andersen:Bollerslev:1998}  started a stream of literature addressing the properties of measuring volatility from intradaily data and using it as an alternative target to the squared returns (a very noisy measure of return variance); for a review of various realized volatility measures, see \citet{Andersen:Bollerslev:Christoffersen:Diebold:2006}; for the care to be exerted in working with ultra--high frequency data, see \citet{Brownlees:Gallo:2006}. While in the GARCH literature measurement and forecasting were steps performed simultaneously, the development of volatility measures left the  question open as of what model would be the most appropriate to derive forecasts. Volatility being non--negatively valued and persistent lends these measures to be represented in some autoregressive fashion; as expressed by market trading activity volatility is accompanied by other indicators which exhibit the stylized facts of volatility clustering: next to realized volatility and durations, we can mention the volume, range, number of trades, bid--ask spread. In \citet{Engle:2002}, a new class of models was suggested for non--negative valued processes, called \textit{Multiplicative Error Models} (MEM), obtained by multiplicatively combining the conditional expectation of a process (deterministically dependent upon an information set at a previous time period) with a random disturbance representing unpredictable news. This approach is different in spirit from the linear and additive modeling suggested by \citet{Corsi:2009} in his Heterogeneous Autoregressive (HAR) model for realized variances, which has the merit of recognizing that long--memory features, often cited in the profile of realized volatility especially, can be mimicked upon suitable aggregation of the lagged variances \citep[see also][for a HAR with jumps]{Andersen:Bollerslev:Diebold:2007}.

In the past 20 years, the class of MEM models has grown in size, at times applying and extending concepts that were originated in the GARCH literature, also as a consequence of the fact that a GARCH model is a MEM for squared returns. Within the \ACD, from the already mentioned seminal contribution of \citet{Engle:Russell:1998}, further developments for the financial durations are documented in \citet{Hautsch:2004}; \citet{Haerdle:Hautsch:Mihoci:2015} provide a local adaptive approach to \ACD's, using \MEM\ and \ACD\ as synonyms; \cite{Perera:Hidalgo:Silvapulle:2016} provide a testing framework for \ACD\ parametric specifications. A \MEM\ for the daily range \citep{Parkinson:1980, Garman:Klass:1980} is called Conditional Autoregressive Range  (CARR) model by \citet{Chou:2005} and a review of this approach is contained in \citet{Chou:Chou:Liu:2015}.

Initial applications of the \MEM\ were favored by the fact that GARCH routines in standard software could expediently be used to estimate the new model, by imposing that the "dependent variable" were the square root of the actual variable of interest, that the mean equation was forced to zero, and that the innovation distribution was set to the Gaussian. This estimation practice was later justified in a more general context by \citet{Engle:Gallo:2006}, who formally proved the result for a Gamma distribution for the multiplicative innovation, and showed that the first order conditions for the ML estimation of the parameters of the conditional expectation in a \MEM\ do not depend upon the unknown scale parameter of the Gamma, establishing QMLE properties for the estimator. Independently developed, the Generalized Autoregressive Score approach of \citet{Creal:Koopman:Lucas:2013} and the Dynamic Conditional Score approach by \citet{Harvey:2013} both address a wider class of models which comprise the \MEM\ as a special case under a parametric assumption of a Gamma distribution for the error term.

For the case in which the series may contain zeros (as with absolute returns or volume at very high frequency with thinly traded assets), \citet{Hautsch:Malec:Schienle:2014} suggest a way to augment the \MEM\ model to accommodate a mass of the density at zero.
In the case of intradaily processes, the \MEM\ gets expressed by several multiplicative components as in \citet{Brownlees:Cipollini:Gallo:2011} who specify a model for volumes whose conditional expectation is the result of the product of a daily component, a periodic intradaily component (essentially capturing time of day) and a non periodic intradaily component.
\citet{Cipollini:Gallo:Otranto:2020} address the issue of measurement error in the realized volatility measures and show that the additive models used by \citet{Bollerslev:Patton:Quaedvlieg:2016} in the univariate context which make use of realized quarticity can be conveniently supplemented by \MEM s which, thanks to their multiplicative structure, better capture the heteroskedastic nature of measurement errors.

\citet{Brownlees:Gallo:2010} adopt a \MEM\  to model several versions of realized volatility in a risk management context where the VaR is used as a basis to evaluate forecast performance. While they find that realized kernel volatility has the most desirable prediction properties, they also establish that the daily range is a good substitute when intradaily data are not available.

\citet{Engle:Gallo:2006} contains an application of a trivariate \MEM\ on three measures of volatility, estimated equation--by--equation, allowing for lagged dynamic interdependence \citep[a triangular structure with contemporaneous dependence is contained in][]{Manganelli:2005}. The same structure is adopted by \citet{Engle:Gallo:Velucchi:2012} in the context of daily range for several East Asian markets, where they test for market interdependence and spillovers in the Asian crisis of 1997. A full--fledged theory for vector \MEM, i.e. \vMEM, is suggested by \citet{Cipollini:Engle:Gallo:2006} in the context of copula functions \citep[later developed in][]{Cipollini:Engle:Gallo:2017} and of Generalized Method of Moments \citep[later developed in][]{Cipollini:Engle:Gallo:2013}. Diagnostics checking for \MEM\ is provided by \citet{Koul:Perera:Silvapulle:2012} and by \citet{Perera:Silvapulle:2017} and for \vMEM\ by \citep{Ng:Li:Yu:2016}. 
An automated procedure for \vMEM\ specification selection is proposed in \citet{Cipollini:Gallo:2010}.
Another development involving a trivariate intradaily market activity process is suggested by \citet{Hautsch:2008} and a dynamic conditional correlation model with \MEM\ is developed by  \citet{Bodnar:Hautsch:2016}.

The feature of a positive valued financial time series showing a slow--moving low frequency component may be addressed adapting the component GARCH by \citet{Engle:Lee:1999}: the dynamics may depend on the sum of two components,  one forced to be persistent and another capturing  more short-term dynamics. In the \MEM\ context, this is suggested by \citet{Brownlees:Cipollini:Gallo:2012} who present an application on univariate volatilities of several stocks. \citet{Cipollini:Gallo:2019} extend the idea to the multivariate context, by keeping the additivity of the components in a \vMEM, but forcing the slow moving volatility to be common to several Euro area market indices.

\citet{Gallo:Otranto:2015} investigate the features of a low--frequency component in a \MEM\ in  the context of Markov Switching (\MSAMEM) and Smooth Transition (\STAMEM) models, suggesting the concept of changing average volatility levels and comparing the behavior of the \MSAMEM\ and of the \STAMEM\ to various versions of the \HAR\ model by \citet{Corsi:2009} in forecasting. Among these models, there is a \HAR\ specification  with jumps: the presence of jumps is addressed within the \MEM\ approach by \citet{Caporin:Rossi:Santucci:2017}.

This paper retraces the logic behind a \MEM, starting from some stylized facts, and focuses on the presence of a component which captures slow--moving behavior is of specific interest in our contribution; specifically, we address the possibility of having components combining multiplicatively, both in  the univariate \citep[extending][for an ACD]{Veredas:Rodriguez:Espasa:2007} and in the multivariate \citep[extending][]{Barigozzi:Brownlees:Gallo:Veredas:2014} cases, the latter with a common low--frequency component. We discuss model specification and inference issues providing details for the Generalized Method of Moments method. The applications are performed in reference to several volatility market indices (US, Europe and East Asia), providing evidence of the relevance of the low--frequency component which varies by market. The multivariate case considers a trivariate system in which dynamic  interdependencies in the short--run components of absolute returns, realized kernel volatility and option--based implied volatility indices are assessed with and without a common low--frequency component. Being a GMM-based strategy, estimation bypasses the specification of a parametric distribution for the error term. Yet, we provide evidence of how this error behaves as an estimation residual, should some applications for volatility--at--risk \citep[in the sense of][]{Caporin:Rossi:Santucci:2017} be needed.
  
The structure of the paper is as follows: in Section \ref{sect:spmem} we propose a modification of the standard specification of a \MEM\ (called \SpMEM). In Section \ref{sect:Estimation-spmem} we discuss inferential properties within a GMM estimation strategy. In the univariate case, an empirical application is performed on series of realized volatility from several markets (Section \ref{sect:uniappl}). For the multivariate case,  (Section \ref{sect:SpvMEM}) we extend the \SpvMEM\ model by \citet{Barigozzi:Brownlees:Gallo:Veredas:2014} to allow for dynamic interdependence across variables, for which the GMM properties of the estimation strategy followed are presented in Section \ref{sect:Inference}. 
The applications in the multivariate case relate to a trivariate model with absolute returns, realized volatility and implied volatility from option--based indices (Section \ref{sect:multiappl}). Concluding remarks follow.

  \section{The Model}
  \label{sect:spmem}
  
  In the univariate case, the general form of a Multiplicative Error Model (\MEM) can be specified as
  \begin{equation}
    \label{eqn:x(t)-univariate}
    x_t = \underbrace{\mu \; \tau_t \; \xi_t}_{\mu_t} \varepsilon_t,
  \end{equation}
  i.e., designed in a way that $\mu$ is introduced as the unconditional expectation of $x_t$ (assumed mean stationary), around which there are three components combining multiplicatively:
  \begin{itemize}
    \item $\tau_t$ is a low--frequency component capturing the slow-moving secular dynamics in the process (when constant and equal to $1$, we have the base \MEM);
    \item $\xi_t$ is a short--run component, parameterized in a way similar to a GARCH process; 
    \item $\varepsilon_t$ is a conditionally unpredictable homoskedastic component. 
  \end{itemize}
  Upon imposition of a unit expectation on the $\varepsilon_t$ term, the expectation of $x_t$ conditional on the information set $\info{I}{t-1}$ is thus $\mu_{t} = \mu \tau_t \xi_t$; moreover, for identification purposes, $E(\tau_t)=E(\xi_t)=1$, which is achieved via suitable normalizations.
  
\begin{figure}[h!]
    \centering
    \caption{DJI Realized kernel volatility $x_t$ (in grey); its unconditional mean $\mu$ (horizontal red line); its low--frequency component $\mu \tau_{t}$ (in brown); its conditional expectation $\mu_{t} = \mu \tau_{t} \xi_{t}$ (in blue). Sample Aug. 2, 2013--Apr. 13, 2021. }
    \label{fig:DJI}
    \includegraphics[width=\textwidth]{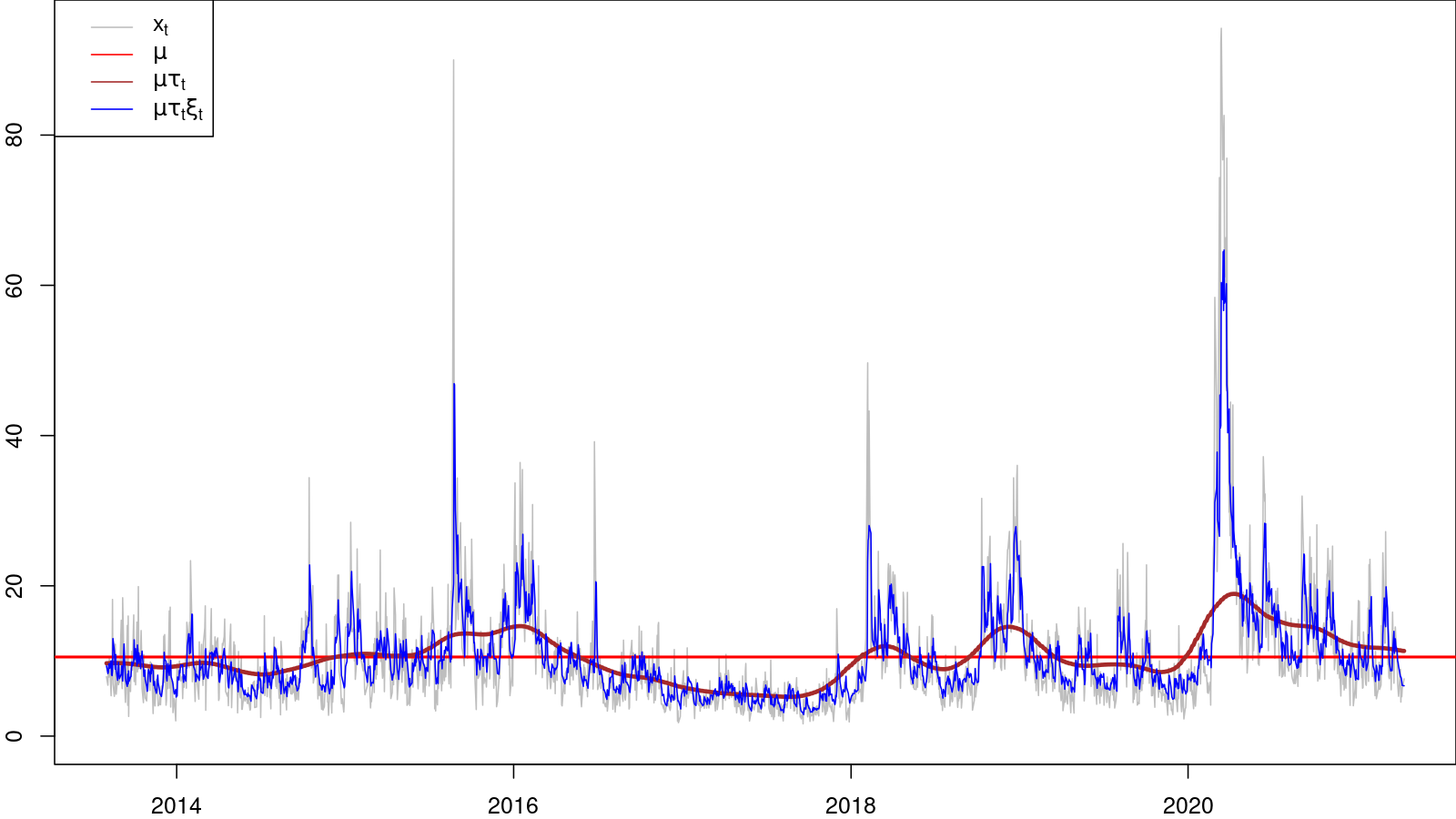}
  \end{figure}

  To motivate this specification, let us refer to Figure~\ref{fig:DJI} of a typical positive valued financial time series, namely the realized kernel volatility measured on the Dow Jones 30 Industrial Index.\footnote{The data span the period August 2nd, 2013 -- April 13th, 2021 and the series is the annualized square root of the realized kernel variance taken from the OMI Realized Library by \citet{Heber:Lunde:Shephard:Sheppard:2009}.} 
 For future reference, and motivation for the subsequent analysis, we have reported what we obtain as the sample estimate of the unconditional mean $\mu$ (the horizontal red line); relative to it, the series exhibits a slow--moving feature which is captured by the $\tau_t$ term (the brown line depicts the estimated product $\mu \tau_{t}$); in turn, around it, we notice the high--frequency, familiar persistent pattern deriving by the well-known property of volatility clustering reproduced by $\xi_{t}$ (the blue line depicts the estimated product $\mu_{t} = \mu \tau_{t} \xi_{t}$); finally, the estimated error component  $\varepsilon_{t}$ would be derived as the ratio between the grey and the blue line. 
  
  As per the first component $\tau_t$, should one decide to do away with it (equivalent to take $\tau_{t} \equiv 1 \; \forall t$), the outcome would be the base \MEM, as suggested by \citet{Engle:2002}, with expectation targeting $\mu$; to model $\tau_t$, one can resort to some smooth function of time \citep[as in, e.g, a spline or in the smooth transition framework of][]{Amado:Terasvirta:2017}, or, even, a Markov Switching approach as in \citet{Gallo:Otranto:2015} based on the concept of average level of volatility within each regime.
  In what follows, also in view of the multivariate specification below, we adopt a non parametric specification derived from \citet{Veredas:Rodriguez:Espasa:2007} in the context of the ACD model, where the only requirement is that $\tau_{t}$ is a sufficiently smooth function of time: we call it semi--nonparametric-\MEM\ or \SpMEM. 
  For practical purposes, at the estimation stage, it takes the form:
  \begin{equation}
    \widehat{\tau}_{t} = 
    \frac{\displaystyle \sum_{s = 1}^{T} x_{t}^{(\tau)} K \left( \frac{z_{t} - z_{s}}{h} \right) }{\displaystyle \sum_{s = 1}^{T} K \left( \frac{z_{t} - z_{s}}{h} \right)} \\  
  \end{equation}
  where 
  \begin{equation*}
    x_{t}^{(\tau)} = \frac{x_{t}}{\mu \xi_{t}},
  \end{equation*}
  $\mu$ and $\xi_{t}$ denote here estimates of the corresponding components, $z_{t} = t / T$, $K(\cdot)$ is a kernel function and $h$ is the bandwidth.
  Further details on model estimation are in Section~\ref{sect:Estimation-spmem}.
  
  The second component $\xi_t$ captures the short--run dynamics with an autoregressive dependence on past observations of $x_t$ and past  $\xi_t$'s, as mentioned, in a GARCH structure of the type
  \begin{equation}
    \label{eqn:xi-1}
    \xi_t = \left[ 1 - \left(\beta_1 + \alpha_1 + \frac{\gamma_1}{2} \right) \right] + \beta_1 \xi_{t-1} + \alpha_1 x_{t-1}^{(\xi)} + \gamma_1 x_{t-1}^{(\xi-)},  
  \end{equation}
  where
  \begin{equation*}
    x_{t}^{(\xi)} = \frac{x_{t}}{\mu \tau_t}
    \qquad{}
    x_{t}^{(\xi-)} = x_{t}^{(\xi)} {D}^{-}_{t},
  \end{equation*}
  ${D}^{-}_{t} = \Ind(r_{t} < 0)$ and $r_{t}$ is the return.
  An equivalent way to express the $\xi_{t}$ dynamics is
  \begin{equation}
    \label{eqn:xi-2}
    \xi_t = \left( 1-\beta_{1}^{*}\right) + \beta_1^{*} \xi_{t-1} + \alpha_1 v_{t-1} + \gamma_1 v_{t-1}^{(-)},  
  \end{equation}
  where
  \begin{equation*}
    v_{t} = x^{(\xi)}_{t} - \xi_{t}
    \qquad{}
    v_{t}^{(-)} = x^{(\xi-)}_{t} - \xi_{t} / 2
  \end{equation*}
  are zero-mean innovations and $\beta_{1}^{*} = \beta_{1} + \alpha_{1} + \gamma_{1} / 2$ is the persistence parameter.%
  \footnote{The equivalence between (\ref{eqn:xi-1}) and (\ref{eqn:xi-2}) may be no longer valid in case additional lags are added to some of the terms.}

  The third component $\varepsilon_t$ is a unit mean, homoskedastic r.v.\ with non-negative support, in symbols
  \begin{equation}
    \label{eqn:eps(t)}
    \varepsilon_{t} \overset{\iid}{\sim} d^{+}(1, \sigma^{2}).
  \end{equation}
  In the univariate case, we will refer to r.v.'s with such characteristics, namely the Gamma, the Log--Normal,  the Beta$^\prime$ and the Log-Logistic described in detail in the Appendix.
  
  \section{Inference Issues}
  \label{sect:Estimation-spmem}
  
  Following \citet{Veredas:Rodriguez:Espasa:2007} and \citet{Barigozzi:Brownlees:Gallo:Veredas:2014}, inference of the \SpMEM\ is obtained using the following strategy.
  $\mu$ is estimated once for all by the sample average of the $\{x_{t}\}$ data;
  regarding the other components, after initialization of all $\xi_{t}$'s to $1$, estimation is done by iterating between these two steps, until convergence:
  \begin{enumerate}
    \item $\tau_{t}$ is estimated using the kernel estimator detailed in Section~\ref{sect:spmem} on the $\{x_{t}^{(\tau)} = x_{t} / (\mu \xi_{t}) \}$ current values;
    \item $\xi_{t}$ parameters and $\sigma^{2}$ are estimated on the $\{x_{t}^{(\xi)} = x_{t} / (\mu \tau_{t}) \}$  current values using the approach detailed below.
  \end{enumerate}
  
  Clearly, in case the model is pure \MEM\ (i.e.\ $\tau_{t} \equiv 1 \ \forall t$), only the second step is needed and inference is obtained by estimating the $\xi_{t}$ parameters on the $\{x_{t}^{(\xi)} = x_{t} / \mu \}$ values.
  
  To describe inference on the $\xi_{t}$ parameters and $\sigma^{2}$, we refer to the framework presented in \citet[][Section~9.2.2]{Brownlees:Cipollini:Gallo:2012}. 
  Assuming that $\xi_{t}$ is correctly specified and indicating with $\bm{\theta}$ the vector of parameters entering it, two estimation strategies are illustrated: Generalized Method of Moments (GMM), and Quasi Maximum Likelihood (QML).

  \subsection{Generalized Method of Moments Inference} 
  \label{sect:GMMInference}
  
  GMM estimation can be conveniently used to estimate the parameters of interest without an explicit choice of the error term distribution of Equation~(\ref{eqn:eps(t)}). 
  Let
  \begin{equation} 
    \label{eq:eps(t)}
    \varepsilon_{t} = \frac{x^{(\xi)}_{t}}{\xi_{t}},
  \end{equation}
  where $x^{(\xi)}_{t} = x_{t} / (\mu \tau_{t})$.
  Under model assumptions, $\varepsilon_{t} - 1$ is a conditionally homoskedastic martingale difference, with conditional expectation zero and conditional variance $\sigma^2$.
  Following \citet[][Section~9.2.2.2]{Brownlees:Cipollini:Gallo:2012}, the \emph{efficient} GMM estimator of $\bm{\theta}$, say $\widehat{\bm{\theta}}_{GMM}$, solves the criterion equation
  \begin{equation}
    \label{eqn:Score:Equation:Gamma}
    \sum_{t = 1}^T ( \varepsilon_{t} - 1 ) \bm{a}_{t} = \bm{0},
  \end{equation}
  and has asymptotic variance matrix 
  \begin{align}
    \label{eqn:Avar:Sandwich}
    \avar(\widehat{\bm{\theta}}_{GMM}) & =  \sigma^{2} \bm{A}^{-1},
  \end{align}
  where
  \begin{equation}
    \label{eq:a(t)}
    \bm{a}_{t} = \frac{1}{\xi_{t}} \nabla_{\bm{\theta}} \xi_{t}
  \end{equation}
  and
  \begin{equation*}
    \bm{A} = \lim_{T \rightarrow \infty} \left[ T^{-1} \sum_{t = 1}^T E \left( \bm{a}_{t}  \bm{a}_{t}^\prime \right) \right].
  \end{equation*}
  As a consequence, a consistent estimator of the asymptotic variance matrix is 
  \begin{equation*}
    \widehat{\avar}(\widehat{\bm{\theta}}_{GMM}) =
    \widehat{\sigma}^2 \widehat{\bm{A}}^{-1},
  \end{equation*}
  where 
  \begin{equation*}
    \widehat{\sigma}^{2} = T^{-1} \sum_{t = 1}^T \left( \widehat{\varepsilon}_{t} - 1 \right)^{2}
  \end{equation*}
  is a Method of Moments estimator of $\sigma^2$, 
  \begin{equation*}
    \widehat{\bm{A}} = 
    T^{-1} \sum_{t = 1}^T \widehat{\bm{a}}_{t} \widehat{\bm{a}}_{t}^\prime,
  \end{equation*}
  $\widehat{\varepsilon}_{t}$ and $\widehat{\bm{a}}_{t}$  correspond to  (\ref{eq:eps(t)})  and (\ref{eq:a(t)}), respectively, evaluated at $\widehat{\bm{\theta}}_{GMM}$.
  
  From a practical point of view, then, the procedure becomes fully feasible in our context upon substituting the unknown $\mu$ and $\tau_t$ in $x_t^{(\xi)}$ by consistent estimates $\widehat{\mu}$ and $\widehat{\tau}_t$.

  \subsection{Quasi Maximum Likelihood Inference}
  \label{sect:QMLInference}
  
  Following \citet{Engle:Gallo:2006}, an alternative approach is Quasi Maximum Likelihood (QML): this is obtained assuming $\varepsilon_{t}$ as Gamma distributed and then estimating $\bm{\theta}$ by Maximum Likelihood.
  
  More specifically, if (\ref{eqn:eps(t)}) is specified as $\varepsilon_{t} | \info{F}{t-1} \sim Gamma \left( \phi, \phi \right)$ (so as $E \left( \varepsilon_{t} | \info{F}{t-1} \right) = 1$ and $V \left( \varepsilon_{t} | \info{F}{t-1} \right) = \sigma^{2} = 1 / \phi$) the \textit{log-likelihood} function is
  \begin{equation*}
    \label{eqn:loglik}
    l_T
    =
    \sum_{t = 1}^T \left[ \phi \ln \phi - \ln \Gamma(\phi) + \phi \ln \varepsilon_{t} - \phi \varepsilon_{t} - \ln x^{(\xi)}_{t} \right].
  \end{equation*}
  This equation shows that its maximization in $\bm{\theta}$ can be done maximizing the quasi-log-likelihood
  \begin{equation*}
    \label{eqn:qloglik}
    \sum_{t = 1}^T \left( \ln \varepsilon_{t} - \varepsilon_{t} \right),
  \end{equation*}
  that does not depend on $\phi$.
  As a consequence, the first order condition for $\bm{\theta}$ is given exactly by the GMM condition (\ref{eqn:Score:Equation:Gamma}).
  Notice that this condition can be rewritten as 
  \begin{equation}
    \sum_{t = 1}^T \frac{\partial \xi_{t}}{\partial \bm{\theta}} \frac{x^{(\xi)}_{t} - \xi_{t}}{\xi_{t}^{2}} = \bm{0},
  \end{equation}
  whose LHS, in case $\xi_{t}$ is correctly specified, has a zero expectation even in case $\varepsilon_{t}$ is not Gamma-distributed.

\section{The Univariate Case: Applications}
 \label{sect:uniappl}
In the more general multivariate case, we will use series for the open--to--close absolute returns (rescaled by $\sqrt{\pi/2}$ so that their mean is expressed in volatility units), for the realized kernel volatility (obtained as the percentage annualized square root of the realized variance with the Parzen kernel) on market indices, both taken from the OMI Realized Library \citep{Heber:Lunde:Shephard:Sheppard:2009}, and for the option--based implied volatility indices \citep[built  for several market indices like the  VIX][]{Whaley:2009}, taken from the site \texttt{Investing.com}\footnote{at the URL \texttt{https://www.investing.com}}. In Table \ref{tab:data} in the Appendix, we report the symbol we will use and the description of the market indices and volatility indices, as well as the start and end of the available sample period. In what follows, we will use the acronyms arVol, rkVol, impVol, respectively, for these series.

In what follows, we limit ourselves to the univariate case, showing the results on just the realized kernel volatility. 

 \begin{sidewaystable}[htbp]
	\centering
	\caption{\MEM: Estimation summary table.}
	\label{tab:inf-mem}
	\input{Table/Tab-Inf-mem.tex}
	\bigskip
	\centering
	\caption{\SpMEM: Estimation summary table.}
	\label{tab:inf-spmem}
	\input{Table/Tab-Inf-spmem.tex}
\end{sidewaystable}

We report the coefficient estimation results in Table \ref{tab:inf-mem} for the \MEM\ and in Table  \ref{tab:inf-spmem} for the \SpMEM. In both cases we adopt a $(2,1)$ specification and the Tables contain the value of the unconditional mean $\mu$, the persistence coefficient $\beta_1^*$ defined above, followed by $\alpha_{1}$, $\alpha_{2}$ and $\gamma_{1}$. The coefficient $\sigma^2$ is the variance of the estimated residuals, while $R^2$ is the squared correlation coefficient between the observed  realized volatility and the estimated conditional mean. In the lower portion of the tables, we also report the p-values of the Ljung--Box test statistics at various lags ($5$, $10$, $15$, and $20$) with degrees of freedom net of the number of the parameters \citep[following the logic of][]{Lutkepohl:2005}.

In reference to Table \ref{tab:inf-mem}, we notice that the lowest persistence is had for the DJI and the IXIC, the other indices being well above $0.96$. Correspondingly, the $\alpha_1$ parameter is higher for the markets with less persistence, while for the other markets it is comprised between $0.15$ and $0.21$.  With just one exception (KS11), the $\gamma_1$ has the right sign and is significant. The measures of fit are comprised in a range between $0.47$ and $0.67$. With the exception of STOXX50E and KS11, the p--values of the Ljung--Box statistics signal the presence of undetected dynamics from this model.
  
In the presence of the slow--moving component $\tau_{t}$, the results of Table \ref{tab:inf-spmem} for the \SpMEM\ show results that are different in two substantial directions: the first is the drastic drop in $\beta_1^*$ which now is well below $0.9$ for most markets, the second is the generalized improvement of the Ljung--Box p--values, showing that the low--frequency component is actually capable of capturing relevant dynamics. 

Figure \ref{fig:DJI-3} shows for all market the estimated pattern of the various components, estimated in the \SpMEM\ specification. In it, we reproduce the overall unconditional mean $\mu$, as a flat line; the thicker red curve is the estimated $\mu\tau_t$ and the product of the three components, $\mu\tau_t\xi_t$, i.e. the estimated conditional expectation, is reproduced in blue. The observed series is in the background in a shade of grey. For most series, the low frequency component absorbs a substantial portion of the volatility behavior.

\begin{figure}
	\caption{\SpMEM\ on rkVol.  The overall unconditional mean $\mu$ is a flat line; the thicker red curve is  $\mu \tau_t$  (with $h = 6$ months) and the conditional expectation $\mu \tau_{t} \xi_t$ is in blue. The observed series is in the background in a shade of grey. For the sample periods, see Table \ref{tab:data}.}
	\label{fig:DJI-3}
	\begin{subfigure}{.43\textwidth}
		\centering
		\includegraphics[width=\linewidth]{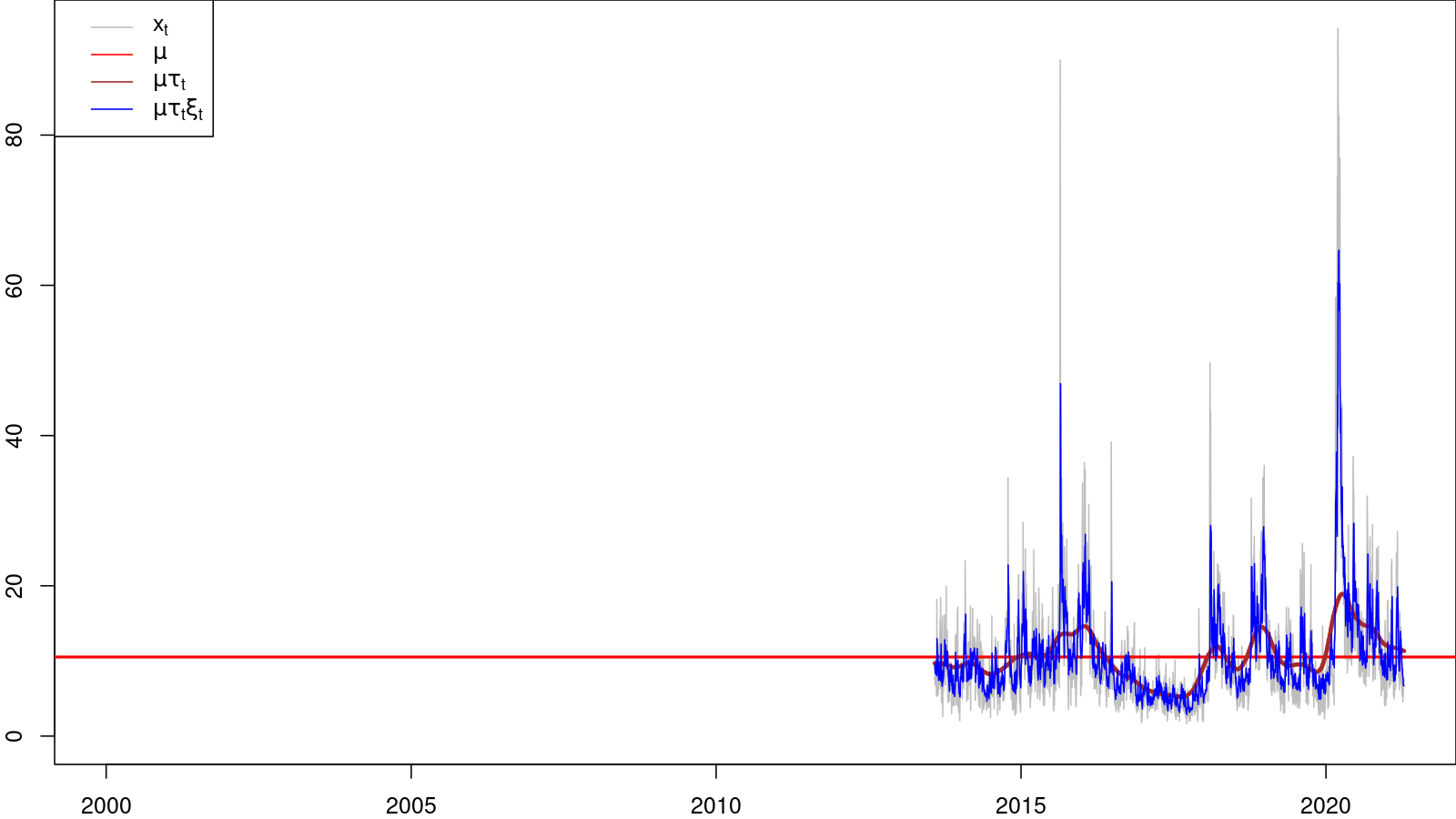}  
		\caption{DJI}
		\label{fig:Uni-DJI}
	\end{subfigure}
	\begin{subfigure}{.43\textwidth}
		\centering 
		\includegraphics[width=\linewidth]{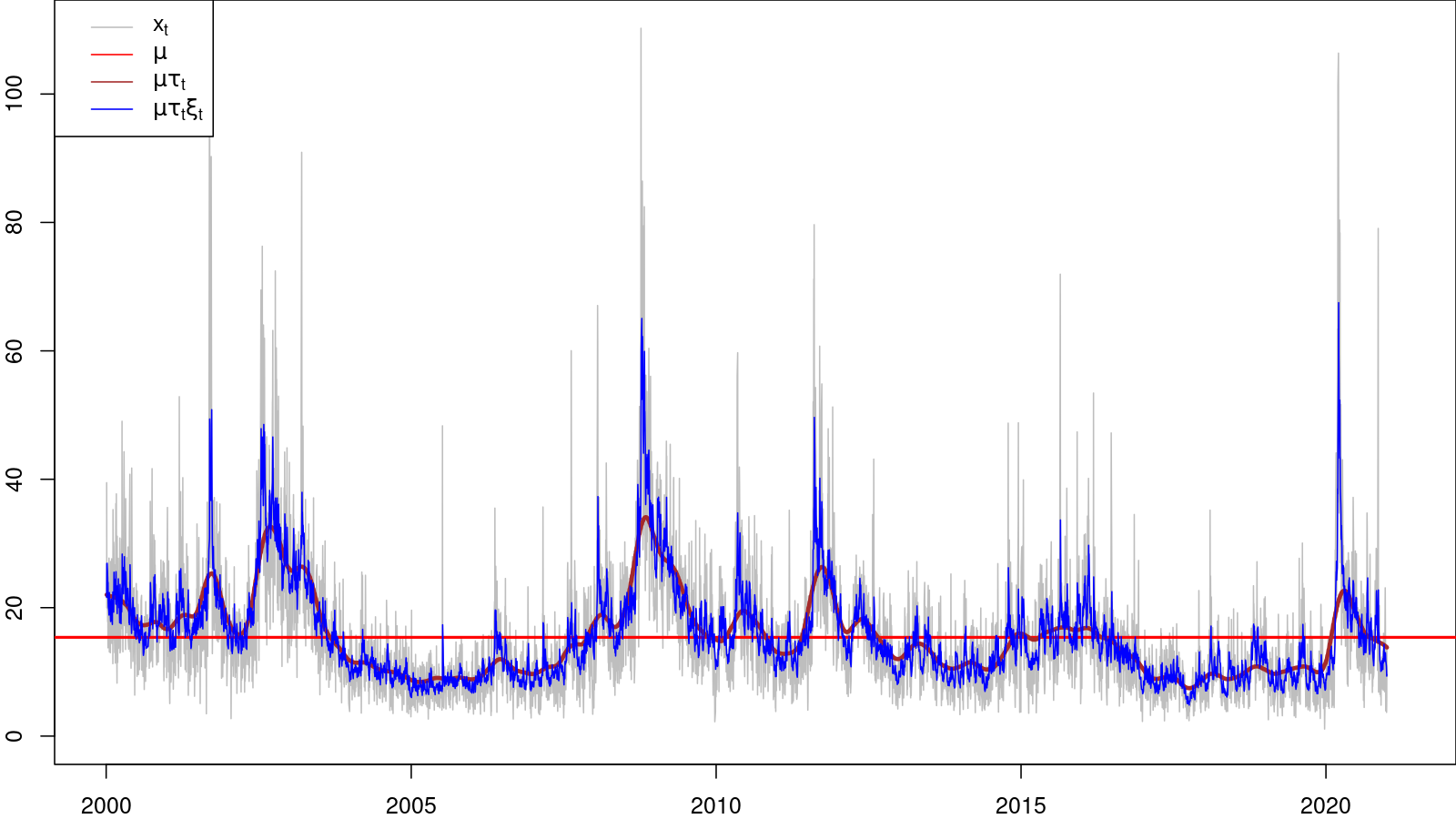}  
		\caption{FCHI}
		\label{fig:Uni-FCHI}
	\end{subfigure}
	\\
	\centering
	\begin{subfigure}{.43\textwidth}
		\centering
		\includegraphics[width=\linewidth]{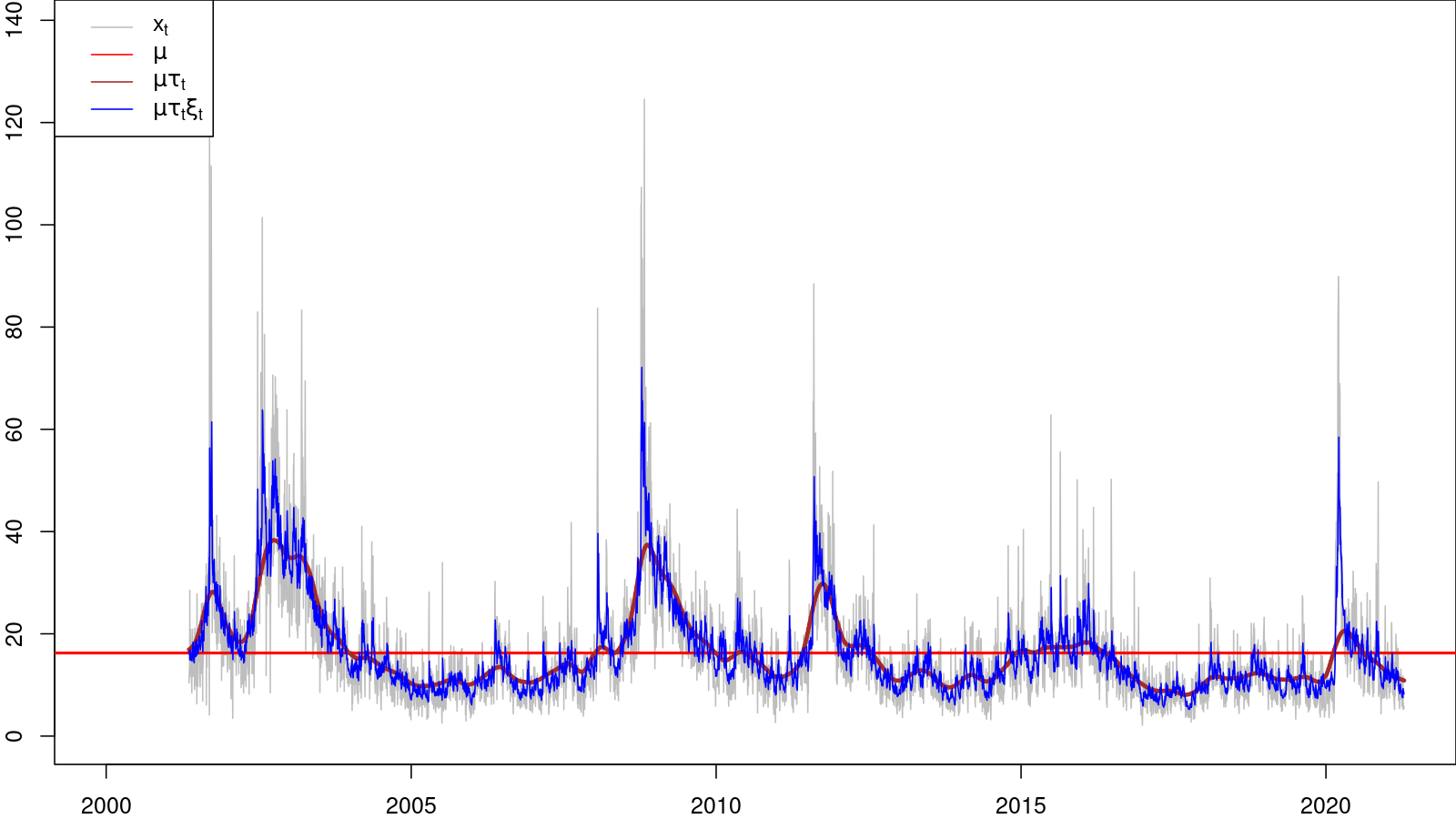}  
		\caption{GDAXI}
		\label{fig:Uni-GDAXI}
	\end{subfigure}
	\begin{subfigure}{.43\textwidth} 
		\centering
		\includegraphics[width=\linewidth]{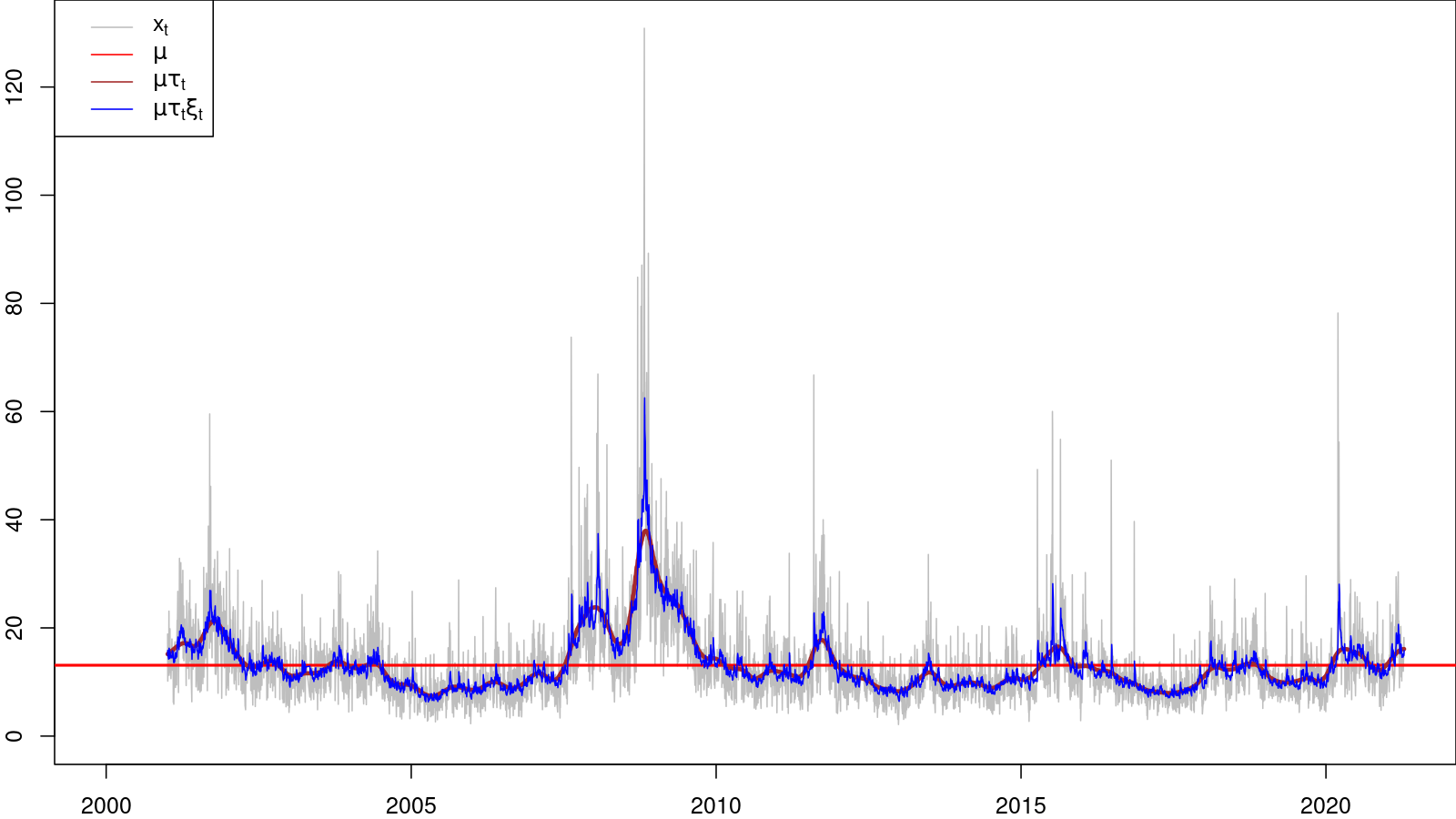}   
		\caption{HSI}
		\label{fig:Uni-HSI}
	\end{subfigure}
	\\
	\centering
	\begin{subfigure}{.43\textwidth} 
		\centering
		\includegraphics[width=\linewidth]{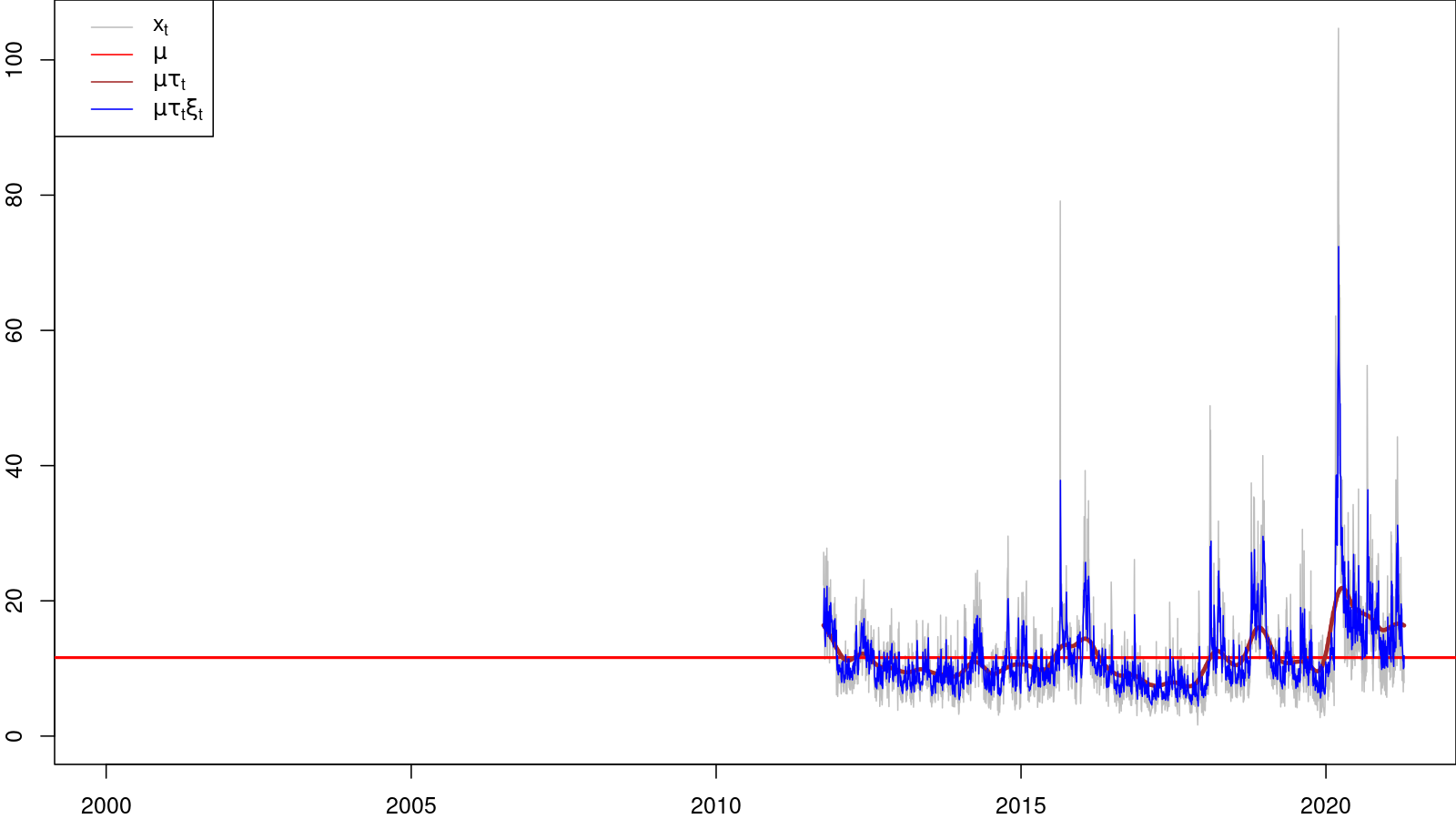}   
		\caption{IXIC}
		\label{fig:Uni-IXIC}
	\end{subfigure}
	\begin{subfigure}{.43\textwidth} 
		\centering
		\includegraphics[width=\linewidth]{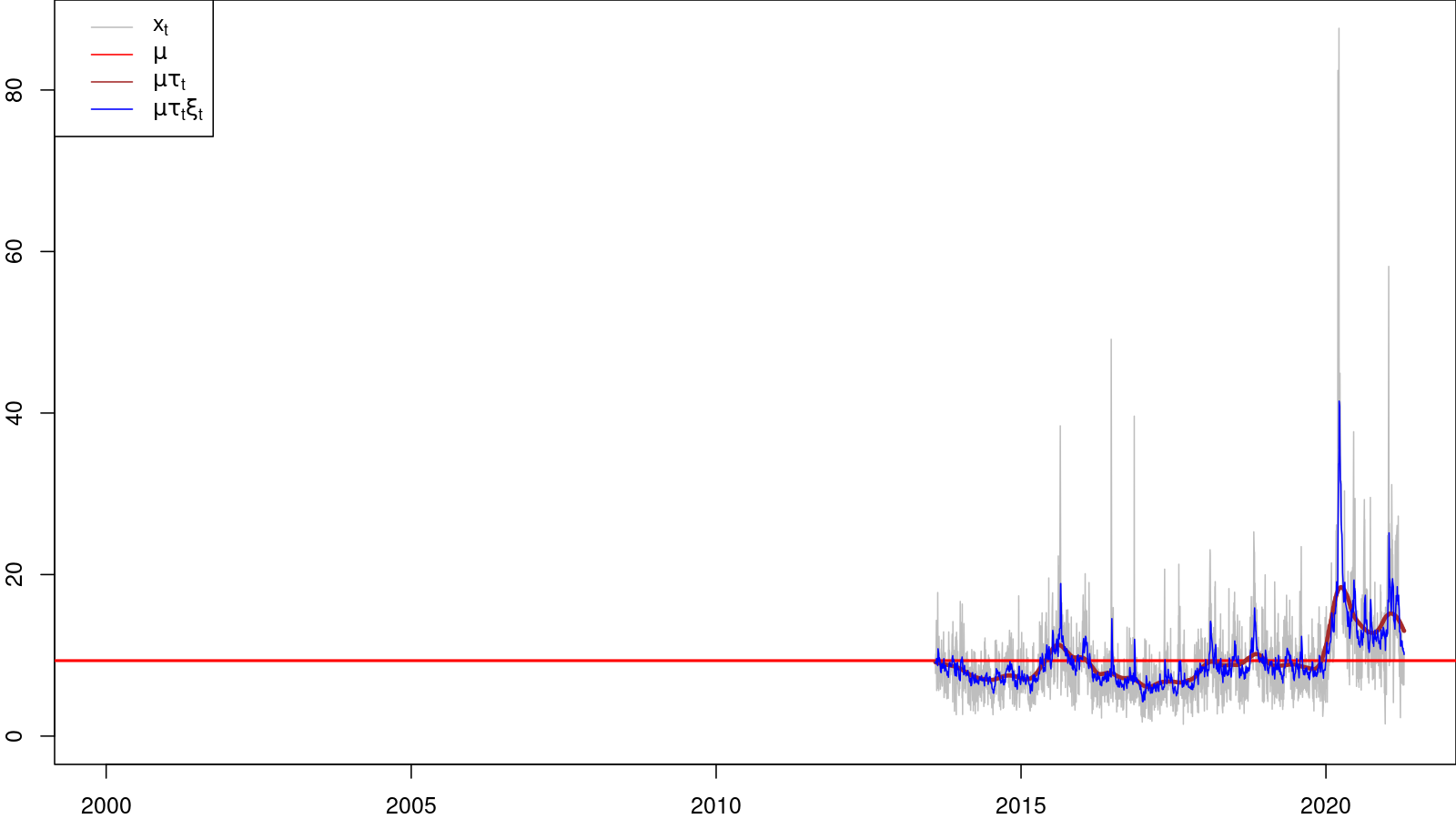}  
		\caption{KS11}
		\label{fig:Uni-KS11}
	\end{subfigure}
	\\
	\centering
	\begin{subfigure}{.43\textwidth}
		\centering
		\includegraphics[width=\linewidth]{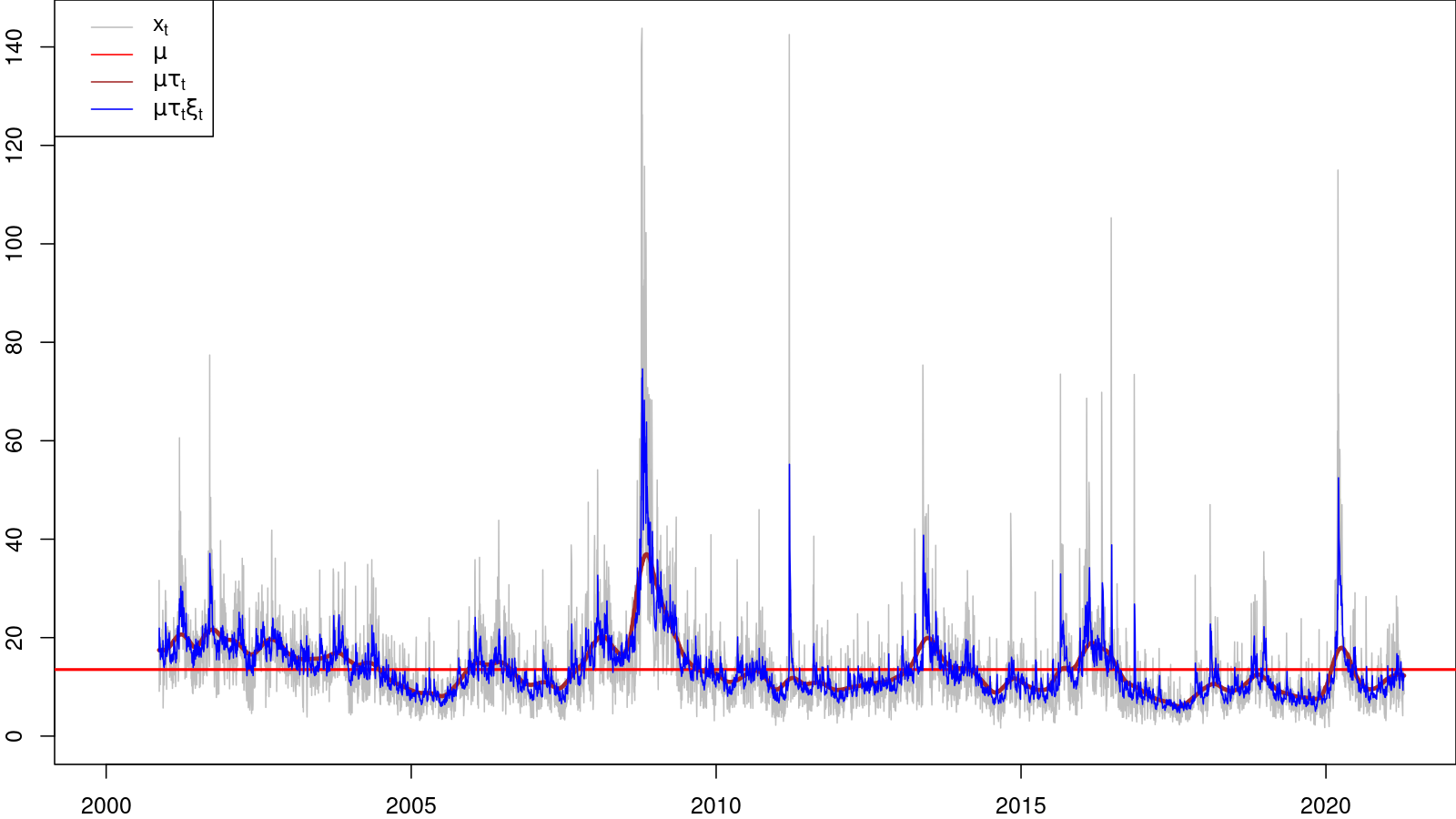}  
		\caption{N225}
		\label{fig:Uni-N225}
	\end{subfigure}
	\begin{subfigure}{.43\textwidth}
		\centering
		\includegraphics[width=\linewidth]{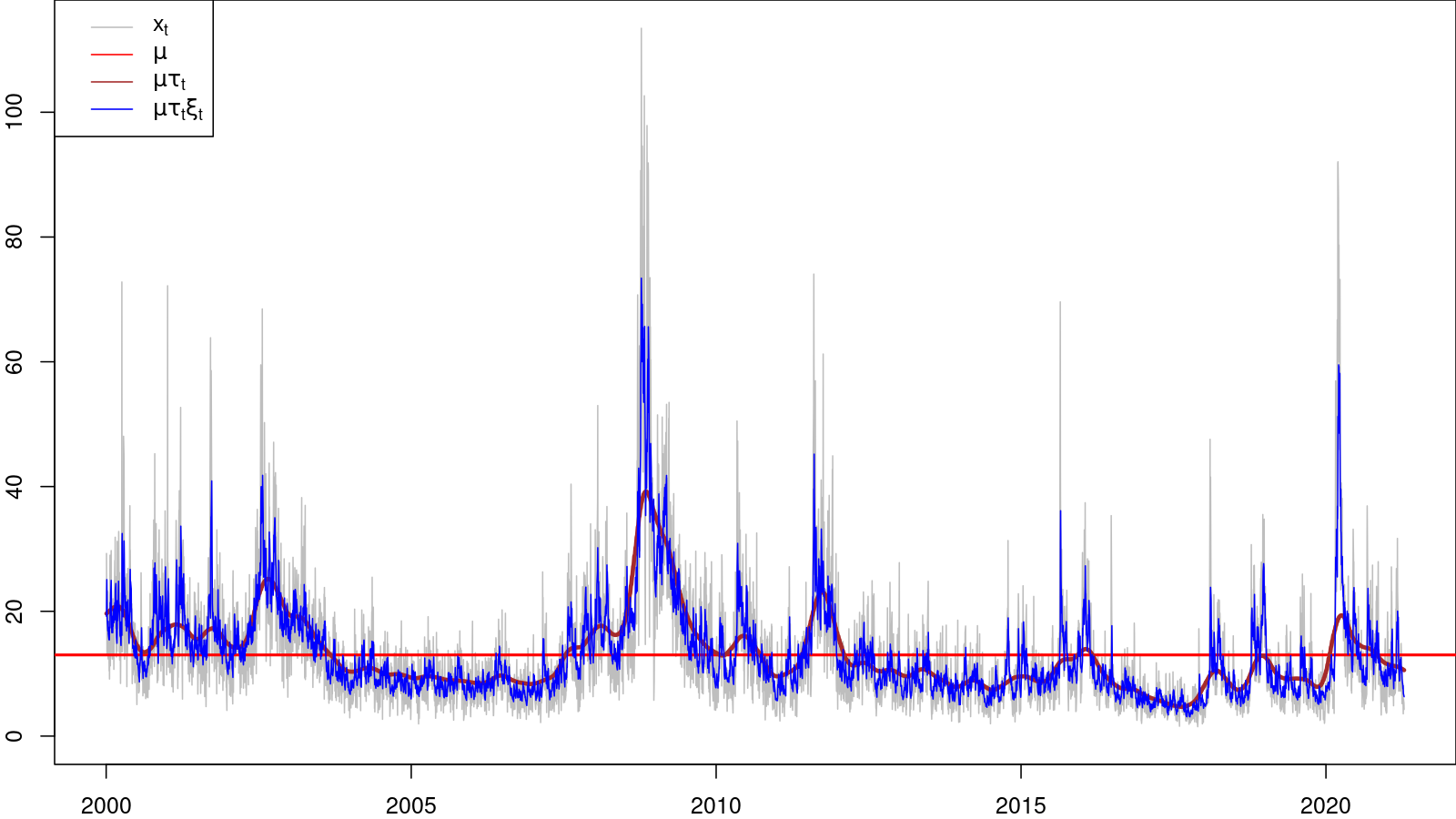}  
		\caption{SPX}
		\label{fig:Uni-SPX}
	\end{subfigure}
	\\
	\centering
	\begin{subfigure}{.43\textwidth}
		\centering
		\includegraphics[width=\linewidth]{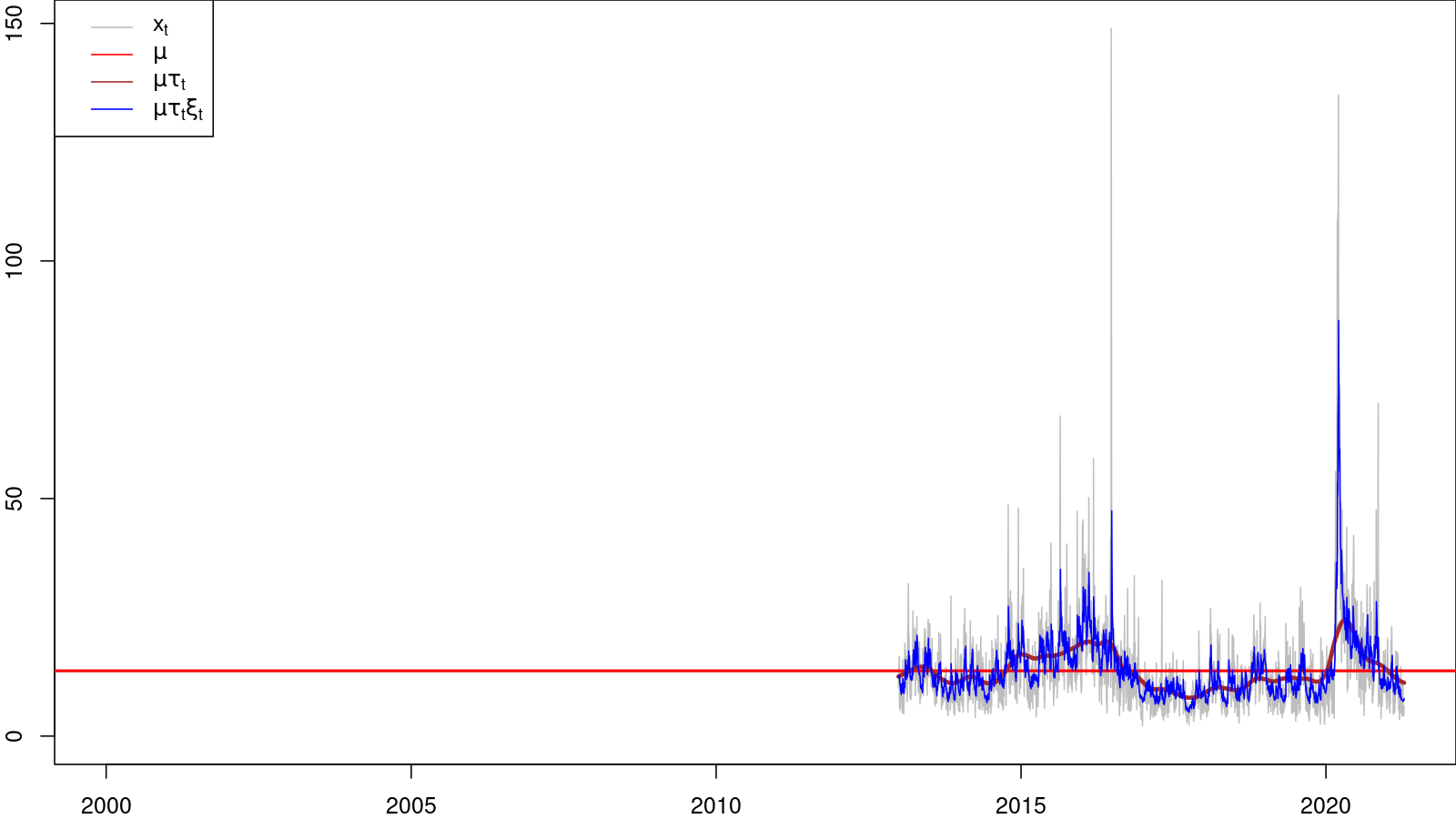}  
		\caption{STOXX50E}
		\label{fig:Uni-STOXX50E}
	\end{subfigure}
\end{figure}

Just to provide further evidence on the features of the \SpMEM\ model relative to the \MEM, we report the behavior of the Autocorrelation Function (ACF), as well as the fitted distribution of the multiplicative residuals with superimposed some known target distributions (Gamma, Log--Normal, Beta$^\prime$, and Log--Logistic, cf. the description in the Appendix), calibrated on the basis of the estimated residual variance.

\begin{figure}
  \caption{Residual correlogram for the rkVol estimated with \SpMEM. 
  	For the sample periods by market, cf. Table \ref{tab:data}.}
  \label{fig:acfSpMEM}
  \centering
  \begin{subfigure}{.43\textwidth}
    \centering
    \includegraphics[width=\linewidth]{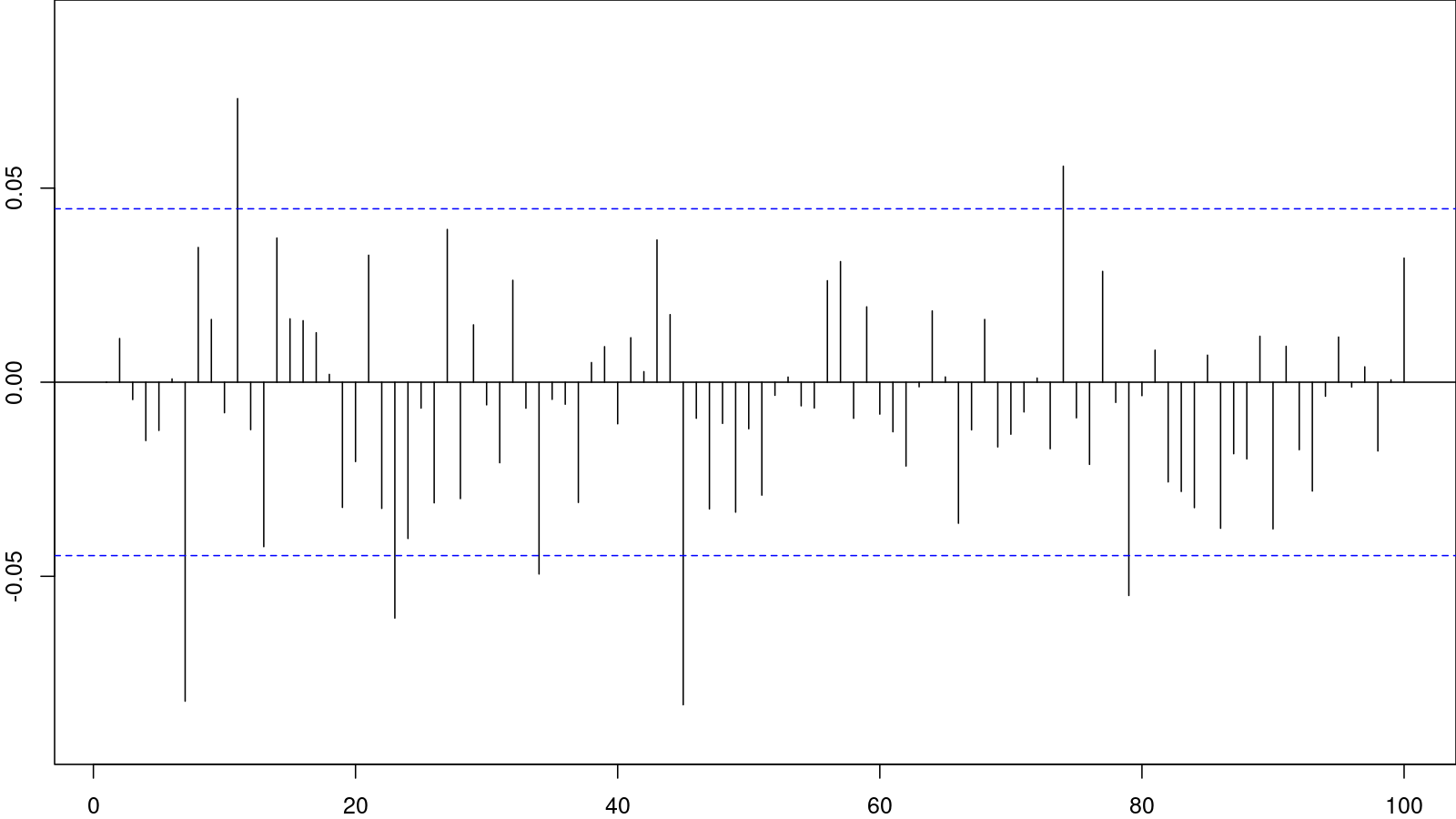}  
    \caption{DJI}
  \end{subfigure}
  \begin{subfigure}{.43\textwidth}
    \centering
    \includegraphics[width=\linewidth]{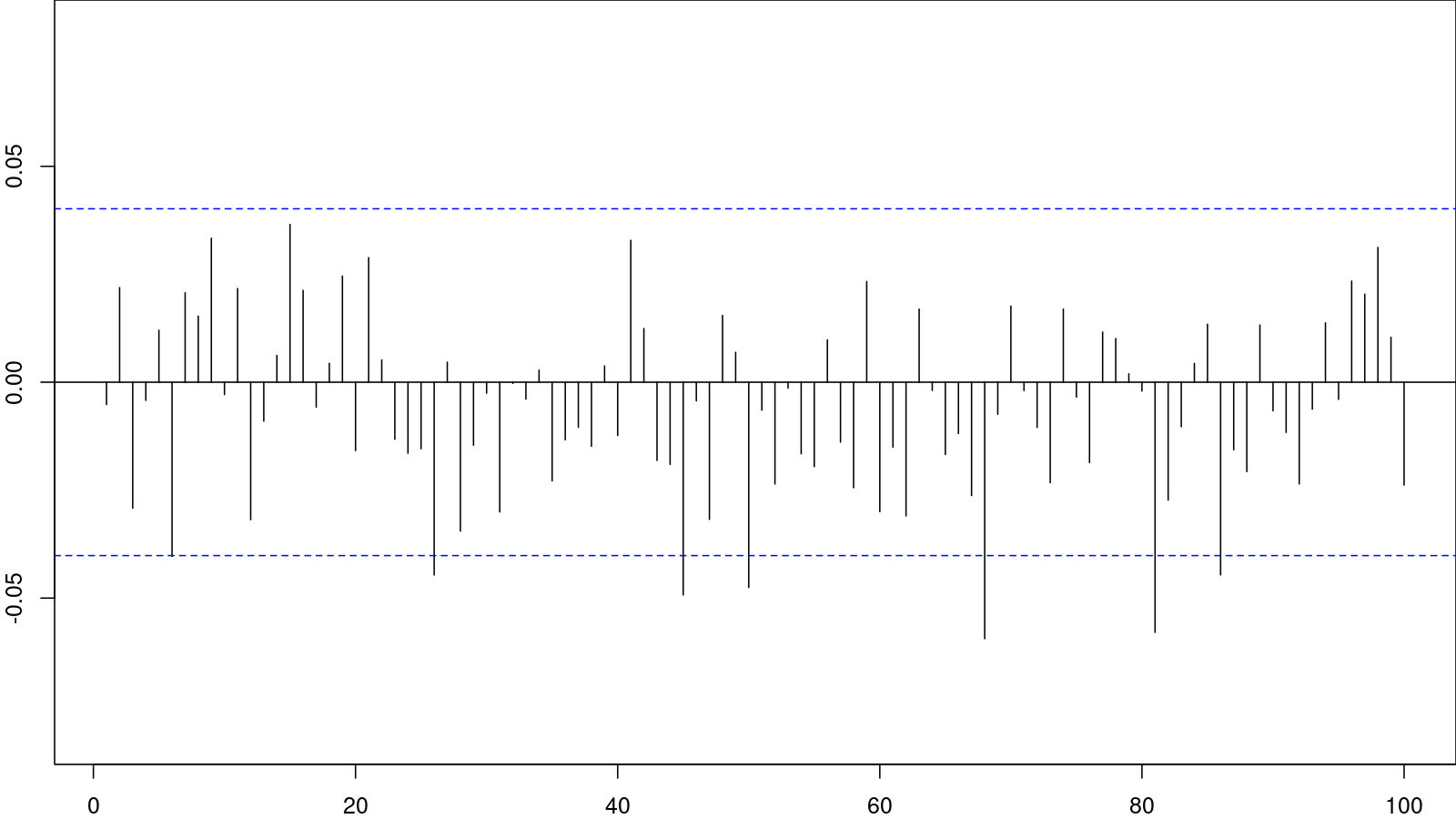}  
    \caption{IXIC}
  \end{subfigure} \\
  \begin{subfigure}{.43\textwidth}
    \centering
    \includegraphics[width=\linewidth]{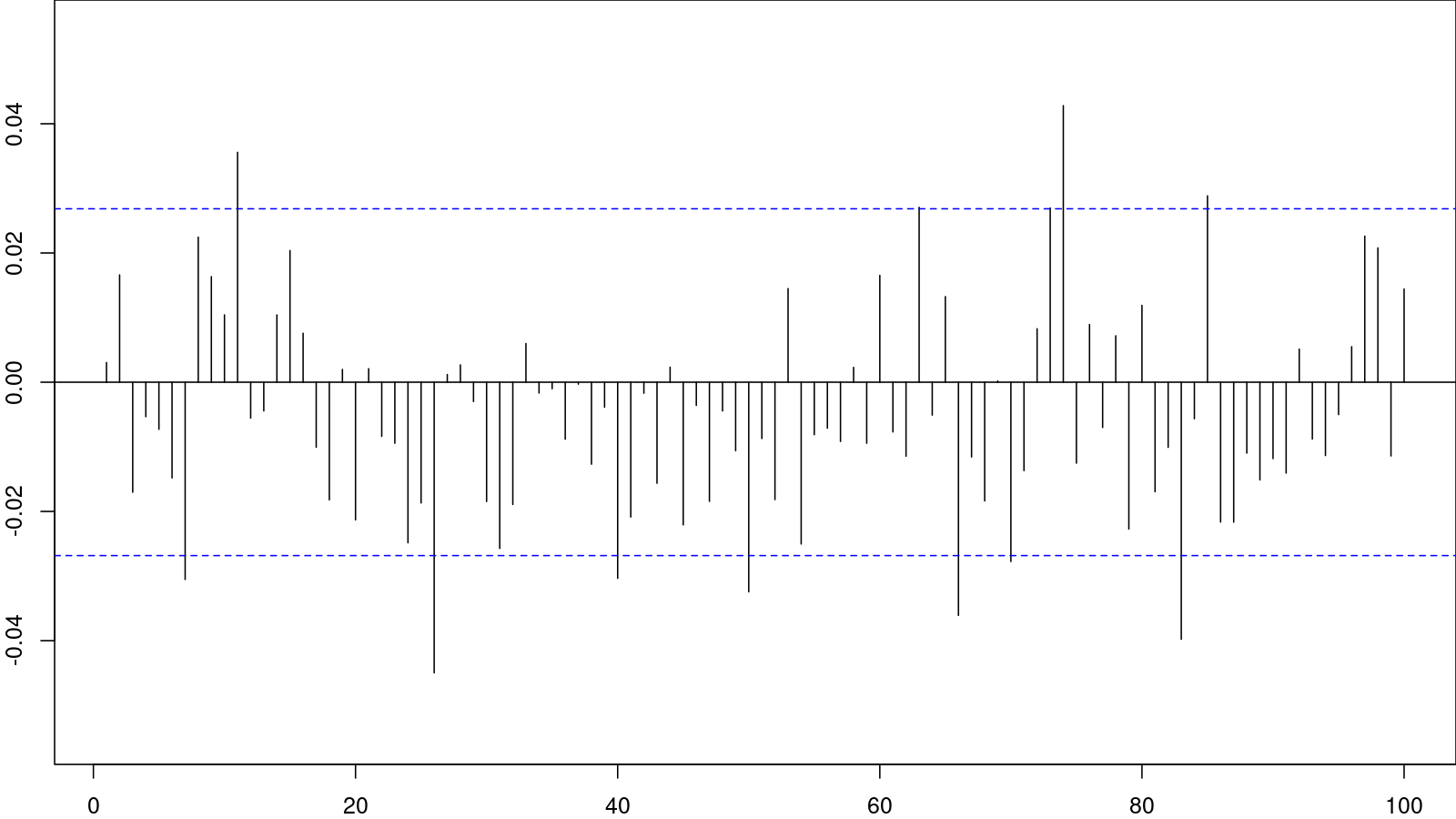}  
    \caption{SPX}
  \end{subfigure}
  \begin{subfigure}{.43\textwidth}
    \centering
    \includegraphics[width=\linewidth]{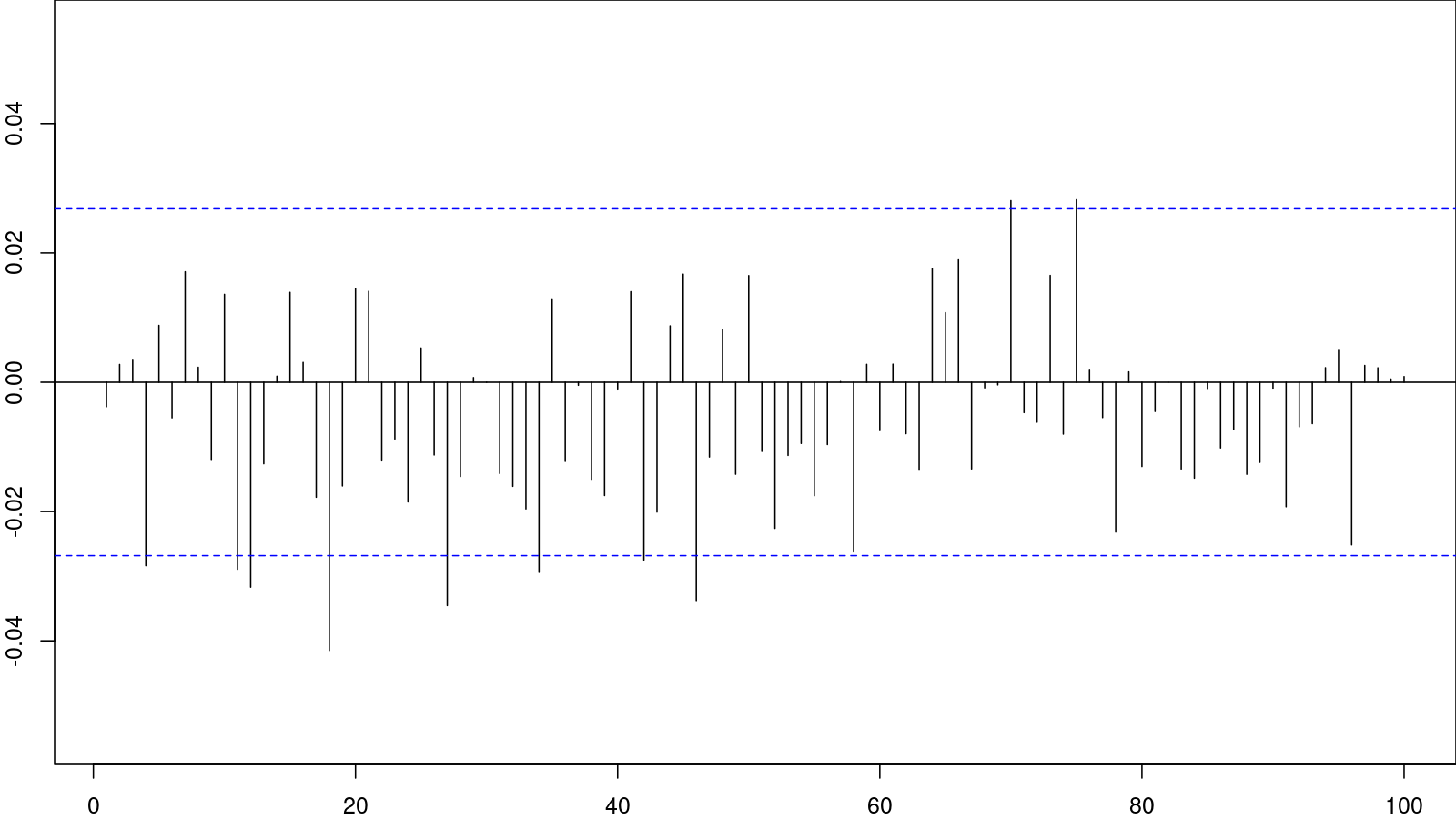}  
    \caption{FCHI}
  \end{subfigure}\\
  \begin{subfigure}{.43\textwidth}
    \centering
    \includegraphics[width=\linewidth]{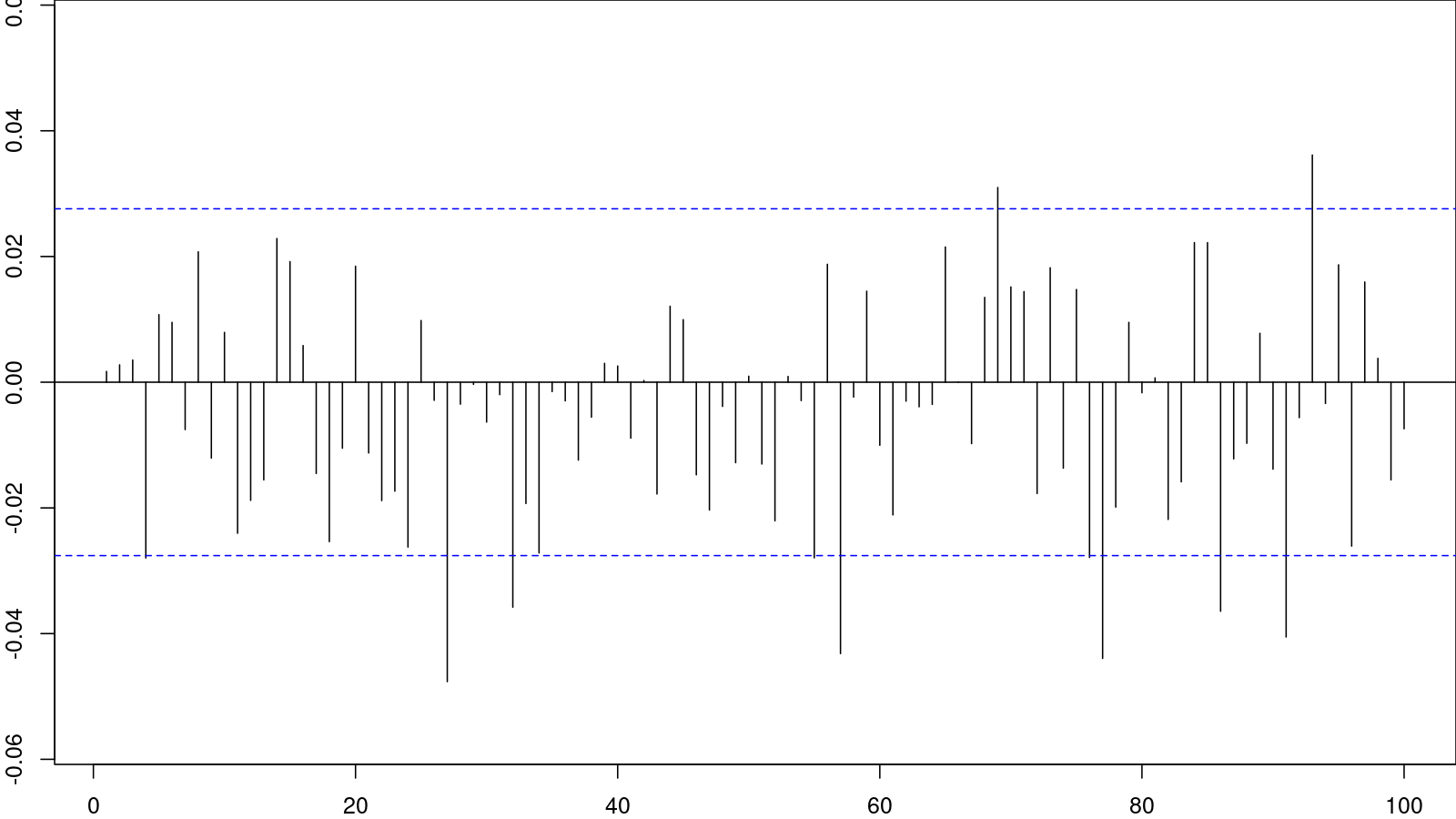}  
    \caption{GDAXI}
  \end{subfigure}
  \begin{subfigure}{.43\textwidth}
	\centering
	\includegraphics[width=\linewidth]{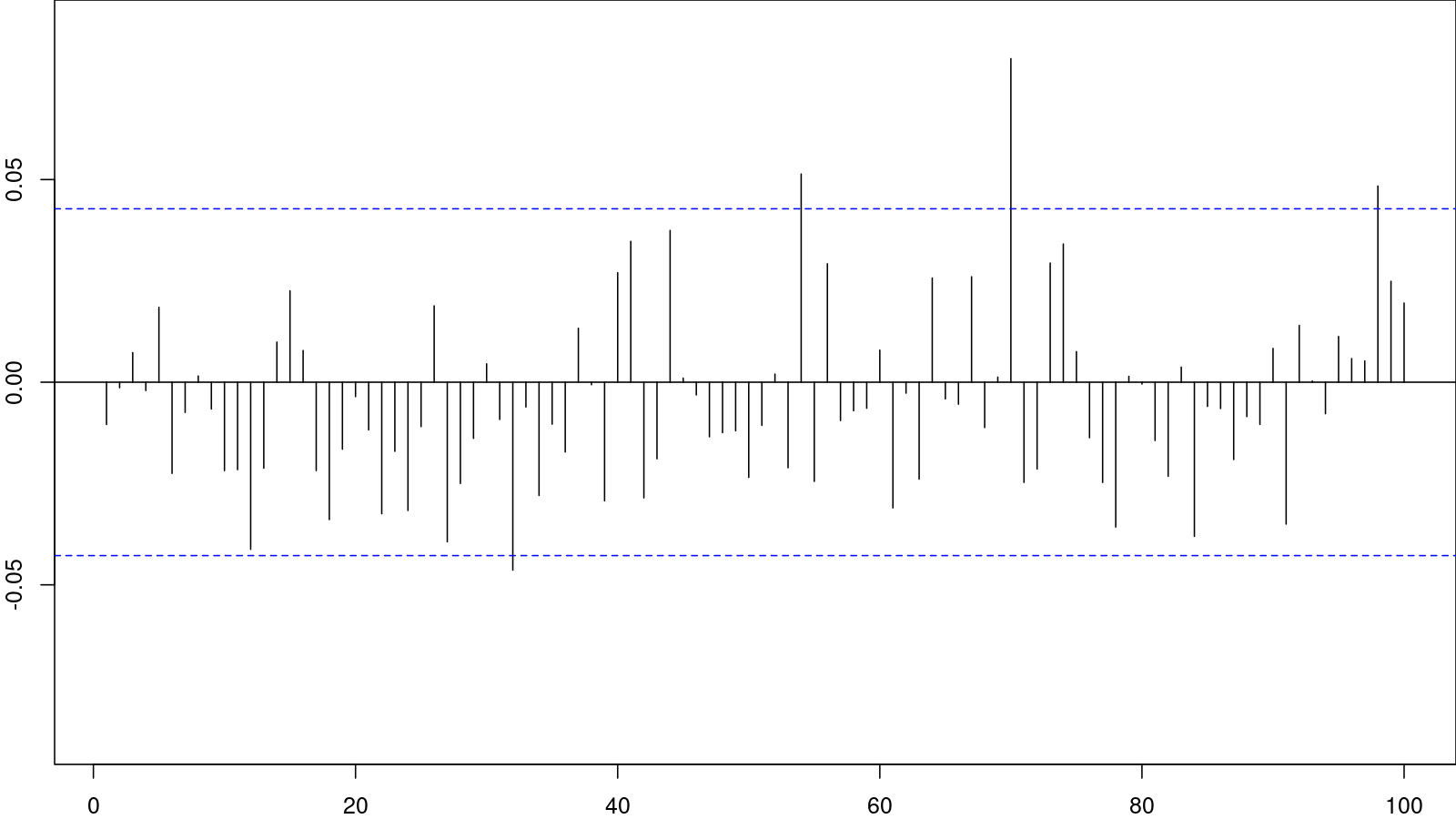}  
	\caption{STOXX50E}
\end{subfigure}\\
  \begin{subfigure}{.43\textwidth}
	\centering
	\includegraphics[width=\linewidth]{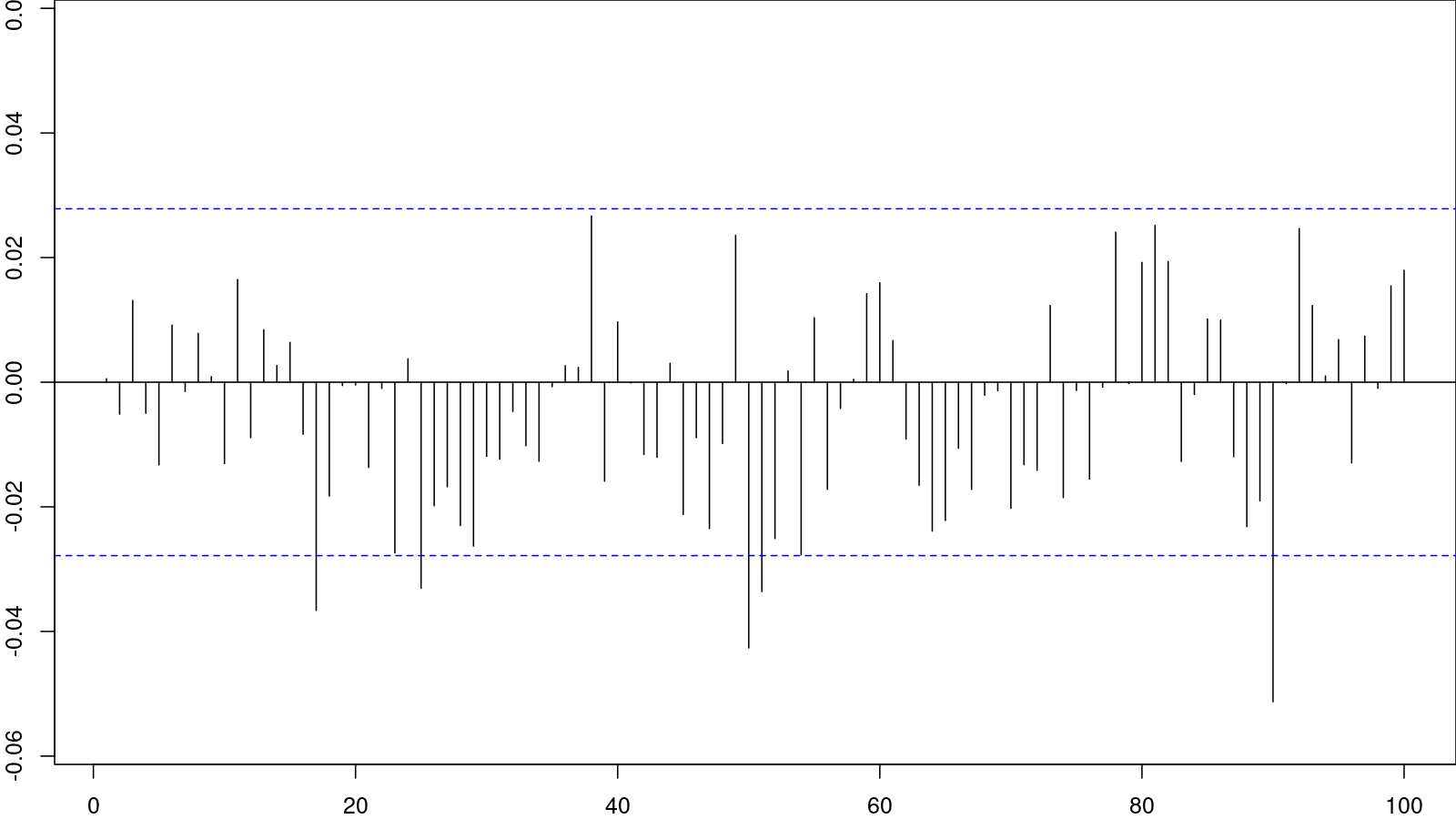}  
	\caption{HSI}
\end{subfigure}
  \begin{subfigure}{.43\textwidth}
	\centering
	\includegraphics[width=\linewidth]{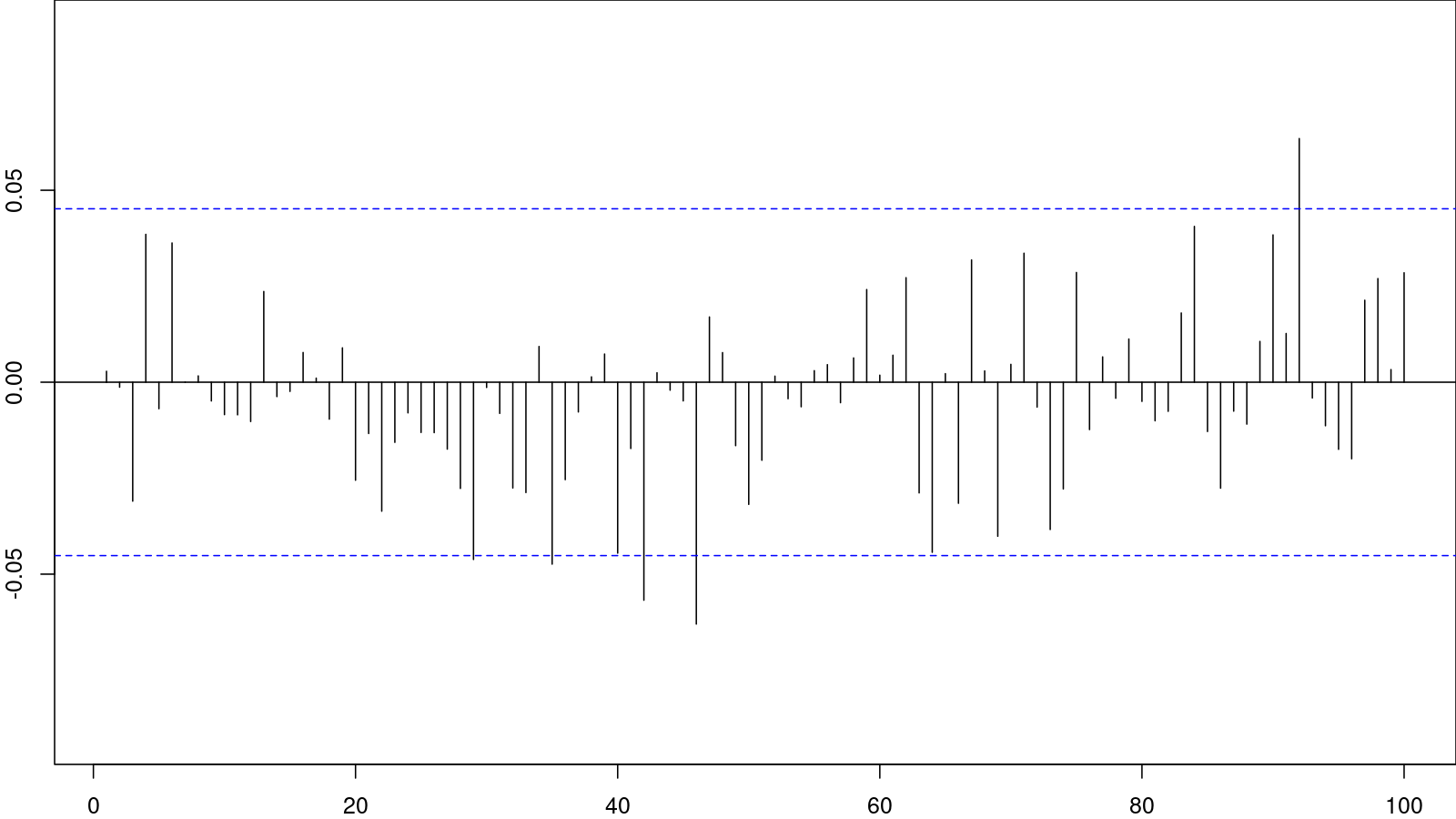}  
	\caption{KS11}
\end{subfigure} 
\\
  \begin{subfigure}{.43\textwidth}
	\centering
	\includegraphics[width=\linewidth]{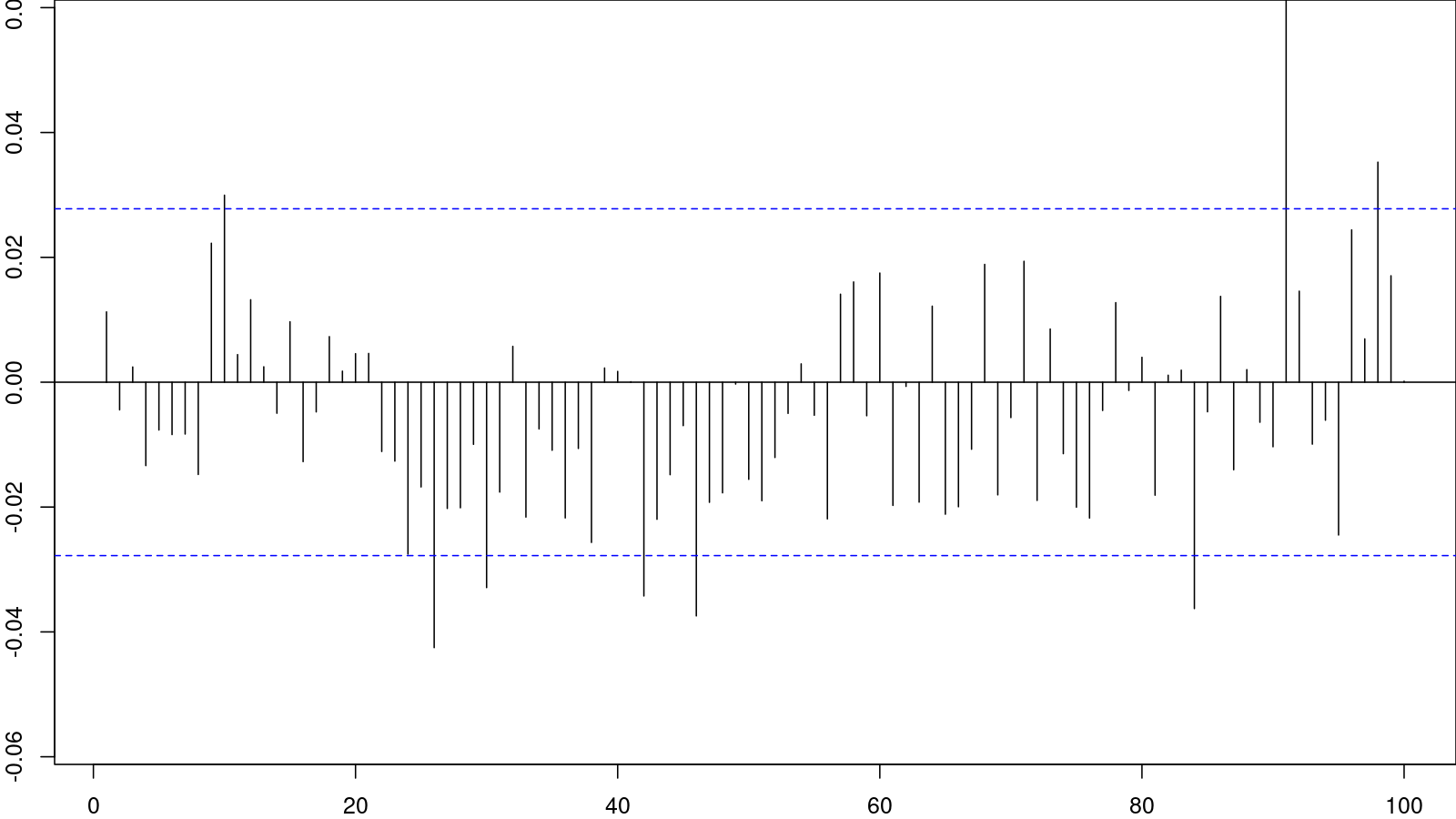}  
	\caption{N225}
\end{subfigure}

\end{figure}

\begin{figure}
  \caption{Histograms of residuals for the rkVol estimated with \SpMEM. 
  	For the sample periods by market, cf. Table \ref{tab:data}.}
  \label{fig:histSpMEM}
  \centering
\begin{subfigure}{.43\textwidth}
	\centering
	\includegraphics[width=\linewidth]{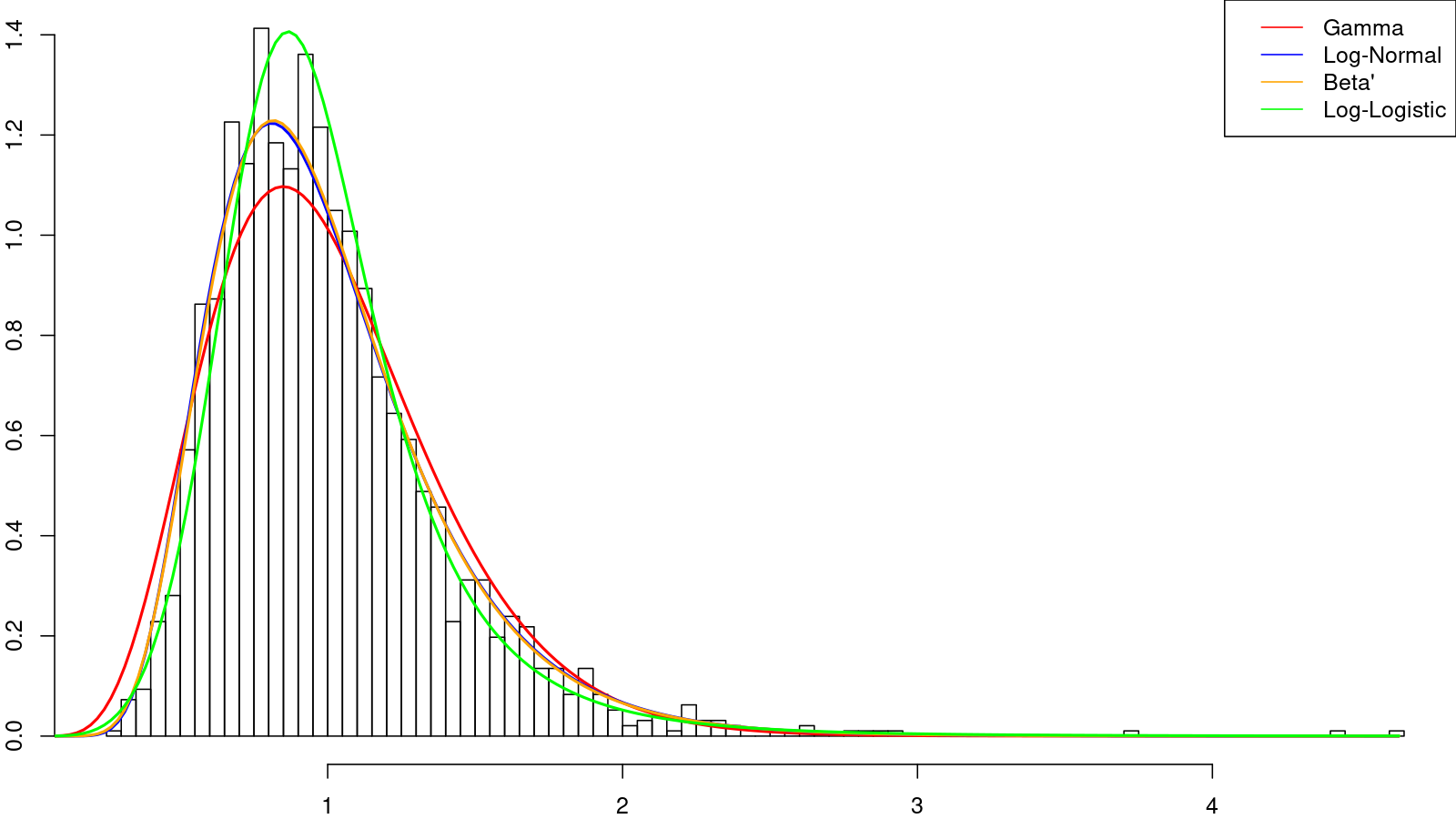}  
	\caption{DJI}
\end{subfigure}
\begin{subfigure}{.43\textwidth}
	\centering
	\includegraphics[width=\linewidth]{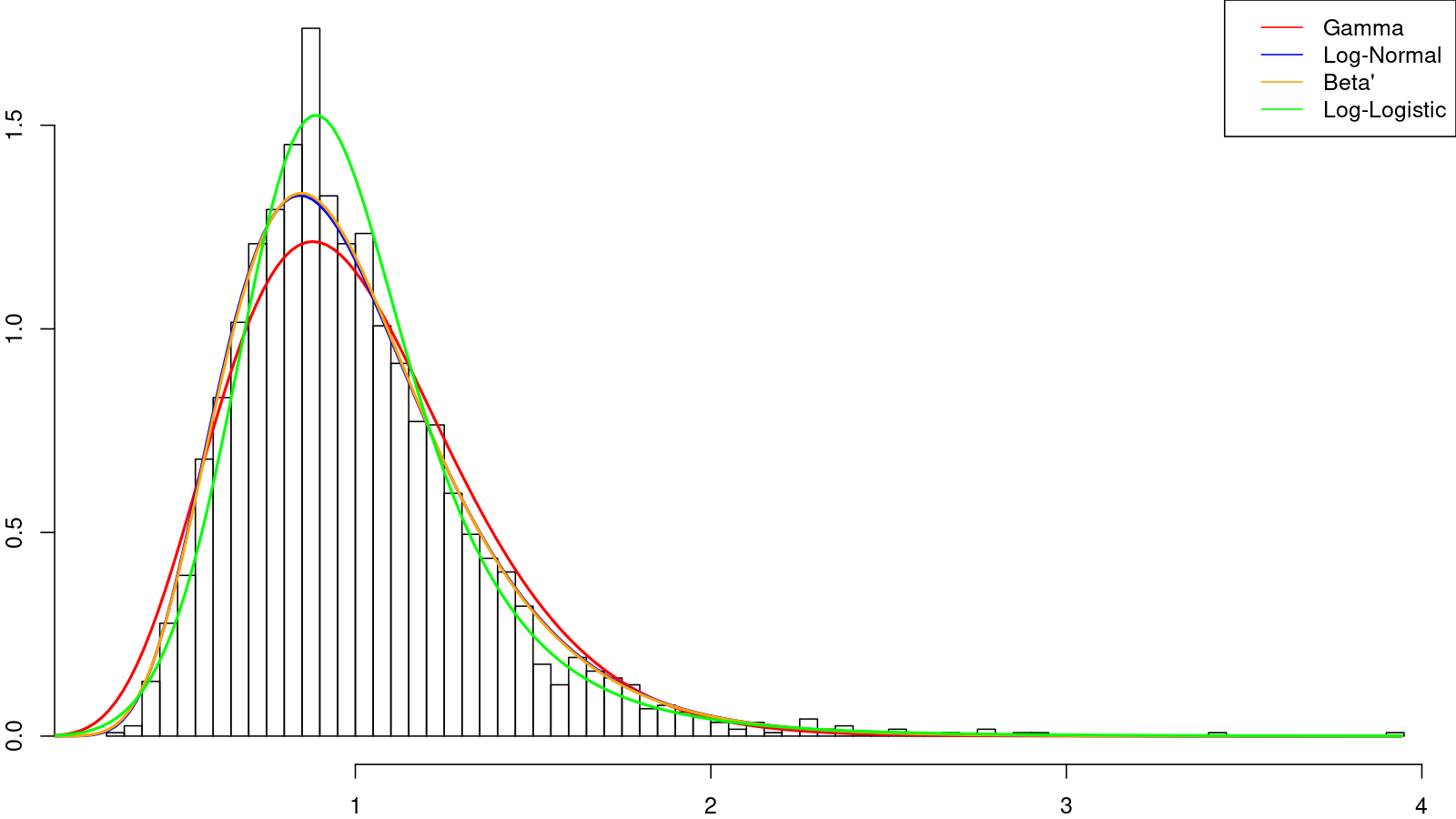}  
	\caption{IXIC}
\end{subfigure}
\\

\begin{subfigure}{.43\textwidth}
	\centering
	\includegraphics[width=\linewidth]{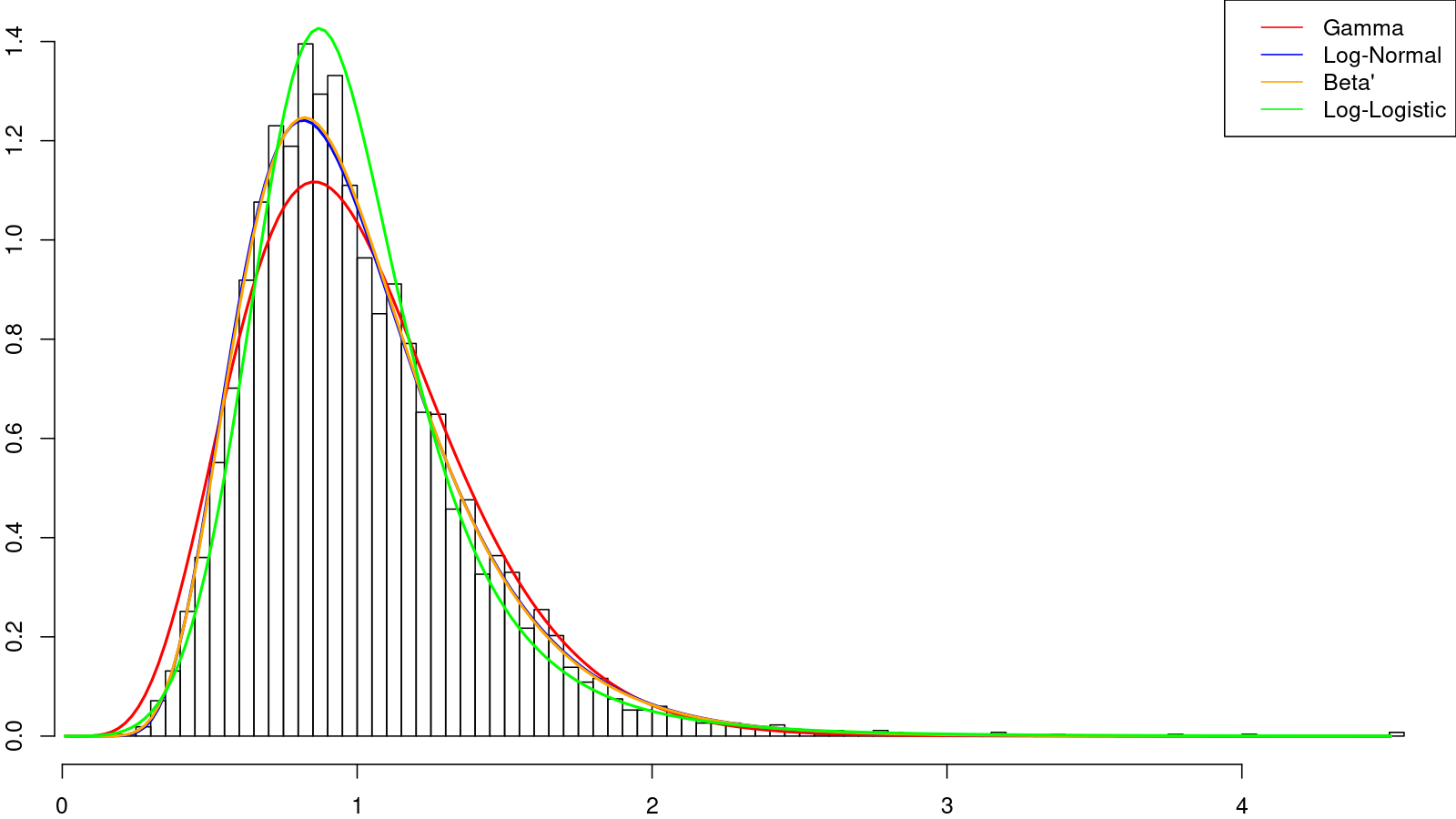}  
	\caption{SPX}
\end{subfigure}
\begin{subfigure}{.43\textwidth}
	\centering
	\includegraphics[width=\linewidth]{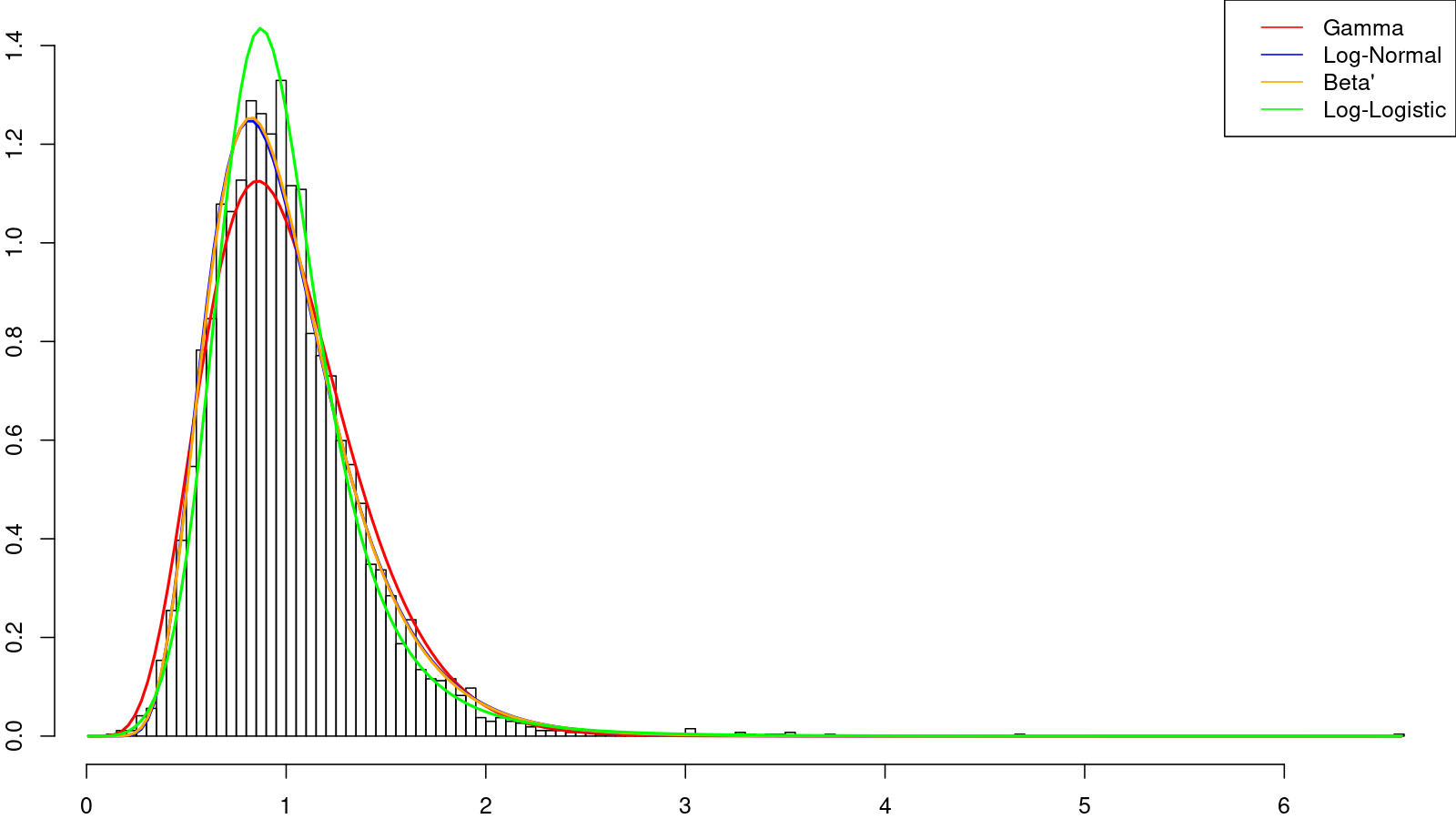}  
	\caption{FCHI}
\end{subfigure}
\\
\begin{subfigure}{.43\textwidth}
	\centering
	\includegraphics[width=\linewidth]{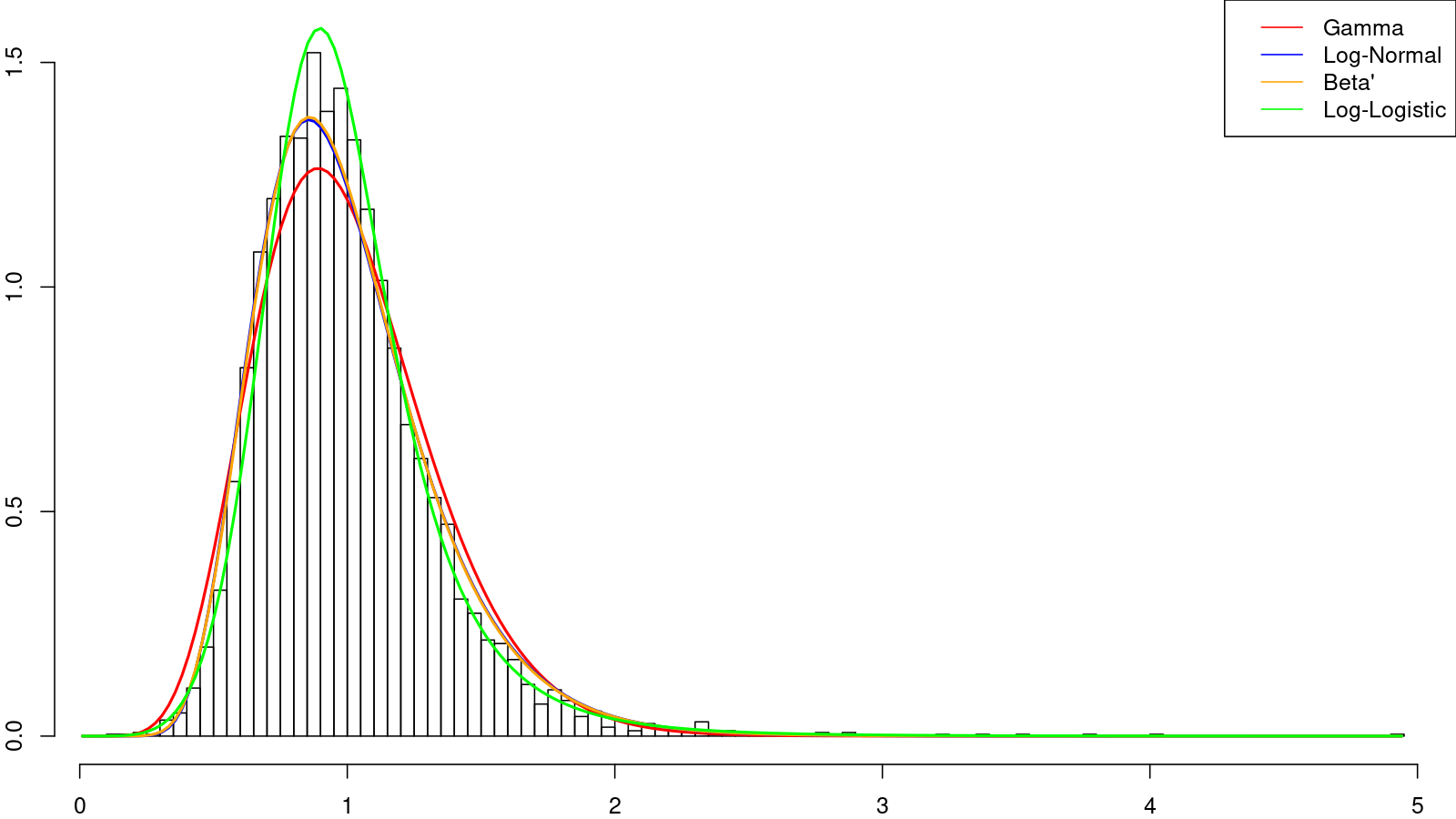}  
	\caption{GDAXI}
\end{subfigure}
  \begin{subfigure}{.43\textwidth}
    \centering
    \includegraphics[width=\linewidth]{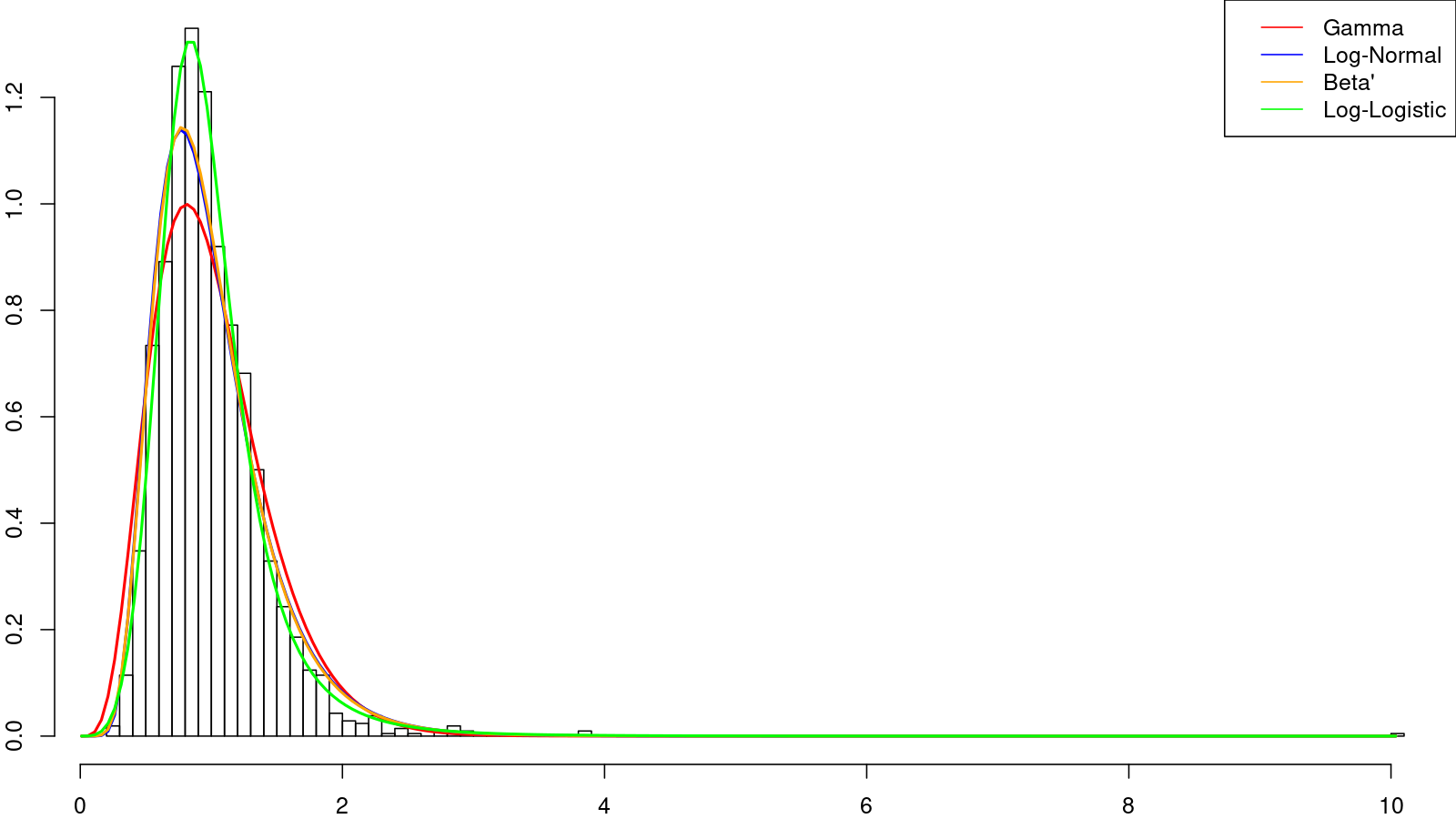}  
    \caption{STOXX50E}
  \end{subfigure}
  \\
  \begin{subfigure}{.43\textwidth}
    \centering
    \includegraphics[width=\linewidth]{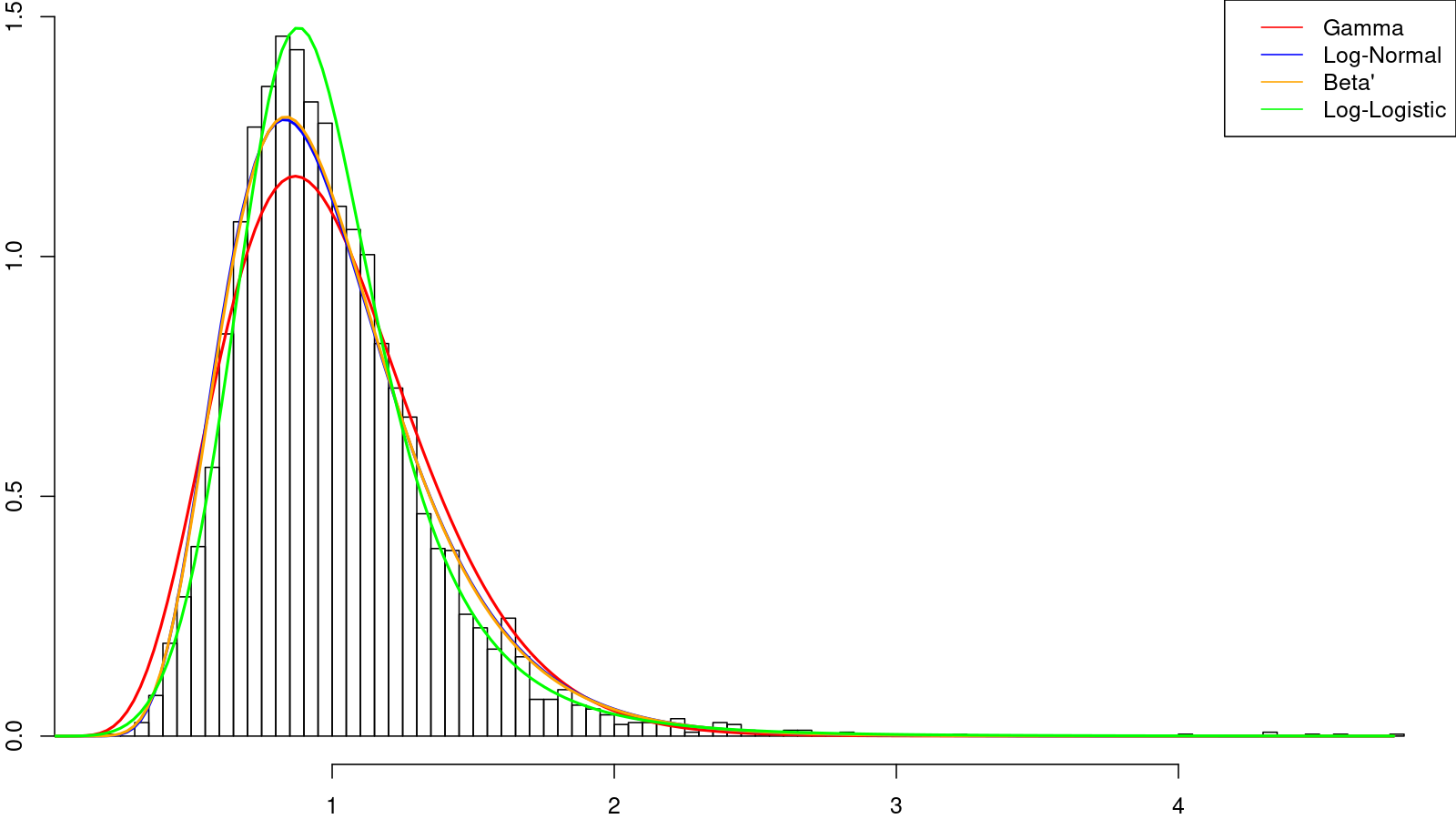}  
    \caption{HSI}
  \end{subfigure}
  \begin{subfigure}{.43\textwidth}
	\centering
	\includegraphics[width=\linewidth]{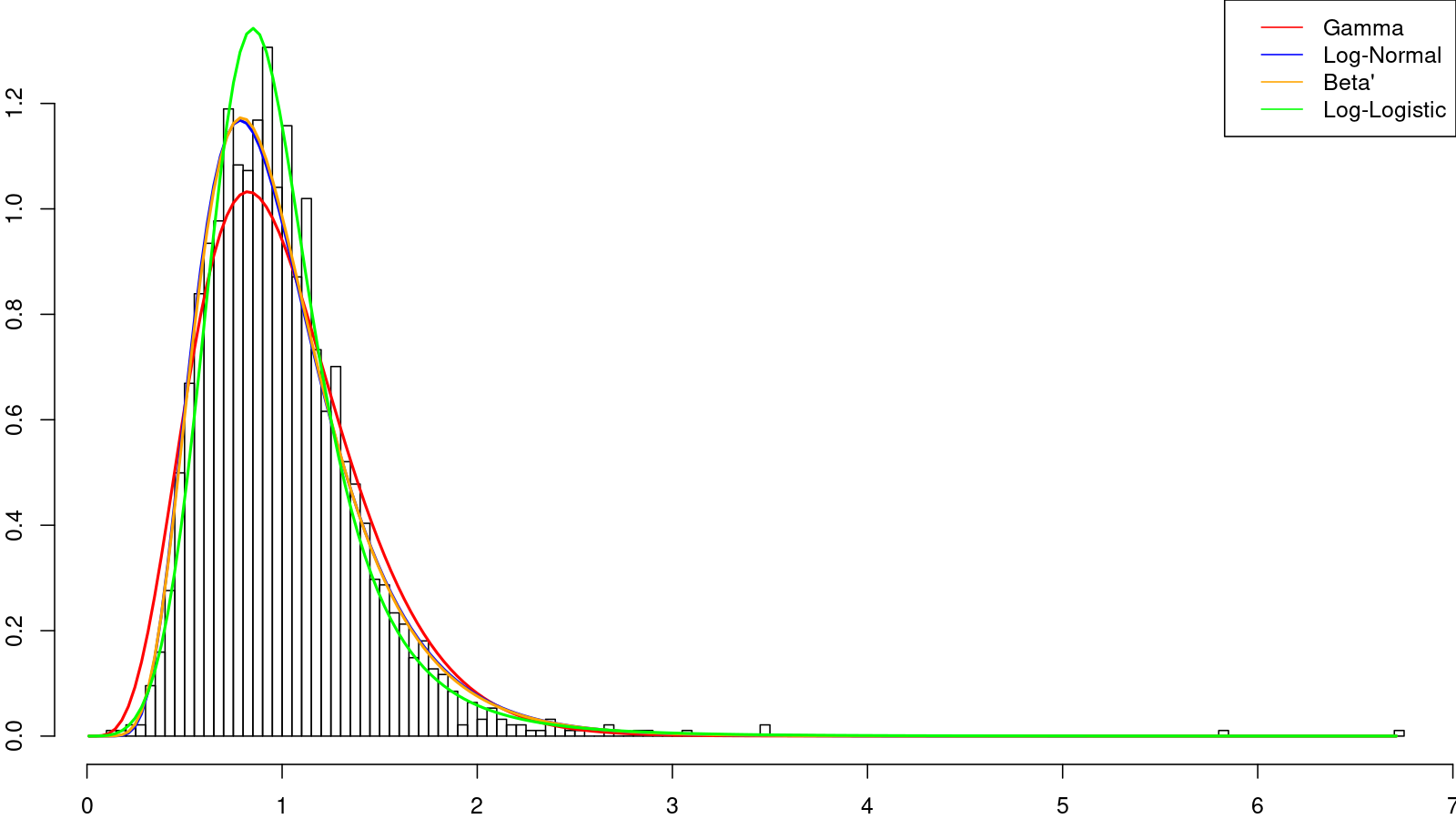}  
	\caption{KS11}
\end{subfigure}  \\
\begin{subfigure}{.43\textwidth}
	\centering
	\includegraphics[width=\linewidth]{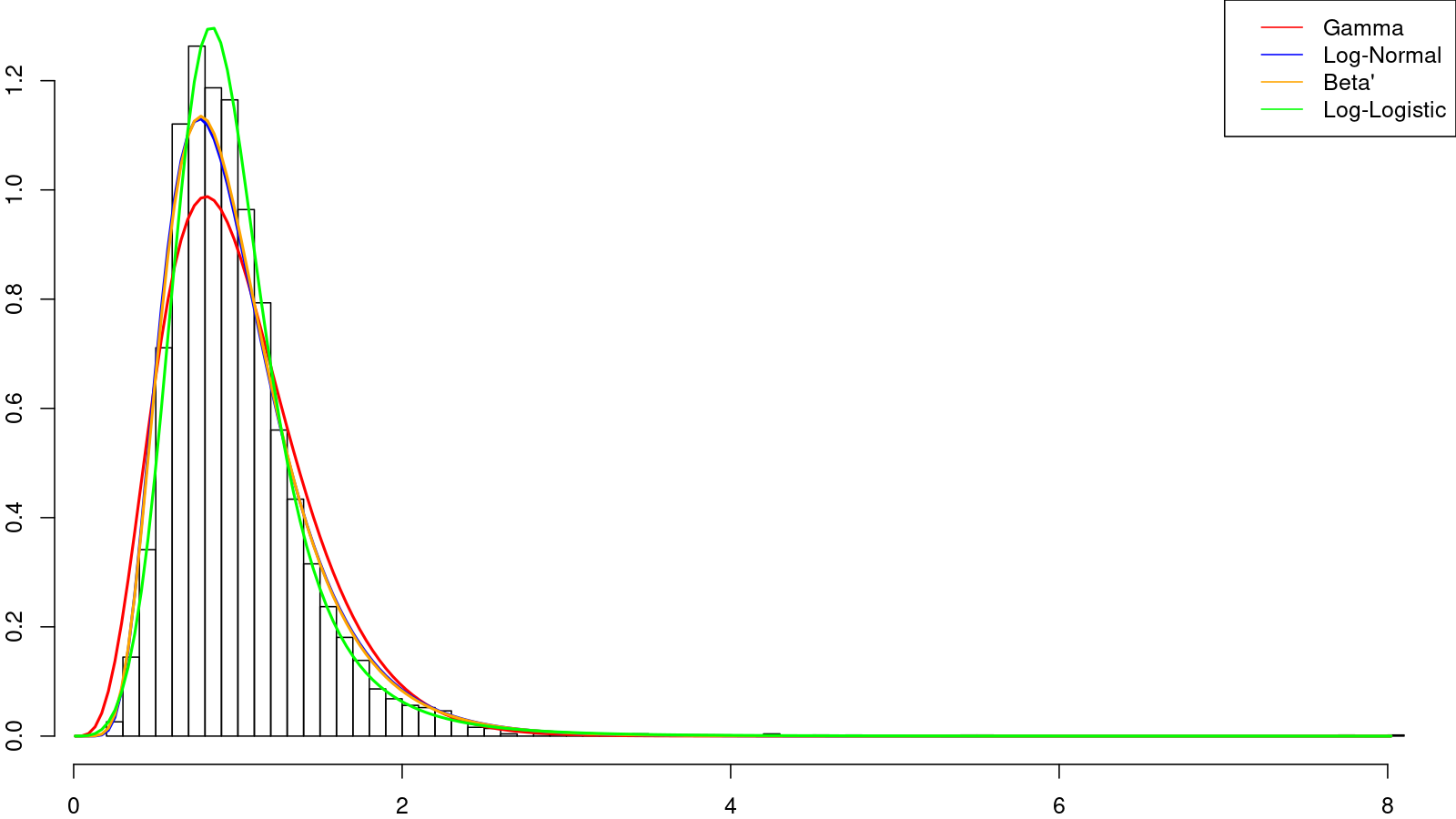}  
	\caption{N225}
\end{subfigure}

\end{figure}

The evidence, shown in the various panels of Figure \ref{fig:acfSpMEM}, indicates that the ACF's do not offer any systematic pattern: there is a low number of significant individual autocorrelations and, when that happens, there is a prevalence of negative over positive values.
As per the fitting of distributions, the first pattern of the varioous panels in Figure \ref{fig:histSpMEM} is that the Gamma seems to systematically miss the peak and hence it overfits the sides of the distribution; by contrast, a generally remarkable performance is had by the Log--Logistic density, which, for all markets manages to reproduce adequately the behavior of the realized kernel volatility residuals. Also we note the almost perfect overlap between the Log--Normal and the Beta$^\prime$, with an intermediate behavior relative to  the other two.

More formally, the distribution fit of these \SpMEM\ residuals can be tested with two popular test statistics, the Anderson--Darling (AD) and Cram\`{e}r-von Mises (CvM) tests, as done in Table \ref{tab:gof} reporting the p--values of the tests, using the Gamma, Log--Normal (Log--N), Beta$^\prime$, Log--Logistic (Log--L) as target densities. While the Gamma density is never supported by the data, the performance of the Log--Logistic is very satisfactory, especially with the CvM test which shows a more conservative behavior, failing to reject the reference null hypothesis in 6 out of 9 cases at 5\%, with an additional case (GDAXI) added at 1\%. The evidence is weaker, but still strong with the AD test. As noted graphically, the performance of the Log--Normal and Beta$^\prime$ densities is somewhat complementary to the Log--L, with p--values higher than 1\% in correspondence to 5 cases out of 9 for both tests. In view of the GMM estimation procedure suggested, this goodness of distributional fit tests help in guiding the \textit{ex post} choice of a parametric distribution, in cases in which \textit{volatility at risk}  \citep[in the sense of][]{Caporin:Rossi:Santucci:2017} evaluation is of interest \citep[for volatility of volatility concerns, see][]{Corsi:Mittnik:Pigorsch:Pigorsch:2008}.

\begin{table}[htbp]
  \centering
  \caption{Realized kernel volatility. P--values of the Anderson--Darling (AD) and Cram\`{e}r-von Mises (CvM) tests for distribution fitting on  \SpMEM\ residuals. Target densities: Gamma, Log--Normal (Log--N), Beta$^\prime$, Log--Logistic (Log--L).}
  \label{tab:gof}
  \input{Table/Tab-spmem-gof.tex}
\end{table}

\section{The \SpvMEM}
\label{sect:SpvMEM}

In view of the fact that financial time series tend to show a similar slow moving pattern (both across measures of market activity and different markets), we extend the model representation of Equation~(\ref{eqn:x(t)-univariate}) to a $K-$dimensional time series with non-negative components $\{\bm{x}_{t}\}$ as
\begin{equation*} 
  \bm{x}_t = \underbrace{\tau_t \; \bm{\mu} \odot  \bm{\xi}_t}_{\bm{\mu}_t} \odot \bm{\varepsilon}_t,
\end{equation*}
where $\odot$ denotes the element--by--element product, and
\begin{itemize}
  \item $\tau_{t}$ is a scalar slow-moving component satisfying $E\left( \tau_{t}\right) = 1$, a common synthesis of low--frequency features of the $K$ series, possessing a smooth behavior, similarly to what was done in Section~\ref{sect:spmem}, in the univariate case. Also in this case, when $\tau_t=1, \ \forall t$, we are in the base \vMEM\ of \cite{Cipollini:Engle:Gallo:2013}. 
  The semi--nonparametric vector MEM \SpvMEM\ was derived in \citet{Barigozzi:Brownlees:Gallo:Veredas:2014}, extending the \citet{Veredas:Rodriguez:Espasa:2007} approach illustrated in Section~\ref{sect:Estimation-spmem} to the multivariate case: here it will be  extended in ways to be seen in Section \ref{sect:Inference};
  \item $\bm{\mu}$ is the $K \times 1$ vector of the unconditional means;
  \item $\bm{\xi}_{t}$ is a $K \times 1$ vector of interconnected short-run components satisfying $E\left(\bm{\xi}_{t}\right) = \mathds{1}$.
  More specifically, the $K$-dimensional counterpart of Equation~(\ref{eqn:xi-1}) is 
  \begin{equation}
    \label{eqn:xi-vec}
    \bm{\xi}_t = \left[ \mathds{1} - \left( \bm{\beta}_1 + \bm{\alpha}_1 + \frac{\bm{\gamma}_1}{2} \right) \mathds{1} \right] + 
    \bm{\beta}_1 \bm{\xi}_{t-1} + \bm{\alpha}_1 \bm{x}_{t-1}^{(\xi)} + \bm{\gamma}_1 \bm{x}_{t-1}^{(\xi-)},  
  \end{equation}
  where
  \begin{equation*}
    x_{t,i}^{(\xi)} = \frac{x_{t,i}}{\mu_{i} \tau_t}
    \qquad{}
    x_{t,i}^{(\xi-)} = x_{t,i}^{(\xi)} \mathbbm{1}^{-}_{t}
    \qquad{}
    i = 1, \ldots, K
  \end{equation*}
  and $\bm{\beta}_1$, $\bm{\alpha}_1$ and $\bm{\gamma}_1$ are in general $K \times K$ matrices.
 In the base formulation, $\bm{\beta}_1$  and $\bm{\gamma}_1$ are diagonal, but $\bm{\alpha}_1$ is allowed to be full in order to capture interdependency structures.
  Paralleling (\ref{eqn:xi-2}), an equivalent specification is
  \begin{equation}
    \label{eqn:xi-v}
    \bm{\xi}_t = (\mathds{1} - \bm{\beta}_1^{*} \mathds{1}) + \bm{\beta}_1^{*} \bm{\xi}_{t-1} + \bm{\alpha}_1 \bm{v}_{t-1} + \bm{\gamma}_1 \bm{v}^{(-)}_{t-1},
  \end{equation}
  where
  \begin{equation*}
    v_{t, i} = x^{(\xi)}_{t, i} - \xi_{t, i},
    \qquad{}
    v_{t, i}^{(-)} = x^{(\xi-)}_{t, i} - \frac{\xi_{t, i}}{2},
    \qquad{}
    i = 1, \ldots, K
  \end{equation*}
  and $\bm{\beta}_1^{*} = \bm{\beta}_1 + \bm{\alpha}_1 + \bm{\gamma}_1 / 2$; 
  
  \item $\bm{\varepsilon}_{t}$ is a $K \times 1 $ conditionally homoskedastic unit mean error term,
  \begin{equation*}
    \bm{\varepsilon}_t \overset{\iid}{\sim} d^{+}(\mathds{1}, \bm{\Sigma});
  \end{equation*}
  the actual form of the multivariate distribution $d^+$ is left unspecified, with the proviso that it is not straighforward to find a suitable parametric specification. As a matter of fact, as pointed out by \cite{Cipollini:Engle:Gallo:2017}, one could resort to a copula function--based solution (more details provided below) or a multivariate Log--Normal, later used also in \citet{Cattivelli:Gallo:2020}.
\end{itemize}

Mirroring the univariate case, under the above conditions, the conditional expectation of $\bm{x}_{t}$ is $\bm{\mu}_{t} = \tau_{t} \bm{\mu} \odot \bm{\xi}_{t}$. 

In the empirical application in the multivariate context below, for each market described in Table \ref{tab:data}, we use the three measures of volatility observed at a daily frequency introduced before: the absolute return, the realized kernel volatility, and a synthetic index derived from implied volatilities in option prices. For illustration purposes, one can refer to Figure~\ref{fig:DJI-3}, where the three series are reproduced for the Dow Jones Index: in each panel we have the observed series $i$ in light grey, its unconditional mean $\mu_i$ as a flat line, the product of $\tau_t\mu_i$ as a smooth line and then the full conditional expectation $\tau_{t}\mu_{i}\xi_{t,i}$ in blue. The reported estimation follows the \SpvMEM\ specification.

\begin{figure}
  \caption{ \SpvMEM\ estimated on DJI absolute returns (arVol -- panel (a)), realized kernel volatility (rkVol -- panel(b)), and implied volatility (impVol -- panel(c)).
  The overall unconditional mean $\bm{\mu}$ is a flat line; the thicker red curve is the estimated (with $h = 3$ months) $\bm{\mu} \tau_t$  and the estimated conditional expectation $\tau_t \bm{\mu} \bm{\xi}_t$ is reproduced in blue. The observed series is in the background in a shade of grey. For sample periods refer to Table \ref{tab:data}.}
  \label{fig:DJI-3}
  \begin{subfigure}{.49\textwidth}
    \centering
    \includegraphics[width=\linewidth]{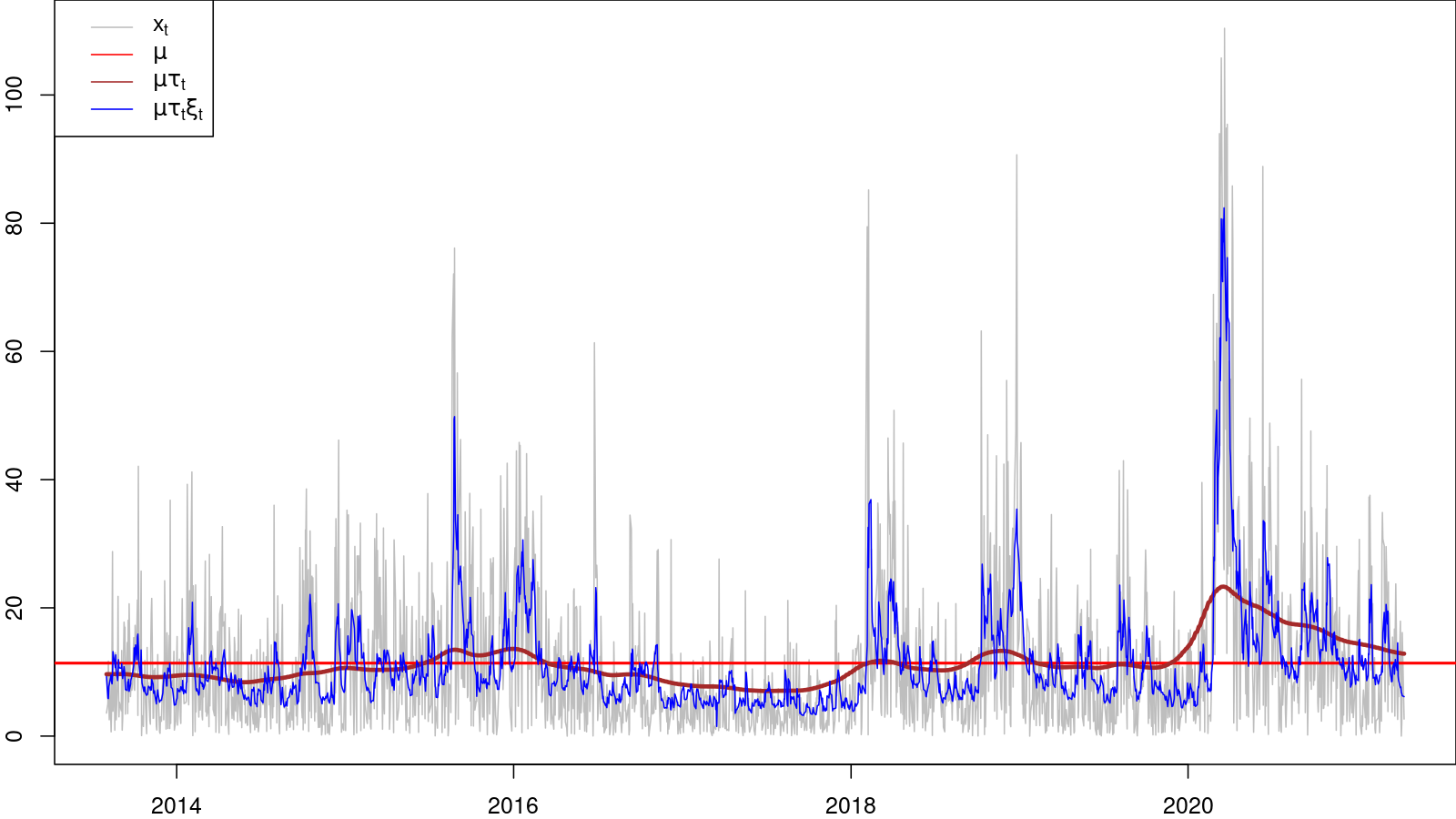}  
    \caption{arVol.}
    \label{fig:DJI-ar}
  \end{subfigure}
  \begin{subfigure}{.49\textwidth}
    \centering
    \includegraphics[width=\linewidth]{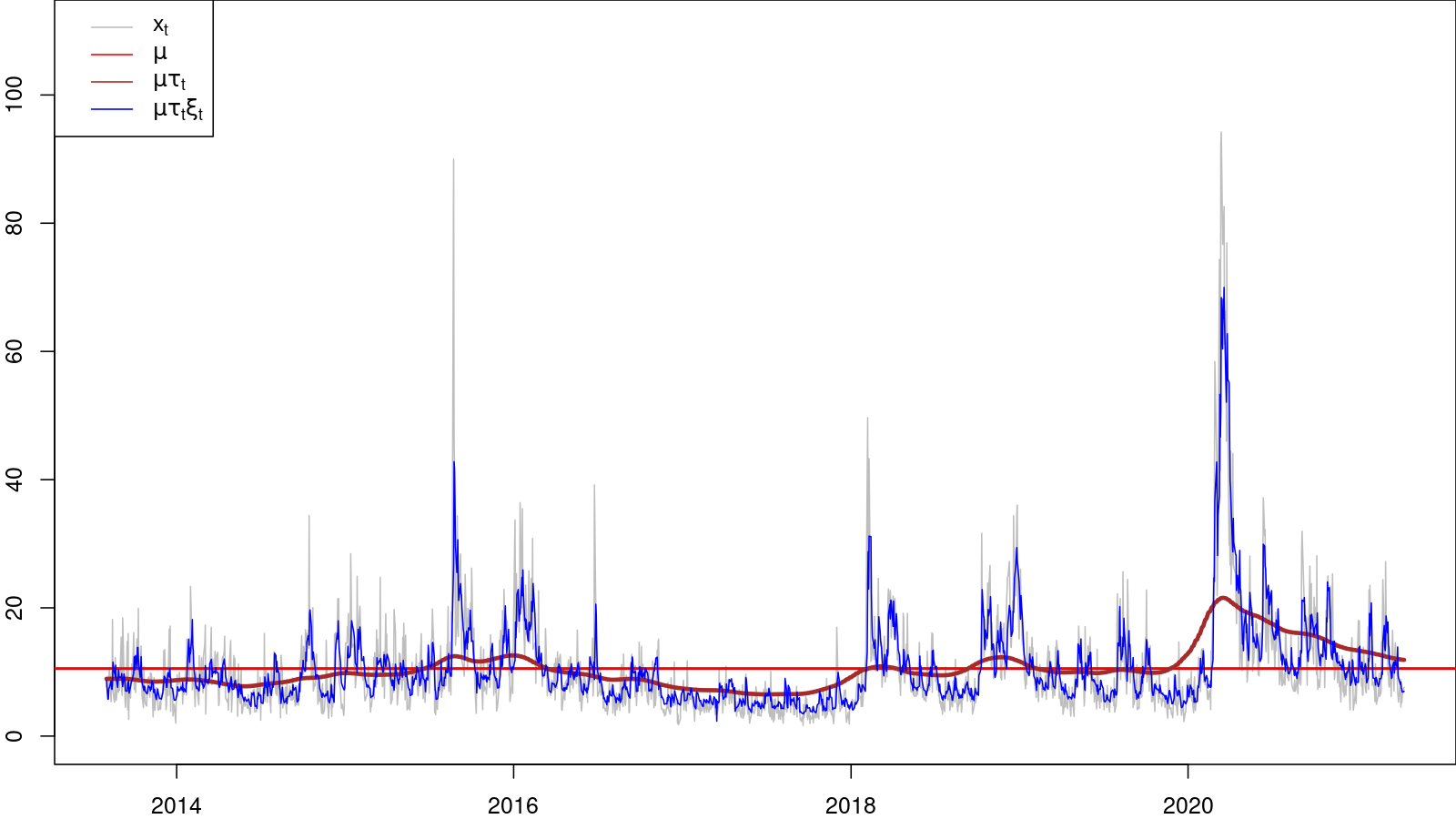}  
    \caption{rkVol.}
    \label{fig:DJI-rk}
  \end{subfigure}
  \\
  \centering
  \begin{subfigure}{.49\textwidth}
    \centering
    \includegraphics[width=\linewidth]{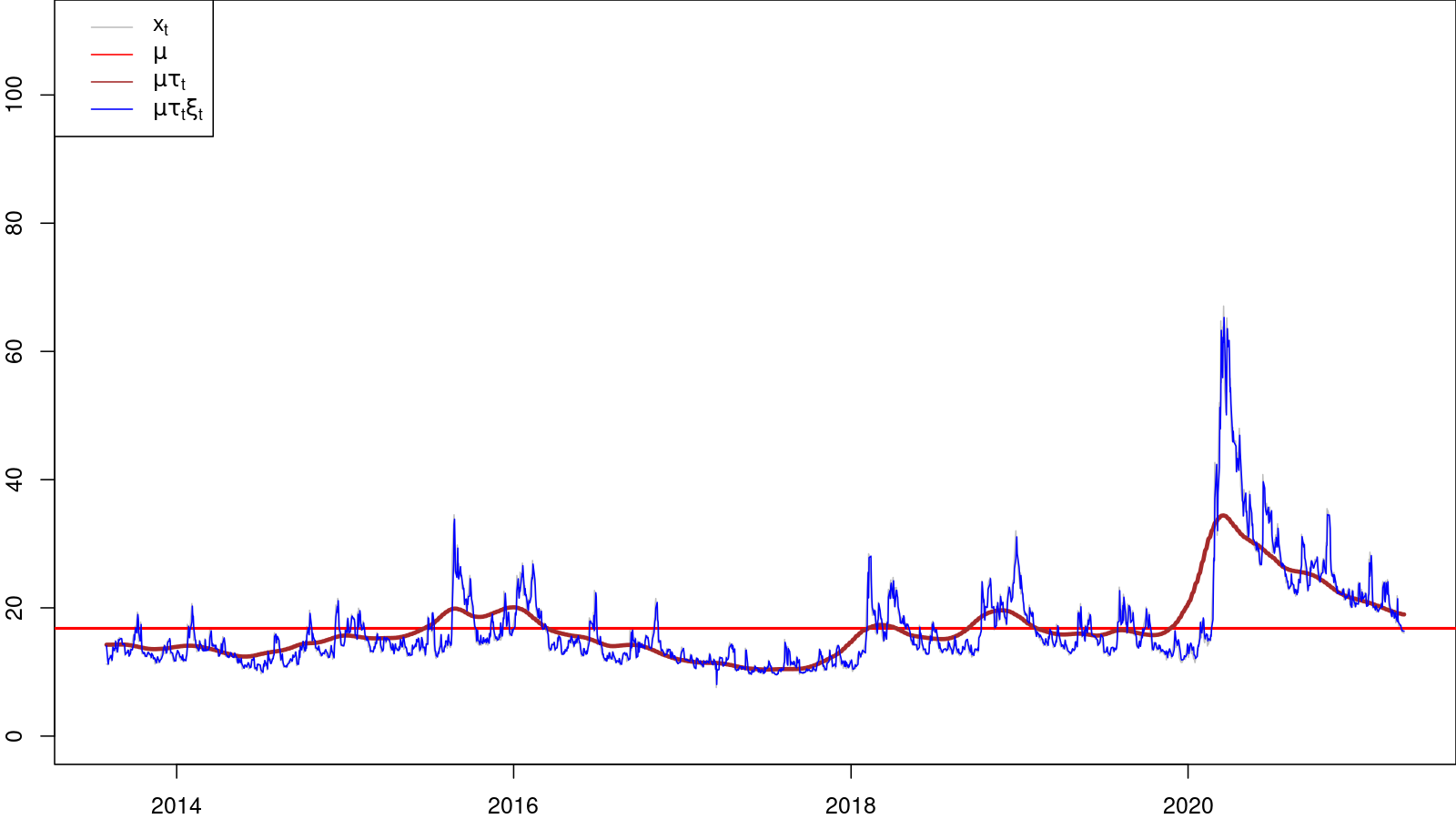}  
    \caption{impVol.}
    \label{fig:DJI-imp}
  \end{subfigure}
\end{figure}

\section{Model Inference} \label{sect:Inference}

The starting point for the estimation procedure is to get $\bm{\mu}$ as the average centroid of the $\{ \bm{x}_{t} \}$ series.
Then, after initialization of all $\bm{\xi}_{t}$'s to $\mathds{1}$, the following two steps are alternated and iterated until convergence:
\begin{enumerate}
  \item $\tau_{t}$ is estimated using the kernel estimator
  \begin{equation}
    \widehat{\tau}_{t} = \frac{\displaystyle \sum_{s = 1}^{T} \bm{x}^{(\tau)}_{t} K \left( \frac{z_{t} - z_{s}}{h} \right)}
    {\displaystyle \sum_{s = 1}^{T} K \left( \frac{z_{t} - z_{s}}{h} \right)} 
  \end{equation}
  where
  \begin{equation*}
    \bm{x}^{(\tau)}_{t} = \sum_{j = 1}^{K} \frac{x_{j, t}}{\mu_{j} \xi_{j, t}} \frac{\sigma_{j}^{-2}}{\sum_{j = 1}^{k} \sigma_{j}^{-2}},
  \end{equation*}
  $\bm{\mu}$, $\bm{\xi}_{t}$ and $\sigma_{j}^{2} = \bm{\Sigma}_{j,j}$ denote current estimates of the corresponding parameters.
  Notice that estimation of $\tau_{t}$ is based on all $K$ elements of $\bm{x}_{t}$, with higher contributions, the lower the individual residual's variance;    
  
  \item the parameters in $\bm{\xi}_{t}$ and in $\bm{\Sigma}$ are estimated from the current values ${x_{t,i}^{(\xi)} = x_{t,i} /(\mu_{i} \tau_{t})}$,  using the approach detailed below.  
\end{enumerate}

Clearly, when $\tau_{t} \equiv 1 \; \forall t$, the model is the base vMEM introduced by \cite{Cipollini:Engle:Gallo:2006}, so that only the second step above is needed, and inference is obtained by estimating the $\bm{\xi}_{t}$ parameters on the $\{ x^{(\xi)}_{t,i} = x_{t,i} / \mu_{i} \}$ values.

To define inference on the $\bm{\xi}_{t}$ parameters and $\bm{\Sigma}$ adopting a parametric specification of the distribution $d^+$ of the error term in a multivariate context has, as a matter of fact, several problems:
\begin{enumerate}
  \item Multivariate distributions deﬁned on the non-negative orthant are often not sufﬁciently ﬂexible: see, for example, the discussion on the multivariate Gamma in~\citet{Cipollini:Engle:Gallo:2006}). 
  The only reasonable choice seems to be the multivariate log-Normal which, on the other hand, does not carry over to QML characteristics similarly to what was seen for the Gamma in the univariate case (as discussed in Section~\ref{sect:QMLInference}).
  \item Considering the one-to-one correspondence between the cumulative distribution function of an absolutely continuous $K-$dimensional random variable and its copula representation, resorting to copulas bypasses some issues: some proposals are in \citet{Cipollini:Engle:Gallo:2017}, where marginals for the components are linked together using Gaussian
  or Student-T copulas. 
  However, there is no guidance for choices of a specific copula function except for convenience. 
  Moreover, using copulas in a vMEM context makes inference considerably complex. 
  \item It is not said that all components of the error term share the same marginal distribution. 
  Even though, in principle, one may be able to select the appropriate pdf one by one, this would require a lengthy model tuning (we will provide some evidence on our three volatility series below).
  \item As a ﬁnal remark, the dynamics of the conditional mean is almost always the main focus of the analysis, so that a full \textit{ex ante} speciﬁcation of the distribution of $\bm{\varepsilon}_{t}$ may be of secondary importance.
\end{enumerate}

To by--pass these specific issues, we advocate basing the inference procedure within a GMM context  \citep[][]{Cipollini:Engle:Gallo:2013,Cipollini:Gallo:2019}. Hence, assuming that $\bm{\xi}_{t}$ is correctly specified and indicating with $\bm{\theta}$ the vector of parameters entering it, we let
\begin{equation} 
  \label{eq:eps(t)-multi}
  \varepsilon_{t, i} = \frac{x^{(\xi)}_{t, i}}{\xi_{t, i}},
\end{equation}
where $x^{(\xi)}_{t, i} = x_{t, i} / (\mu_{i} \tau_{t})$.
Under model assumptions, $\bm{\varepsilon}_{t} - \mathds{1}$ is a conditionally homoskedastic martingale difference, with conditional expectation the zero vector and conditional variance matrix $\bm{\Sigma}$.
Then, the \emph{efficient} GMM estimator of $\bm{\theta}$, say $\widehat{\bm{\theta}}_{GMM}$, solves the criterion equation
\begin{equation}
  \label{eqn:Score:Equation:Multi}
  \sum_{t = 1}^T \bm{A}_{t} ( \bm{\varepsilon}_{t} - \mathds{1} ) = \bm{0},
\end{equation}
and has asymptotic variance matrix 
\begin{align}
  \label{eqn:Avar:Sandwich:Multi}
  \avar(\widehat{\bm{\theta}}_{GMM}) & = \left( \lim_{T \rightarrow \infty} \left[ T^{-1} \sum_{t = 1}^T E \left( \bm{A}_{t} \bm{\Sigma} \bm{A}_{t}^\prime \right) \right]\right)^{-1},
\end{align}
where
\begin{equation}
  \label{eq:A(t)}
  \bm{A}_{t} = \frac{\partial \bm{\xi}_{t}^{\prime} }{\partial \bm{\theta}} \diag(\bm{\xi}_{t})^{-1} \bm{\Sigma}^{-1}.
\end{equation}

As a consequence, a consistent estimator of the asymptotic variance matrix is 
\begin{equation*}
  \widehat{\avar}(\widehat{\bm{\theta}}_{GMM}) =
  \left(T^{-1} \sum_{t = 1}^T \widehat{\bm{A}}_{t} \widehat{\bm{\Sigma}} \widehat{\bm{A}}_{t}^\prime \right)^{-1},
\end{equation*}
where
\begin{equation*}
  \widehat{\bm{\Sigma}} = T^{-1} \sum_{t = 1}^T \left( \widehat{\bm{\varepsilon}}_{t} - \mathds{1} \right) \left( \widehat{\bm{\varepsilon}}_{t} - \mathds{1} \right)^{\prime}
\end{equation*}
is the Method of Moments estimator of $\bm{\Sigma}$, 
$\widehat{\bm{\varepsilon}}_{t}$ and $\widehat{\bm{A}}_{t}$  correspond to (\ref{eq:eps(t)-multi})   and  (\ref{eq:A(t)}), respectively, evaluated at $\widehat{\bm{\theta}}_{GMM}$. As with the univariate case, the feasible GMM estimator is obtained upon replacement of $\tau_t\mu_i$ with the system estimated counterparts.

\subsection{Forecasting}

Although, we will not provide empirical evidence of these models in forecasting (for which we refer to the original contributions, notably QQQ), forecasts from the \SpvMEM\ can be obtained via a general expression of the conditional expectation at time $t+h$ given the information available at time $t$ which exploits the fact that forecasts of the components $\tau_{t+h | t}$ and $\bm{\xi}_{t+h | t}$ can be obtained separately:
  \begin{equation*}
    E\left( \bm{x}_{t+h} | \info{I}{t}\right) 
    =
    \bm{\mu}_{t+h|t}
    =
    \tau_{t+h | t} \; \bm{\mu} \odot \bm{\xi}_{t+h | t}. 
  \end{equation*}
Regarding the slow-moving component, in view of the way it is estimated and its high persistence, it is reasonable to take $\tau_{t+h | t} \equiv \tau_{t}$ \citep[at least for small $h$; cf.~][]{Cattivelli:Gallo:2020}.
  As per the short-run component, Equation~(\ref{eqn:xi-vec}) implies 
  \begin{equation*}
    \bm{\xi}_{t+h|t} =
    \left\{
    \begin{array}{ll}
      \displaystyle \left( \mathds{1} - \bm{\beta}_1^{*} \mathds{1} \right) + 
      \bm{\beta}_1^{*} \bm{\xi}_{t} + \bm{\alpha}_1 \bm{x}_{t}^{(\xi)} + \bm{\gamma}_1 \bm{x}_{t}^{(\xi-)} & \text{if } h = 1 \\[2mm]
      \displaystyle \left( \mathds{1} - \bm{\beta}_1^{*} \mathds{1} \right) + 
      \bm{\beta}_1^{*} \bm{\xi}_{t+h-1|t} & \text{if } h > 1.
    \end{array}
    \right.
  \end{equation*}
Other refinements in the formulas for $\bm{\xi}_{t+h|t}$ over horizons $h>1$ are needed when further lags occur in any of the components of   the RHS of Equation~(\ref{eqn:xi-vec}). 
  
  \section{The Multivariate Case: Applications}
  \label{sect:multiappl}
  
As mentioned at various stages so far, in order to illustrate the behavior of the \vMEM\ in the multivariate case, we refer to three series for each market, absolute returns ($i=1$), realized kernel volatility ($i=2$) and implied volatility ($i=3$). The general behavior of the series was presented, in the case of the Dow Jones index in Figure \ref{fig:DJI-3}. Graphs for the other markets reproduce the same features by variable and are left for supplementary material. 
  
In Table \ref{tab:mean}, we report the sample means of the series across markets, to provide some reference about the overall level $\bm{\mu}$ around which the estimated components will move. We can notice that, in a generalized manner, the absolute returns (which include zeros) have mean values similar to those of the realized volatility, and that average implied volatilities are higher.

\begin{table}[htbp]
  \centering
  \caption{Sample averages of the three volatility indicators.  For sample periods refer to Table \ref{tab:data}. }
  \label{tab:mean}
  \input{Table/Tab-Means.tex}
\end{table}

The first set of results relates to the base \vMEM\ in the absence of the low--frequency component, where we have extended Equation (\ref{eqn:xi-vec}) to accommodate a $(1,1)$ structure with diagonal $\bm{\gamma}_1$ and $\bm{\beta}_1$ and $\bm{\alpha}_1$ is a full  matrix: estimated parameters are presented, by market, in Table \ref{tab:inf-vmem}, where, in view of the choices of diagonality just mentioned, we report just the diagonal elements of  $\bm{\beta}_1^*$, namely, $\beta_{i,i,1}^*= \beta_{i,i,1} + \alpha_{i,i,1} + \gamma_{i,i,1}/2$.

\begin{sidewaystable}[htbp]
	\centering
	\caption{\vMEM: Estimation summary table.}
	\label{tab:inf-vmem}
	\input{Table/Tab-Inf-vmem.tex}
\end{sidewaystable}

The first element to comment on are the parameters on the diagonal of $\bm{\beta}_1^*$: the  degree of persistence is very high, and it increases moving from the absolute returns to the realized volatility, to the implied volatility which has the highest, in excess of $0.97$. 

As far as the relevance of the off-diagonal elements of  $\bm{\alpha}_1$, we can point to the generalized significance of the parameters that feed into the absolute returns from the other measures (the $\alpha_{1,j,1}$'s, $j=2,3$, the only exception is for KS11 for realized volatility); the same is true for the $\alpha_{2,j,1}$'s, $j=1,3$, with an exception for the coefficient from lagged absolute returns for STOXX50E; by contrast, the coefficients $\alpha_{3,j,1}$'s, $j=1,2$ are generally not significant with some rare exceptions (both for FCHI and the coefficient from lagged realized volatility for N225). These results point to the autonomous, so--to--speak, dynamics of the implied volatility, but also (and foremost) to a full interdependence in what concerns the other two measures. As a matter of fact, if one were to extend the logic of the HEAVY model by \citet{Shephard:Sheppard:2010} to this framework, the question would be whether absolute returns can be considered irrelevant in their own dynamics (while keeping the lagged dependence on the other measures) and also in the dynamics of the other two  measures. We address this question formally by jointly testing the significance of the four parameters $\alpha_{1,1,1}$, $\gamma_{1,1,1}$, $\alpha_{2,1,1}$, $\alpha_{3,1,1}$, obtaining the results in the first panel of Table~\ref{tab:test-heavy}. For the \vMEM\ these implied `HEAVY' restrictions are strongly rejected, with the DJI providing significant evidence at 5\% but not at 1\%. 

\begin{table}[htbp]
	\centering
	\caption{Joint hypothesis  testing on the `HEAVY' zero restriction of $\alpha_{1,1,1}$, $\gamma_{1,1,1}$, $\alpha_{2,1,1}$, $\alpha_{3,1,1}$, corresponding to  the irrelevance of absolute returns in their own dynamics and in the dynamics of the other two series. Wald statistics and p--values reported for \vMEM\ and for \SpvMEM.}
	\label{tab:test-heavy}
	\input{Table/Tab-vmem-heavy.tex}
\end{table}

The evidence on asymmetric effects is also mixed: for the absolute returns almost all parameters are significant (except HSI and KS11); for the realized volatility we have substantially the same results as in the univariate case (the only insignificant cases are IXIC, HSI and KS11); finally, for the implied volatility only one (FCHI) is significantly positive, while 6 of 9 are significantly negative, 2 are not significant.

\begin{figure}
	\caption{ \vMEM\ estimated on DJI absolute returns (arVol -- panel (a)), realized kernel volatility (rkVol -- panel(b)), and implied volatility (impVol -- panel(c)).
		The individual unconditional mean ${\mu}_i$ is a flat line; the estimated conditional expectation $\mu_i {\xi}_{i,t}$ is reproduced in blue. 
		The observed series is in the background in a shade of grey. For sample periods, refer to Table \ref{tab:data}.}
	\label{fig:DJI-vmem}
	\begin{subfigure}{.49\textwidth}
		\centering
		\includegraphics[width=\linewidth]{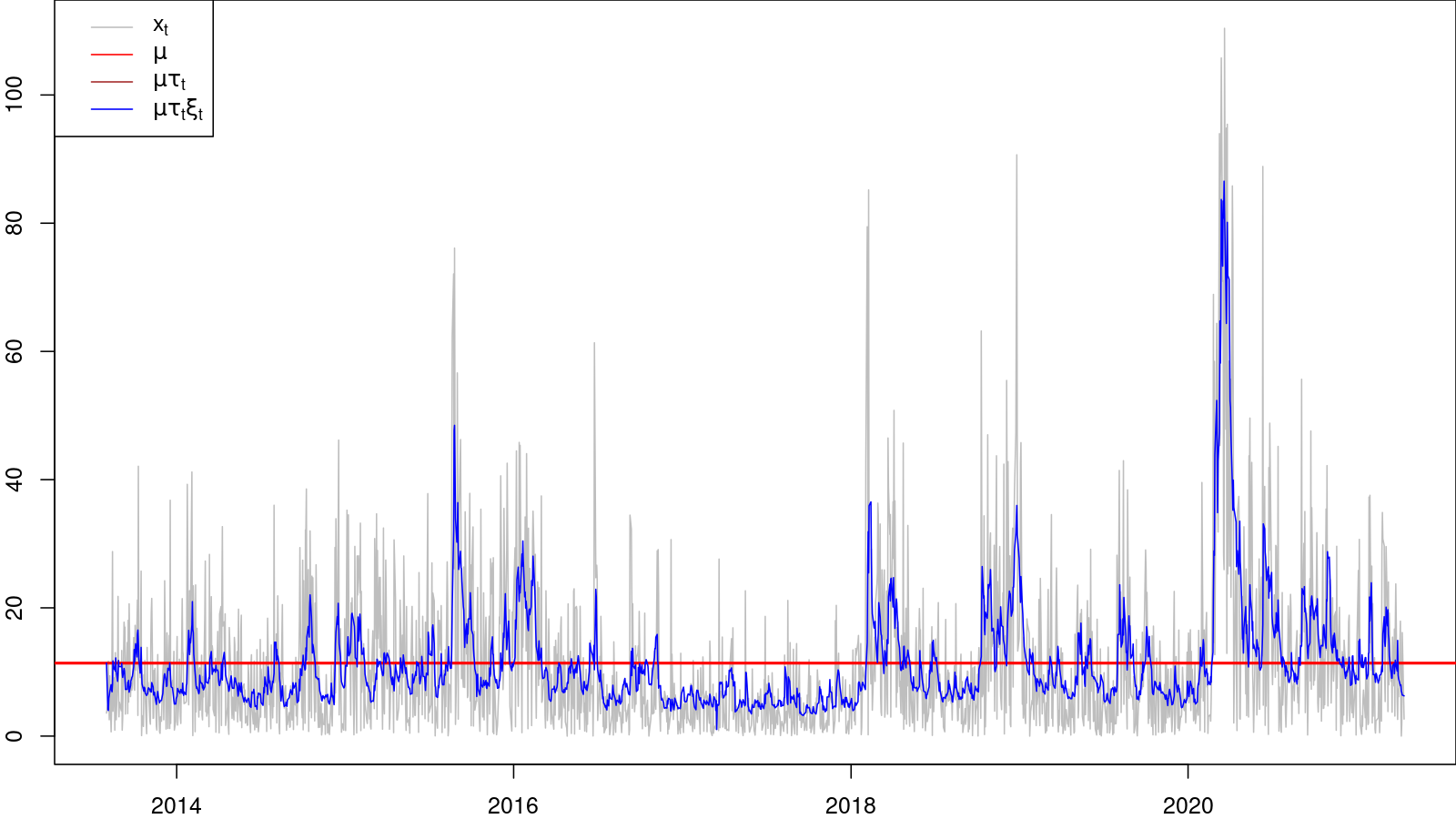}  
		\caption{arVol.}
		\label{fig:DJI-ar}
	\end{subfigure}
	\begin{subfigure}{.49\textwidth}
		\centering
		\includegraphics[width=\linewidth]{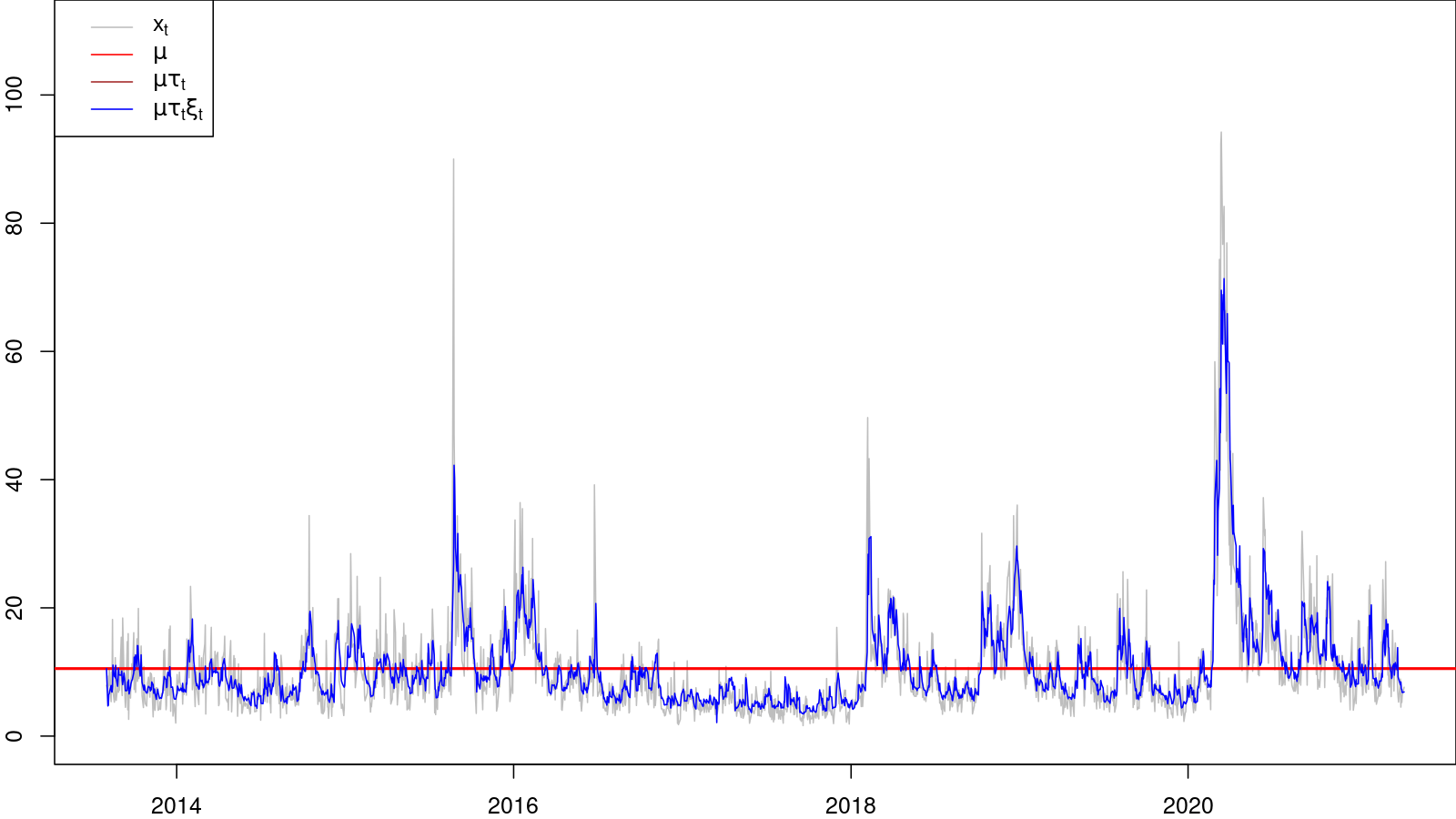}  
		\caption{rkVol.}
		\label{fig:DJI-rk}
	\end{subfigure}
	\\
	\centering
	\begin{subfigure}{.49\textwidth}
		\centering
		\includegraphics[width=\linewidth]{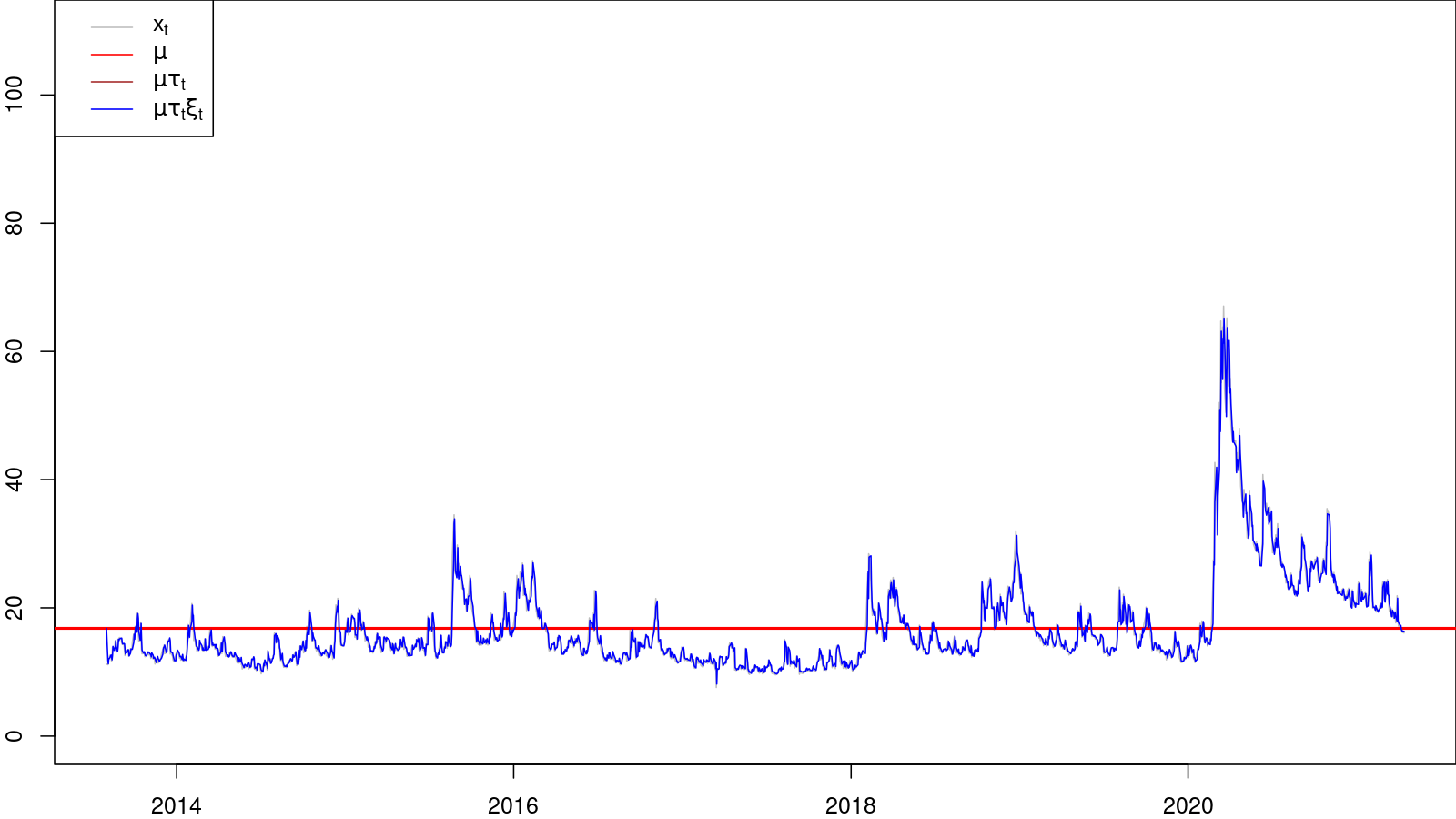}  
		\caption{impVol.}
		\label{fig:DJI-imp}
	\end{subfigure}
\end{figure}

The overall behavior of the estimated components can be graphically illustrated in Figure \ref{fig:DJI-vmem} with reference to the DJI as an example. In it, we have reported the unconditional means $\mu_i$ as a flat line, and the conditional expectations of the three volatility measures, keeping the original series in the background with a shade of grey. 

The estimated covariance matrix for the residuals $\bm{\Sigma}$ is represented using standard deviations  $\sigma_{i}\equiv {\Sigma}_{i,i}$, ($i=1,2,3$), which shows a larger uncertainty surrounding absolute returns, with realized volatility in the middle position and a very low value for the implied volatilities. Off-diagonal, the derived estimated correlations $\rho_{i,j}$ ($i> j$) show a higher correlation between the residuals of the absolute returns and those of the realized volatility, followed by the correlations between realized volatility and implied volatilities and last, but in a range between $0.24$ and $0.33$ for the remaining pair.

As in the univariate case, we report under the label $R^2$, the squared correlation coefficient between the observed values and the estimated conditional expectations for each variable. As one would expect, the fit is lower for absolute returns ($R^2(1)$ generally in the $0.20$'s), for the realized volatility the performance is higher, but somewhat mixed ($R^2(2)$ between $0.50$ and $0.71$); finally for the implied volatility we get a very close fit ($R^2(3)$ in excess of $0.94$). 

The autocorrelation results show some generalized presence of residual dynamics with very low p-values of the Ljung--Box statistics. 

The second set of results relates to the \SpvMEM\ model, presented in Section \ref{sect:SpvMEM} the estimation results of which are reported in Table \ref{tab:inf-spvmem} with the same choice of  specification  before: $\bm{\gamma}_1$ and $\bm{\beta}_1$ are both diagonal matrices and $\bm{\alpha}_1$ is a full matrix.  

By and large, the most striking result is the generalized reduction in the persistence parameters $\bm{\beta}^*$, in some cases even a drastic one, especially for arVol and rkVol which go below $0.9$. This is to be expected, since the $\tau_t$ term in the \SpvMEM\ characterizes part of the persistence in the series as a low--frequency component. 

\begin{figure}[tbp]
	\centering
	\caption{Estimated slow moving component $\tau_t$ for all tickers. Vertical bars mark the start of the sample by ticker. Colors represent geographical areas;  black for the US, red for Europe and blue for East Asia. }
	\label{fig:slow-all}
	\includegraphics[width=\textwidth]{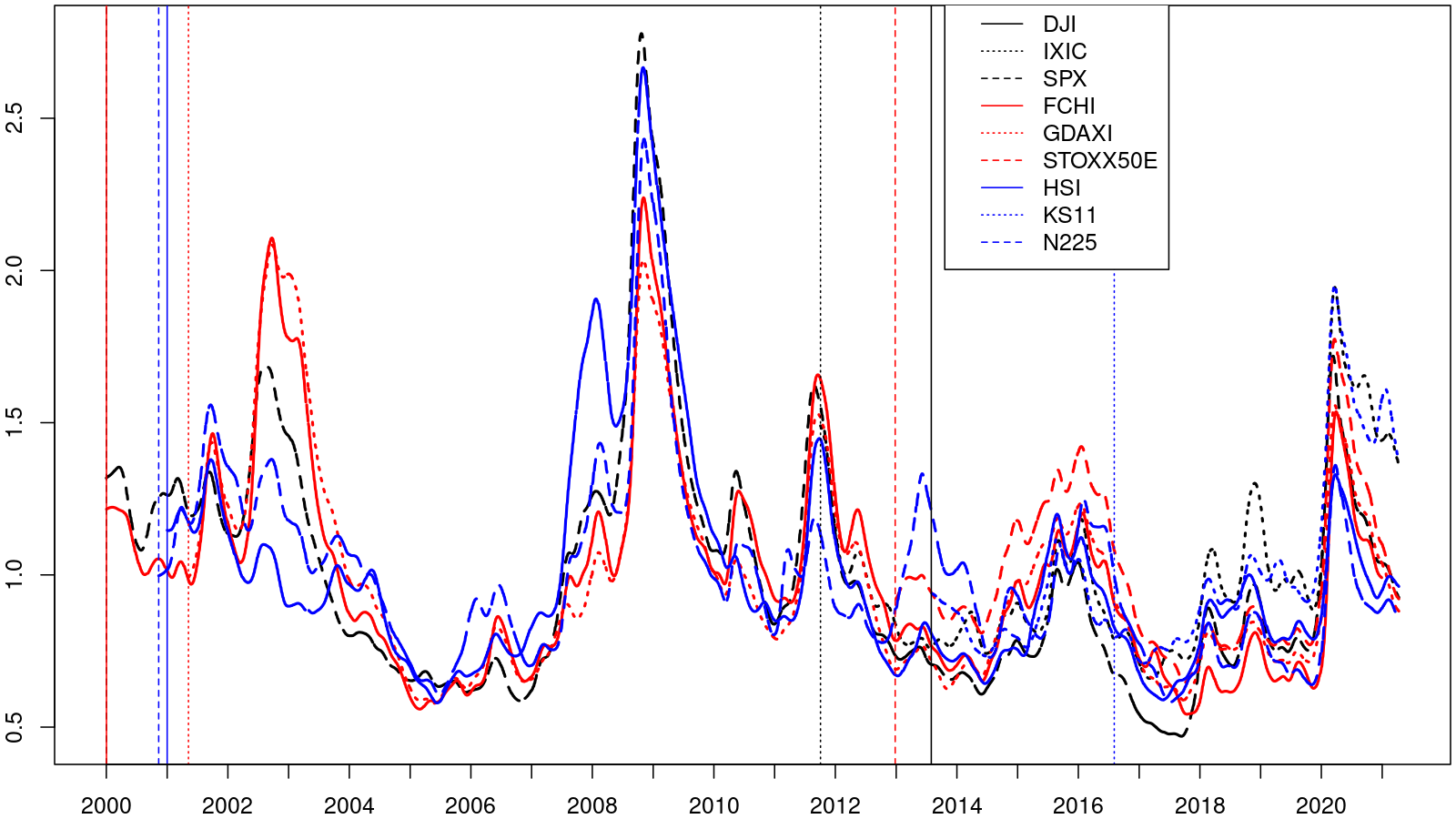}
\end{figure}

While the analysis is carried out by market on the three volatility measures, it is nevertheless interesting to compare the behavior of the estimated $\tau_t$ across markets, as done in Figure \ref{fig:slow-all}, where we have reproduced the various series, with a color reference for ease of recognition of the geographic area the market is in, black for the US, red for Europe and blue for East Asia (we also superimpose vertical bars for the start of the sample for the corresponding ticker). The graphical appraisal allows us to pinpoint that this common volatility component has a strikingly similar behavior, but there is also a great deal of different behavior across markets with bursts of volatility affecting some areas at different rates. For example, commonality in the reaction is seen when the COVID-19 health emergency erupted and when the Great Financial Crisis had its worse episode in September 2008; a similar common episode coincides with 9/11. By the same token, the acceleration of the consequences of the dot com bubble burst were felt especially in Europe in the second half of 2002 (with the DAX and CAC40 indices hitting historical lows), the Euro area sovereign debt crisis spread out to all areas, with a sharp and progressive reduction in market fears induced  by the July 2012 \textit{Whatever it takes} speech by Mario Draghi. These results point out to a warning that attempts to model a single common low--frequency component across these market volatilities by an \SpvMEM\ would probably be frustrated by the presence of market--specific, slow--moving idiosyncratic behavior.

\begin{sidewaystable}[htbp]
  \centering
  \caption{\SpvMEM: Estimation summary table.}
  \label{tab:inf-spvmem}
  \input{Table/Tab-Inf-spvmem.tex}
\end{sidewaystable}

\begin{figure}
  \caption{\SpvMEM\ on DJI ($h = 3$ months).}
  \label{fig:DJI-3-acf-hist}
  \begin{subfigure}{.48\textwidth}
    \centering
    \includegraphics[width=\linewidth]{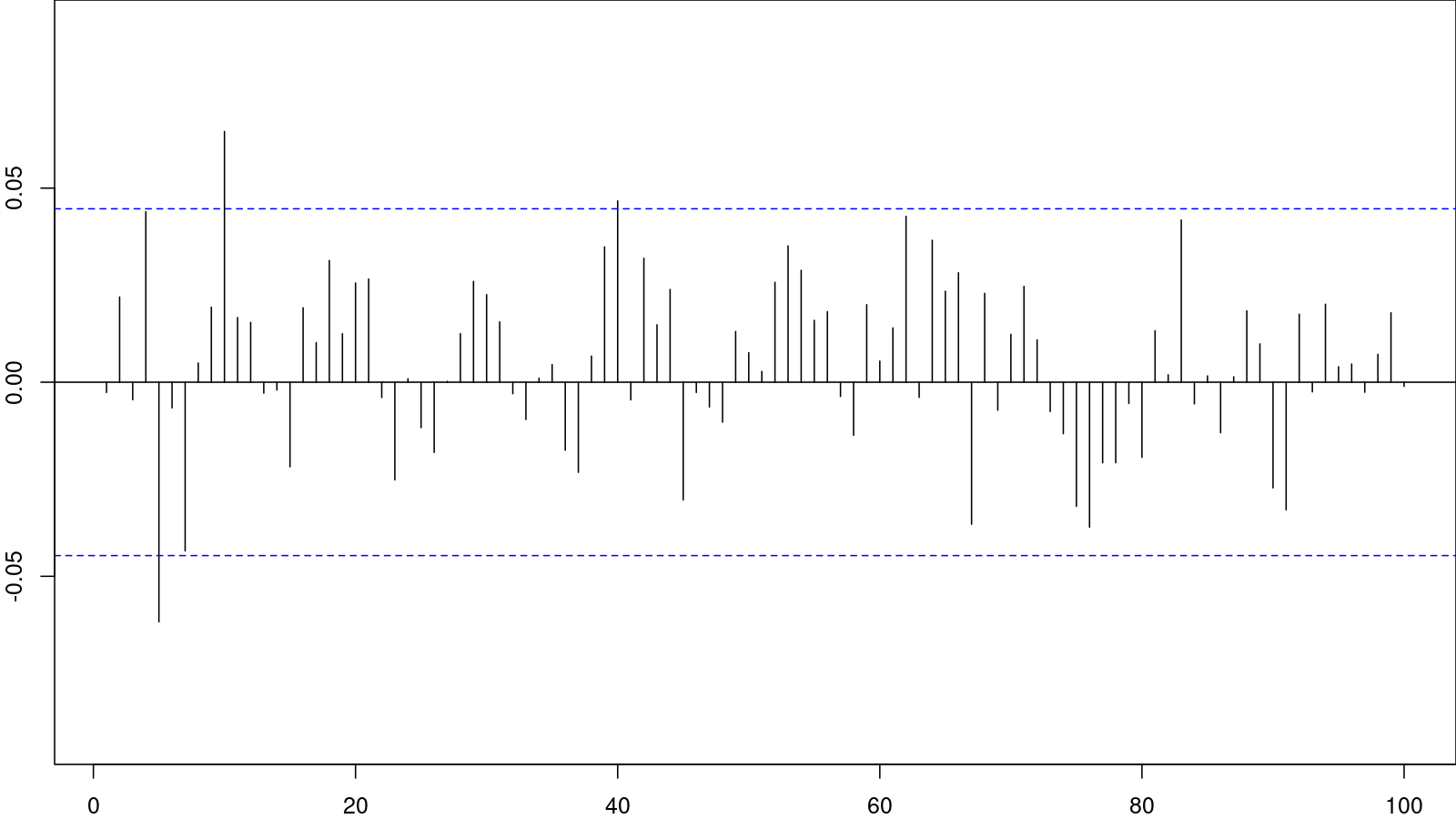}  
    \caption{Arvol, acf}
    \label{fig:dji-arVol-acf}
  \end{subfigure}
  \begin{subfigure}{.48\textwidth}
    \centering
    \includegraphics[width=\linewidth]{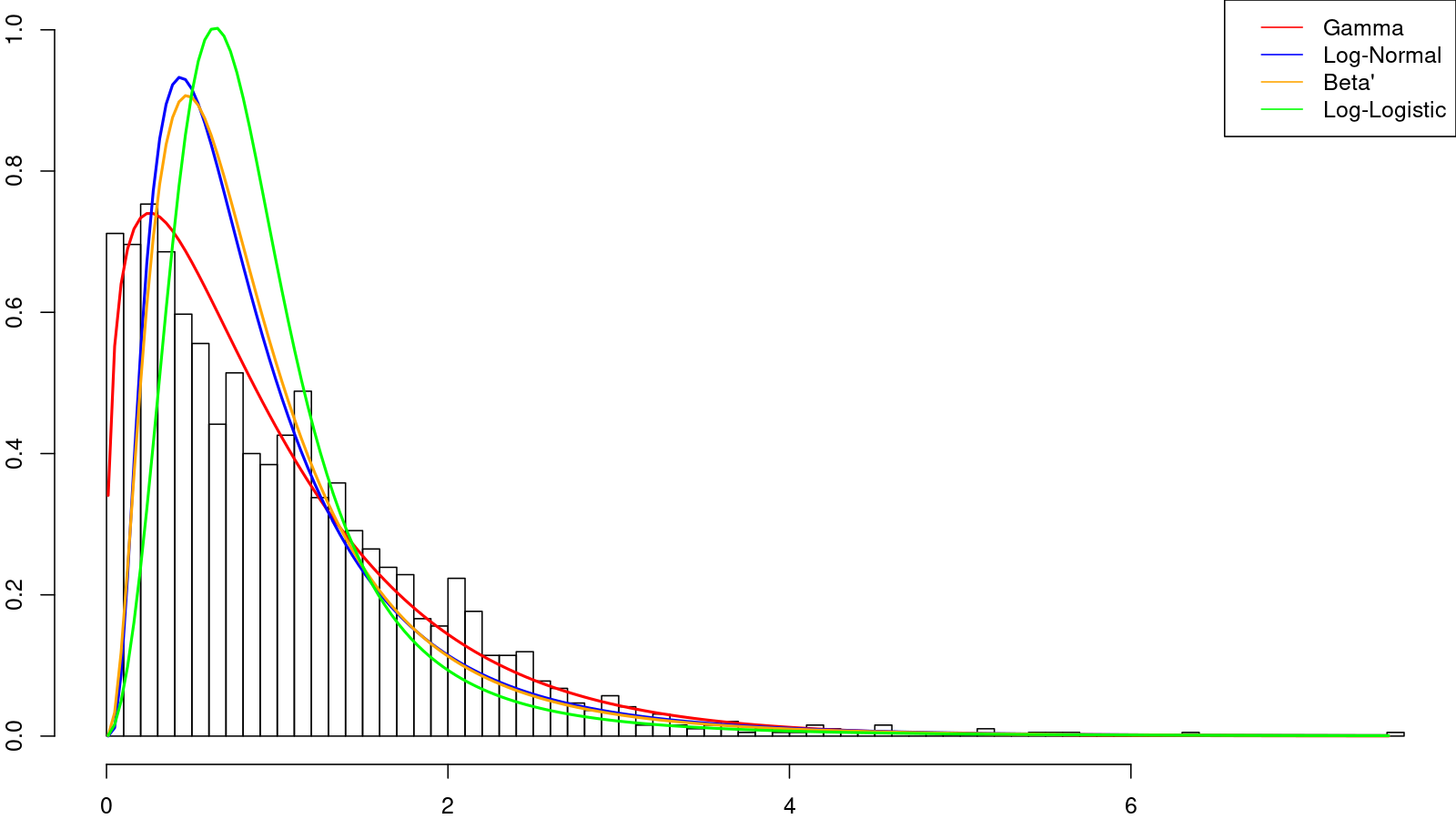}  
    \caption{Arvol, hist}
    \label{fig:dji-arVol-hist}
  \end{subfigure}
  \\
  \begin{subfigure}{.48\textwidth}
    \centering
    \includegraphics[width=\linewidth]{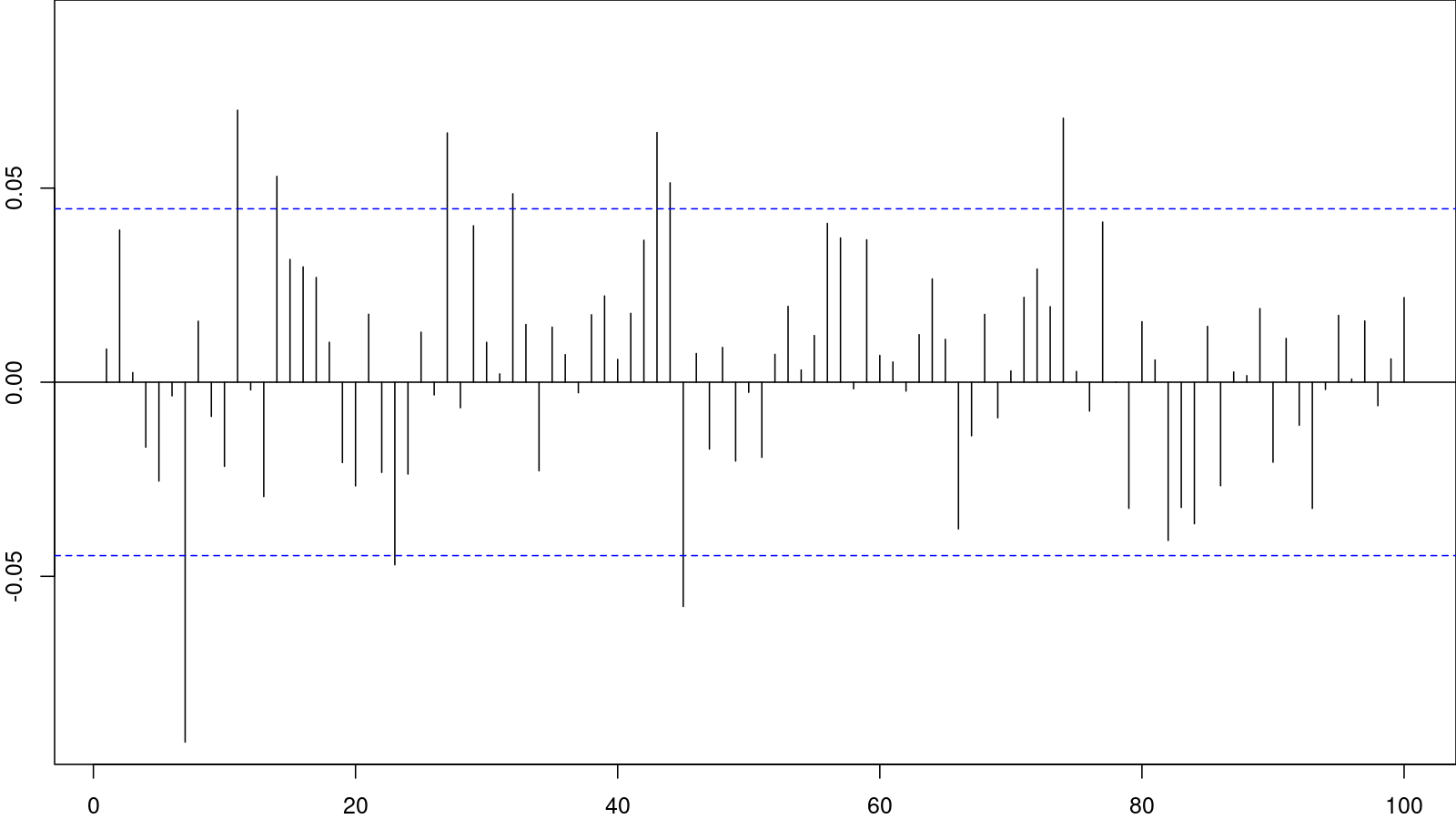}  
    \caption{rkVol, acf}
    \label{fig:dji-rkVol-acf}
  \end{subfigure}
  \begin{subfigure}{.48\textwidth}
    \centering
    \includegraphics[width=\linewidth]{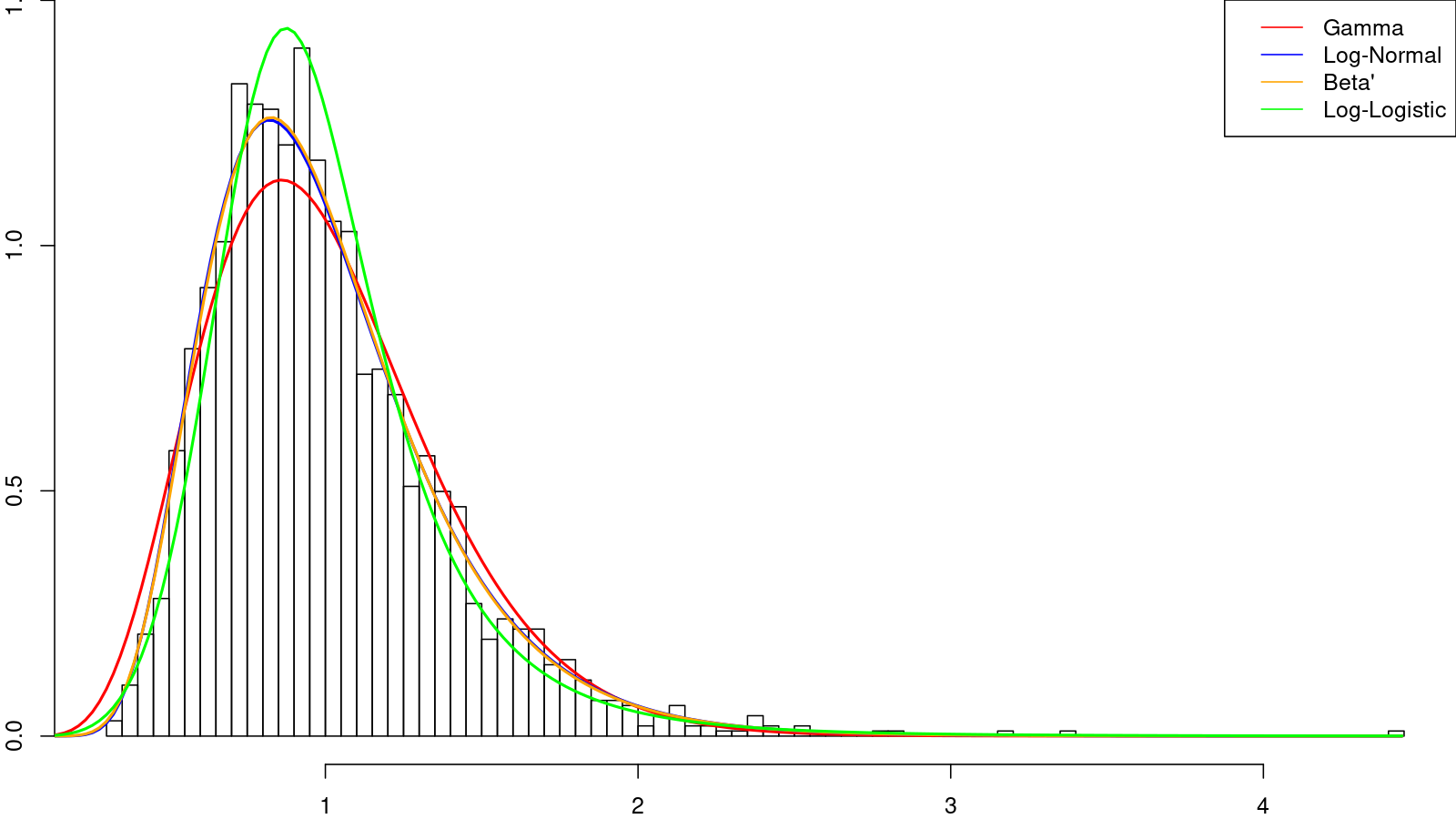}  
    \caption{rkVol, hist}
    \label{fig:dji-rkVol-hist}
  \end{subfigure}
  \\
  \begin{subfigure}{.48\textwidth}
    \centering
    \includegraphics[width=\linewidth]{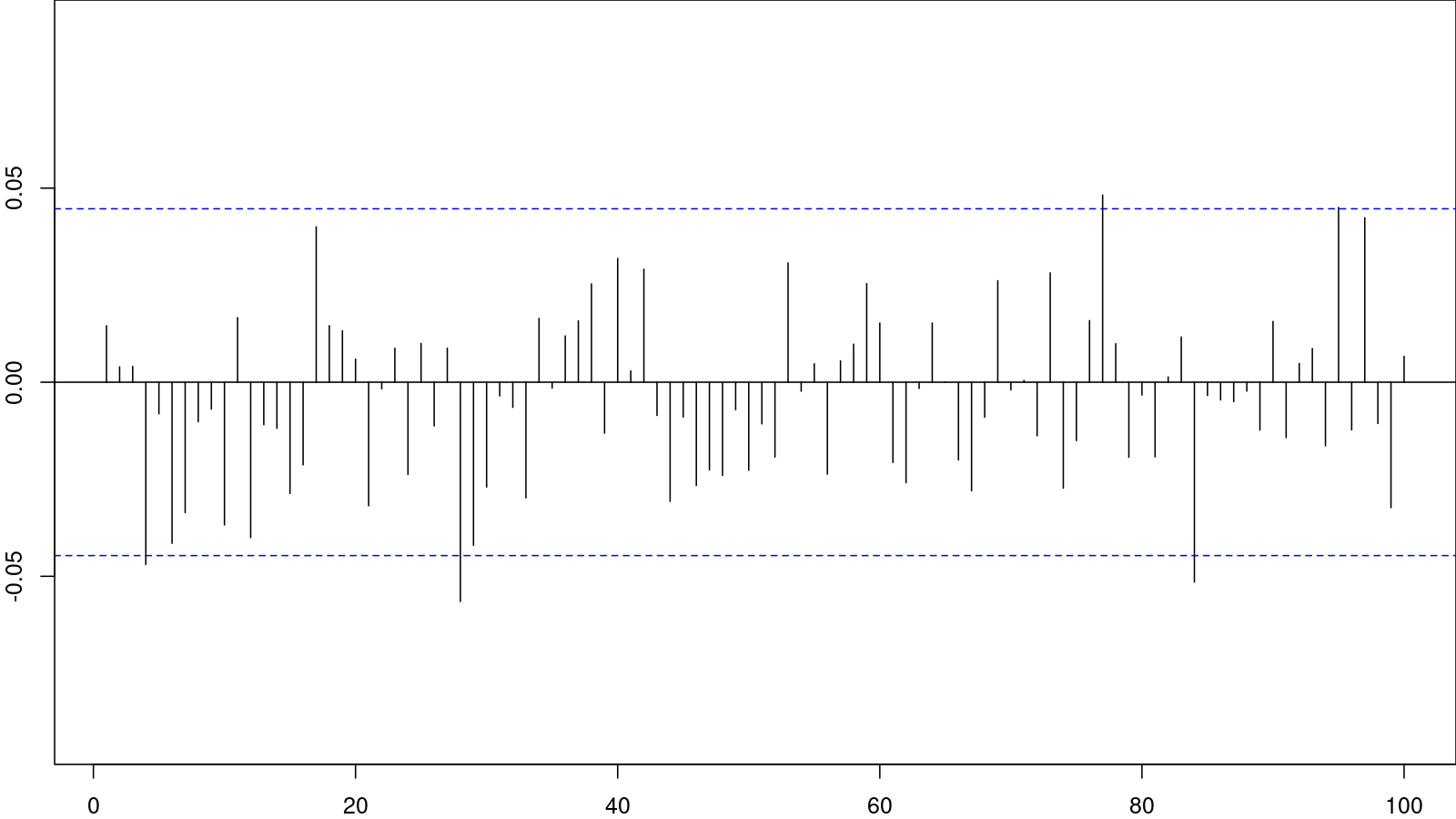}  
    \caption{impVol, acf}
    \label{fig:dji-impVol-acf}
  \end{subfigure}
  \begin{subfigure}{.48\textwidth}
    \centering
    \includegraphics[width=\linewidth]{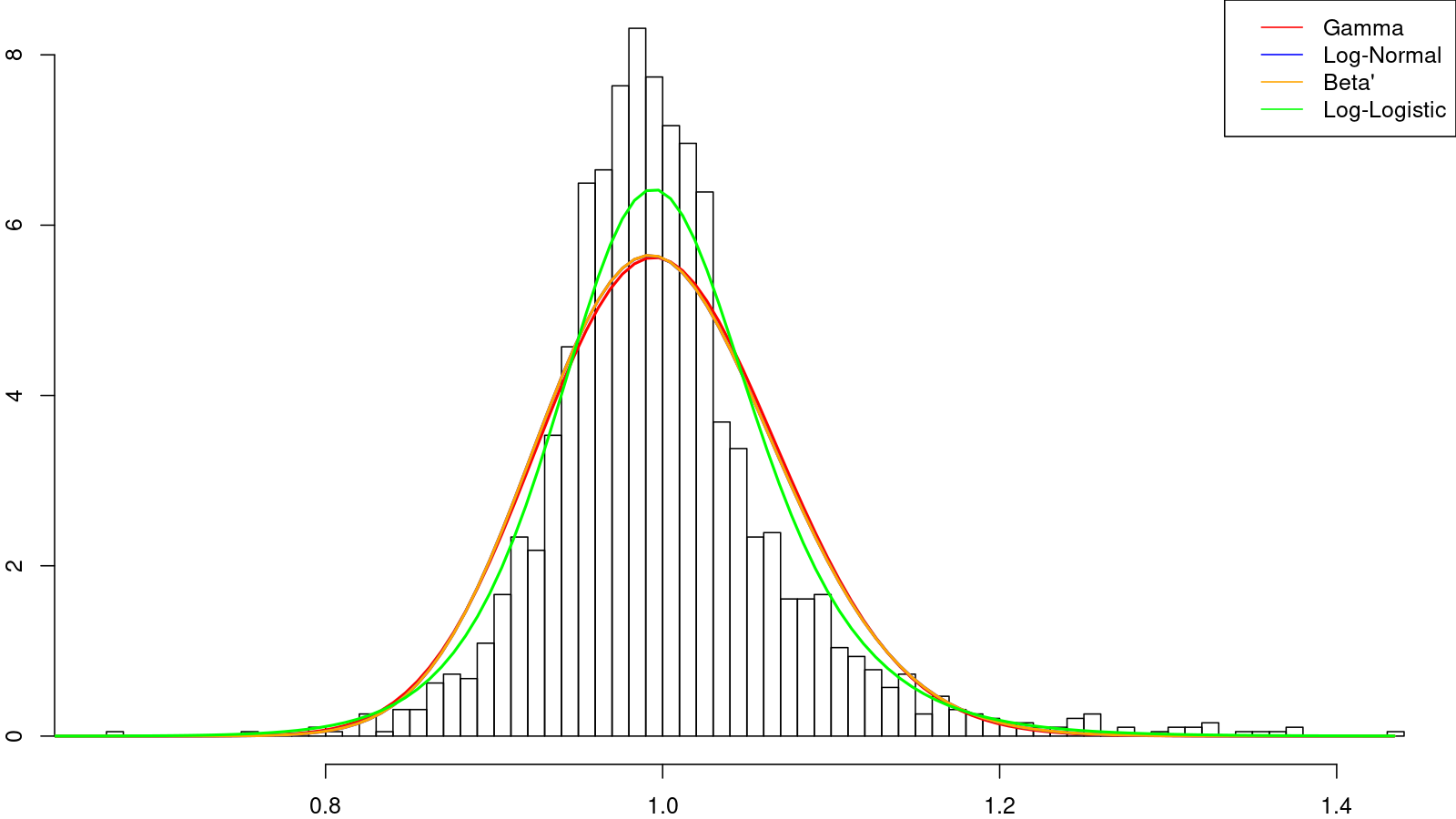}  
    \caption{impVol, hist}
    \label{fig:dji-impVol-hist}
  \end{subfigure}
\end{figure}

\begin{table}[htbp]
  \centering
  \caption{Absolute returns (arVol), Realized kernel volatility (rkVol) and Implied volatility (impVol). P--values of the Anderson--Darling (AD) and Cram\`{e}r-von Mises (CvM) tests for distribution fitting  on  \SpMEM\ residuals. Target densities: Gamma, Log--Normal (Log--N), Beta$^\prime$, Log--Logistic (Log--L).}
  \label{tab:gof}
  \input{Table/Tab-spvmem-gof.tex}
\end{table}

\section{Concluding Remarks}
In reviewing theoretical and empirical aspects of the Multiplicative Error Model, the paper has two main sections, devoted to the univariate, respectively, multivariate versions of the \MEM. Its contribution can be synthesized as follows: we have proposed a modification of the standard specification of a \MEM\ to highlight the composition of the conditional expectation of a positive variable of interest, in a way inspired by \citet{Veredas:Rodriguez:Espasa:2007} (\SpMEM). From a theoretical point of view  we have discussed how GMM applies to this model and the properties of the estimator, showing an equivalence with the Quasi Maximum Likelihood estimator under the Gamma assumption. In this respect, we have proceeded to an original set of applications in the univariate case on series of realized volatility from several markets, presenting estimation results both in tables and in the graphical decomposition achieved among the various components. Moreover, we have discussed the properties of estimated residuals graphically (correlogram and histograms), and more formally with distribution  fitting  tests. In the multivariate context,  we have introduced the extension to the \SpvMEM model by \citet{Barigozzi:Brownlees:Gallo:Veredas:2014} to accommodate interdependence in the multivariate framework,  showing, again, our preference for the GMM estimation strategy. The applications in the multivariate case relate to a trivariate model with absolute returns, realized volatility and implied volatility from option--based indices. Estimation results both for the \vMEM\ and the \SpvMEM are complemented by some tests for HEAVY--type dynamics \citep{Shephard:Sheppard:2010} which show that the impact of absolute returns on realized volatility is relevant. Once again, attention is paid to the properties of estimated residuals, showing how substantially different their profile is in terms of underlying distributions, to the point that while a Gamma often fits the behavior for absolute returns, and a Beta$^\prime$ or a Log--Normal that of realized volatility, none of the distributions considered is suitable for implied volatility. 

This paper marks 20 years of developments within the area of the Multiplicative Error Models, starting from the seminal paper of \citet{Engle:2002}. Overall, the conceptual advantages of using a \MEM\ lie in the fact that positive processes can be modeled directly without a log--transformation, and that GMM inference is widely available avoiding the need for a parametric specification of the error distribution. Residual diagnostics typically improves relative to non--multiplicative specifications, especially in what concerns autocorrelation. In forecasting, the various versions of the \MEM\ have proved successful, especially in reproducing the dynamic behavior of realized volatility \citep[cf., for example]{Cipollini:Gallo:Otranto:2020}, but also of intradaily volumes \citep{Brownlees:Cipollini:Gallo:2011} for Volume Weighted Average Price (VWAP) applications \citep[for volumes and number of trades, see][]{Cipollini:Engle:Gallo:2017}.

When the number of variables grows, the \vMEM\ is somewhat affected by a curse of dimensionality problem. In the context of a factor approach, one should explore the possibility that several variables (e.g. realized  volatilities on a large number of assets) share some common component beyond the single nonparametric one suggested by \citet{Barigozzi:Brownlees:Gallo:Veredas:2014}. On the other hand, the matrix $\bm{\alpha}_1$ which we envisaged to be full both in theory and in the empirical applications may contain some zeros, and estimating them affects the precision of the estimates. A first attempt  was made by \citet{Cattivelli:Gallo:2020} with an adaptive Lasso approach which imposes some penalization on making parameters different from zero at the estimation stage. While promising, there are further refinements in the penalized ML approach, and Lasso is not the only tool available among regularization techniques (cf. ridge--type and/or elastic-net): many of these techniques were successfully introduced in the Vector Autoregression \citep[for recent overviews of the approaches and of the results obtained, cf.][]{Nicholson:Matteson:Bien:2017, Nicholson:Wilms:Bien:Matteson:2020}, and, given the similarity of the models could be explored in the context of the \vMEM\ world.


\input{CG-EcoSta-01.bbl.tex}
\newpage
\appendix{}

\noindent{\LARGE\bf Appendices}

\section{Data Details}

\setcounter{table}{0}
\renewcommand{\thetable}{A\arabic{table}} 

 \begin{table}[htbp]
	\centering
	\caption{Data reference for realized kernel volatility and volatility indices.}
	\label{tab:data}
	\input{Table/Tab-data.tex}
\end{table}

\section{Univariate Distributions of the Error Term}

One of the maintained assumptions of the \MEM\  (Section~\ref{sect:spmem}) is that the multiplicative error term $\varepsilon_{t}$ has non-negative support and unit mean.
In this appendix we list four  possible parametric specifications of the $\varepsilon_{t}$ distribution, reporting the corresponding probability density functions and the parameter constraints needed so as the mean and the variance are one and $\sigma^{2}$, respectively.
In the application, density estimates  are retrieved from the GMM estimate of the $\sigma^{2}$.

\subsection{Gamma}

\begin{equation*}
  f_{\varepsilon}(\varepsilon) = \frac{\beta^{\alpha}}{\Gamma(\alpha)} \varepsilon^{\alpha-1} \exp(-\beta \varepsilon)
\end{equation*}
where $\beta = \alpha = \sigma^{-2}$.

\subsection{Log-Normal}

\begin{equation*}
  f_{\varepsilon}(\varepsilon) = \frac{1}{\varepsilon \sqrt{2 \pi V}} \exp\left(-\frac{1}{2} \frac{\left( \ln \varepsilon - m \right)^{2}}{V} \right)
\end{equation*}
where $V = \ln \left( \sigma^{2} + 1 \right)$ and $m = -V/2$.

\subsection{Beta Prime}

\begin{equation*}
  f_{\varepsilon}(\varepsilon) = \frac{\Gamma(\alpha + \beta)}{\Gamma(\alpha)\Gamma(\beta)} \varepsilon^{\alpha - 1} \left( 1 + \varepsilon \right)^{-(\alpha + \beta)}
\end{equation*}
where $\beta = 2 + 2 \sigma^{-2}$ and $\alpha = \beta - 1$.

\subsection{Log-Logistic}

\begin{equation*}
  f_{\varepsilon}(\varepsilon) =  \frac{\beta}{\alpha} \left( \frac{\varepsilon}{\alpha} \right)^{\beta-1} \left[ 1 + \left( \frac{\varepsilon}{\alpha} \right)^{\beta} \right]^{-2}
\end{equation*} 
where $\beta$ satisfies $\sigma^{2} = \tan (\pi/\beta)/(2 \pi/\beta) - 1$ and $\alpha =  \sin(\pi / \beta)/(\pi / \beta)$.

\end{document}

%% file: Table/Tab-Inf-mem.tex
\scalebox{0.8}{
\begin{tabular}{lrrrrrrrrrrrrrrrrrr}
   & \multicolumn{2}{c}{DJI} & \multicolumn{2}{c}{IXIC} & \multicolumn{2}{c}{SPX} & \multicolumn{2}{c}{FCHI} & \multicolumn{2}{c}{GDAXI} & \multicolumn{2}{c}{STOXX50E} & \multicolumn{2}{c}{HSI} & \multicolumn{2}{c}{KS11} & \multicolumn{2}{c}{N225}\\
 & est & zstat & est & zstat & est & zstat & est & zstat & est & zstat & est & zstat & est & zstat & est & zstat & est & zstat \\ 
  \hline
$\beta^{*}_{1}$ & 0.9459 & 97.70 & 0.9140 & 77.03 & 0.9750 & 281.41 & 0.9787 & 304.29 & 0.9763 & 288.26 & 0.9587 & 124.64 & 0.9798 & 253.40 & 0.9650 & 109.32 & 0.9674 & 187.69 \\ 
  $\alpha_{1}$ & 0.2721 & 11.04 & 0.3726 & 14.93 & 0.1621 & 14.07 & 0.1416 & 13.66 & 0.1808 & 15.85 & 0.1273 & 6.39 & 0.1602 & 15.46 & 0.1859 & 9.44 & 0.1797 & 14.05 \\ 
  $\gamma_{1}$ & 0.1328 & 8.80 & 0.0684 & 4.92 & 0.1238 & 16.21 & 0.0962 & 14.23 & 0.0865 & 13.61 & 0.1273 & 10.05 & 0.0170 & 2.58 & 0.0023 & 0.18 & 0.0818 & 8.95 \\ 
   \hline
$\sigma$ & 0.3977 &  & 0.3568 &  & 0.3941 &  & 0.3854 &  & 0.3374 &  & 0.4453 &  & 0.3756 &  & 0.4220 &  & 0.4581 &  \\ 
   \hline
$R^2$ & 0.6758 &  & 0.6372 &  & 0.6576 &  & 0.5791 &  & 0.6738 &  & 0.5114 &  & 0.5160 &  & 0.4771 &  & 0.4761 &  \\ 
   \hline
$LB(5)$ & 0.1820 &  & 0.0039 &  & 0.0046 &  & 0.0071 &  & 0.0169 &  & 0.5719 &  & 0.0012 &  & 0.0373 &  & 0.0000 &  \\ 
  $LB(10)$ & 0.0028 &  & 0.0031 &  & 0.0009 &  & 0.0436 &  & 0.0554 &  & 0.8704 &  & 0.0006 &  & 0.1938 &  & 0.0000 &  \\ 
  $LB(15)$ & 0.0000 &  & 0.0027 &  & 0.0013 &  & 0.0015 &  & 0.0043 &  & 0.5858 &  & 0.0078 &  & 0.4742 &  & 0.0001 &  \\ 
  $LB(20)$ & 0.0001 &  & 0.0012 &  & 0.0049 &  & 0.0007 &  & 0.0075 &  & 0.7550 &  & 0.0111 &  & 0.7305 &  & 0.0005 &  \\ 
   \hline
\end{tabular}
}

%% file: Table/Tab-Inf-spmem.tex
\scalebox{0.8}{
\begin{tabular}{lrrrrrrrrrrrrrrrrrr}
   & \multicolumn{2}{c}{DJI} & \multicolumn{2}{c}{IXIC} & \multicolumn{2}{c}{SPX} & \multicolumn{2}{c}{FCHI} & \multicolumn{2}{c}{GDAXI} & \multicolumn{2}{c}{STOXX50E} & \multicolumn{2}{c}{HSI} & \multicolumn{2}{c}{KS11} & \multicolumn{2}{c}{N225}\\
 & est & zstat & est & zstat & est & zstat & est & zstat & est & zstat & est & zstat & est & zstat & est & zstat & est & zstat \\ 
  \hline
$\beta^{*}_{1}$ & 0.8643 & 52.25 & 0.7974 & 38.59 & 0.8837 & 102.35 & 0.8798 & 80.13 & 0.8822 & 87.58 & 0.8972 & 67.31 & 0.8005 & 30.14 & 0.8749 & 32.24 & 0.8140 & 45.40 \\ 
  $\alpha_{1}$ & 0.2243 & 8.98 & 0.3363 & 13.09 & 0.0996 & 8.39 & 0.1041 & 9.14 & 0.1393 & 11.41 & 0.0626 & 3.34 & 0.1354 & 10.00 & 0.1560 & 7.02 & 0.1520 & 10.32 \\ 
  $\gamma_{1}$ & 0.1518 & 9.96 & 0.0876 & 6.25 & 0.1587 & 19.85 & 0.1134 & 15.04 & 0.1012 & 14.79 & 0.1498 & 11.67 & 0.0289 & 3.60 & 0.0106 & 0.79 & 0.1075 & 10.31 \\ 
   \hline
$\sigma$ & 0.3888 &  & 0.3461 &  & 0.3811 &  & 0.3778 &  & 0.3310 &  & 0.4360 &  & 0.3620 &  & 0.4186 &  & 0.4409 &  \\ 
   \hline
$R^2$ & 0.6842 &  & 0.6493 &  & 0.6724 &  & 0.5973 &  & 0.6870 &  & 0.5275 &  & 0.5410 &  & 0.4948 &  & 0.4978 &  \\ 
   \hline
$LB(5)$ & 0.4788 &  & 0.1884 &  & 0.0028 &  & 0.0466 &  & 0.0932 &  & 0.5518 &  & 0.0752 &  & 0.0232 &  & 0.2124 &  \\ 
  $LB(10)$ & 0.0127 &  & 0.0999 &  & 0.0035 &  & 0.2043 &  & 0.2755 &  & 0.8269 &  & 0.4399 &  & 0.1625 &  & 0.0927 &  \\ 
  $LB(15)$ & 0.0005 &  & 0.0818 &  & 0.0031 &  & 0.0472 &  & 0.0843 &  & 0.5525 &  & 0.7020 &  & 0.4415 &  & 0.3036 &  \\ 
  $LB(20)$ & 0.0021 &  & 0.1647 &  & 0.0059 &  & 0.0063 &  & 0.0748 &  & 0.6230 &  & 0.4299 &  & 0.6789 &  & 0.5765 &  \\ 
   \hline
\end{tabular}
}

%% file: Table/Tab-spmem-gof.tex
\scalebox{0.85}{
\begin{tabular}{ccrrrrrrrrr}
  Test                 & Distr.          & DJI & IXIC & SPX & FCHI & GDAXI & STOXX50E & HSI & KS11 & N225 \\ 
  \hline{}
  \multirow{4}{*}{AD}  & Gamma           & 0.0000 & 0.0000 & 0.0000 & 0.0000 & 0.0000 & 0.0000 & 0.0000 & 0.0000 & 0.0000 \\ 
                       & Log-N           & 0.0764 & 0.0270 & 0.1278 & 0.0081 & 0.0005 & 0.0000 & 0.0000 & 0.0185 & 0.0000 \\ 
                       & Beta$^{\prime}$ & 0.1055 & 0.0369 & 0.2194 & 0.0184 & 0.0011 & 0.0000 & 0.0000 & 0.0325 & 0.0000 \\ 
                       & Log-L           & 0.0496 & 0.0705 & 0.0000 & 0.0000 & 0.0218 & 0.4204 & 0.1194 & 0.0513 & 0.0524 \\ 
 \multirow{4}{*}{CvM}  & Gamma           & 0.0004 & 0.0001 & 0.0000 & 0.0000 & 0.0000 & 0.0000 & 0.0000 & 0.0008 & 0.0000 \\ 
                       & Log-N           & 0.1578 & 0.0482 & 0.1650 & 0.0237 & 0.0034 & 0.0001 & 0.0002 & 0.0430 & 0.0000 \\ 
                       & Beta$^{\prime}$ & 0.2082 & 0.0628 & 0.2486 & 0.0450 & 0.0059 & 0.0002 & 0.0005 & 0.0699 & 0.0000 \\ 
                       & Log-L           & 0.0708 & 0.0883 & 0.0001 & 0.0007 & 0.0340 & 0.7570 & 0.1034 & 0.0833 & 0.0580 \\ 
\hline
\end{tabular}
}

%% file: Table/Tab-Means.tex
\scalebox{1}{
\begin{tabular}{rrrrrrrrrr}
  & DJI & IXIC & SPX & FCHI & GDAXI & STOXX50E & HSI & KS11 & N225 \\ 
  \hline
arVol & 11.38 & 12.64 & 14.80 & 15.94 & 16.84 & 14.38 & 13.97 & 10.12 & 14.97 \\ 
  rkVol & 10.53 & 11.61 & 13.01 & 15.38 & 16.26 & 13.73 & 13.10 & 9.34 & 13.52 \\ 
  impVol & 16.80 & 19.84 & 19.98 & 22.37 & 22.02 & 19.96 & 22.70 & 16.06 & 24.78 \\ 
   \hline
\end{tabular}
}

%% file: Table/Tab-Inf-vmem.tex
\scalebox{0.8}{
\begin{tabular}{lrrrrrrrrrrrrrrrrrr}
   & \multicolumn{2}{c}{DJI} & \multicolumn{2}{c}{IXIC} & \multicolumn{2}{c}{SPX} & \multicolumn{2}{c}{FCHI} & \multicolumn{2}{c}{GDAXI} & \multicolumn{2}{c}{STOXX50E} & \multicolumn{2}{c}{HSI} & \multicolumn{2}{c}{KS11} & \multicolumn{2}{c}{N225}\\
 & est & zstat & est & zstat & est & zstat & est & zstat & est & zstat & est & zstat & est & zstat & est & zstat & est & zstat \\ 
  \hline
$\beta^{*}_{1,1,1}$ & 0.9558 & 142.28 & 0.9182 & 92.66 & 0.9639 & 280.12 & 0.9587 & 245.30 & 0.9638 & 245.42 & 0.9560 & 156.19 & 0.9722 & 267.54 & 0.9210 & 66.23 & 0.9567 & 198.87 \\ 
  $\alpha_{1,1,1}$ & -0.0022 & -0.20 & -0.0096 & -0.83 & 0.0001 & 0.01 & -0.0149 & -2.47 & -0.0069 & -1.19 & -0.0359 & -4.15 & 0.0071 & 1.32 & 0.0164 & 1.19 & 0.0140 & 1.78 \\ 
  $\alpha_{1,2,1}$ & 0.1425 & 5.20 & 0.1923 & 5.46 & 0.0813 & 5.44 & 0.1022 & 7.15 & 0.1244 & 8.56 & 0.0603 & 3.30 & 0.0381 & 3.25 & 0.0249 & 0.83 & 0.0946 & 5.81 \\ 
  $\alpha_{1,3,1}$ & 1.2051 & 11.33 & 0.9998 & 9.05 & 1.3461 & 18.62 & 0.5885 & 24.27 & 1.1816 & 14.58 & 1.0537 & 9.43 & 0.6913 & 11.90 & 1.1183 & 7.98 & 0.6517 & 11.46 \\ 
  $\gamma_{1,1,1}$ & 0.0229 & 2.05 & 0.0364 & 2.93 & 0.0170 & 3.07 & 0.0432 & 6.84 & 0.0130 & 2.44 & 0.0432 & 4.34 & 0.0062 & 1.47 & 0.0014 & 0.13 & 0.0190 & 3.06 \\ 
  $\beta^{*}_{2,2,1}$ & 0.9603 & 165.01 & 0.9242 & 110.20 & 0.9679 & 319.61 & 0.9628 & 293.96 & 0.9677 & 298.06 & 0.9570 & 162.04 & 0.9711 & 272.17 & 0.9418 & 109.71 & 0.9569 & 212.34 \\ 
  $\alpha_{2,1,1}$ & 0.0156 & 2.19 & 0.0203 & 3.09 & 0.0161 & 4.27 & 0.0112 & 2.88 & 0.0117 & 3.17 & -0.0025 & -0.43 & 0.0155 & 4.54 & 0.0272 & 3.90 & 0.0269 & 5.14 \\ 
  $\alpha_{2,2,1}$ & 0.1046 & 5.75 & 0.1424 & 7.00 & 0.0463 & 4.83 & 0.0812 & 8.42 & 0.0963 & 9.78 & 0.0364 & 2.60 & 0.0388 & 4.71 & 0.0241 & 1.46 & 0.0779 & 6.81 \\ 
  $\alpha_{2,3,1}$ & 0.9404 & 15.66 & 0.8860 & 15.42 & 1.0202 & 28.47 & 0.5022 & 20.64 & 0.9797 & 22.35 & 0.9395 & 15.12 & 0.6211 & 19.00 & 0.7539 & 11.98 & 0.5609 & 15.89 \\ 
  $\gamma_{2,2,1}$ & 0.0200 & 2.24 & 0.0045 & 0.50 & 0.0089 & 2.22 & 0.0404 & 8.53 & 0.0132 & 3.14 & 0.0331 & 4.14 & 0.0038 & 1.01 & 0.0012 & 0.17 & 0.0146 & 2.93 \\ 
  $\beta^{*}_{3,3,1}$ & 0.9773 & 269.96 & 0.9653 & 229.42 & 0.9809 & 519.88 & 0.9692 & 362.80 & 0.9805 & 446.38 & 0.9713 & 219.26 & 0.9822 & 408.62 & 0.9665 & 190.15 & 0.9762 & 384.95 \\ 
  $\alpha_{3,1,1}$ & 0.0051 & 1.45 & 0.0033 & 1.03 & 0.0023 & 1.13 & 0.0064 & 3.15 & 0.0034 & 2.28 & 0.0045 & 1.37 & 0.0031 & 1.65 & -0.0001 & -0.03 & -0.0008 & -0.32 \\ 
  $\alpha_{3,2,1}$ & 0.0125 & 1.42 & 0.0003 & 0.03 & 0.0032 & 0.60 & 0.0166 & 3.58 & 0.0047 & 1.15 & 0.0006 & 0.08 & -0.0086 & -1.78 & 0.0020 & 0.21 & 0.0153 & 2.82 \\ 
  $\alpha_{3,3,1}$ & 0.9013 & 30.92 & 0.9197 & 32.75 & 0.9293 & 51.10 & 0.8507 & 75.81 & 0.9946 & 56.25 & 0.9499 & 33.47 & 0.9769 & 58.58 & 0.9493 & 35.04 & 0.7995 & 48.02 \\ 
  $\gamma_{3,3,1}$ & -0.0066 & -2.05 & -0.0065 & -2.17 & -0.0055 & -2.79 & 0.0042 & 2.10 & -0.0045 & -2.65 & -0.0034 & -0.98 & -0.0039 & -2.54 & -0.0076 & -2.48 & -0.0015 & -0.82 \\ 
   \hline
$\sigma_{1}$ & 0.8608 &  & 0.8595 &  & 0.8426 &  & 0.8245 &  & 0.8421 &  & 0.8673 &  & 0.8189 &  & 0.8252 &  & 0.8611 &  \\ 
  $\sigma_{2}$ & 0.3744 &  & 0.3331 &  & 0.3683 &  & 0.3774 &  & 0.3245 &  & 0.4077 &  & 0.3507 &  & 0.3870 &  & 0.4368 &  \\ 
  $\sigma_{3}$ & 0.0719 &  & 0.0720 &  & 0.0740 &  & 0.0728 &  & 0.0575 &  & 0.0734 &  & 0.0557 &  & 0.0694 &  & 0.0684 &  \\ 
  $\rho_{1,2}$ & 0.5042 &  & 0.4546 &  & 0.5382 &  & 0.5352 &  & 0.5016 &  & 0.5627 &  & 0.5697 &  & 0.5851 &  & 0.6490 &  \\ 
  $\rho_{1,3}$ & 0.3187 &  & 0.3205 &  & 0.3153 &  & 0.2299 &  & 0.2401 &  & 0.2859 &  & 0.2492 &  & 0.2738 &  & 0.3331 &  \\ 
  $\rho_{2,3}$ & 0.4319 &  & 0.5300 &  & 0.3952 &  & 0.3004 &  & 0.3587 &  & 0.3542 &  & 0.3128 &  & 0.3344 &  & 0.4111 &  \\ 
   \hline
$R^2(1)$ & 0.3602 &  & 0.2372 &  & 0.3380 &  & 0.2346 &  & 0.2613 &  & 0.2150 &  & 0.2251 &  & 0.2633 &  & 0.2408 &  \\ 
  $R^2(2)$ & 0.7087 &  & 0.6729 &  & 0.6969 &  & 0.5983 &  & 0.6927 &  & 0.5505 &  & 0.5711 &  & 0.5441 &  & 0.5008 &  \\ 
  $R^2(3)$ & 0.9596 &  & 0.9432 &  & 0.9590 &  & 0.9597 &  & 0.9718 &  & 0.9425 &  & 0.9718 &  & 0.9543 &  & 0.9547 &  \\ 
   \hline
$LB(5)$ & 0.0040 &  & 0.0000 &  & 0.0000 &  & 0.0000 &  & 0.0000 &  & 0.0014 &  & 0.0000 &  & 0.0019 &  & 0.0000 &  \\ 
  $LB(10)$ & 0.0028 &  & 0.0000 &  & 0.0000 &  & 0.0000 &  & 0.0000 &  & 0.1022 &  & 0.0000 &  & 0.0158 &  & 0.0000 &  \\ 
  $LB(15)$ & 0.0246 &  & 0.0003 &  & 0.0000 &  & 0.0000 &  & 0.0000 &  & 0.3265 &  & 0.0000 &  & 0.0121 &  & 0.0000 &  \\ 
  $LB(20)$ & 0.0750 &  & 0.0003 &  & 0.0000 &  & 0.0000 &  & 0.0000 &  & 0.3307 &  & 0.0000 &  & 0.0524 &  & 0.0000 &  \\ 
   \hline
\end{tabular}
}

%% file: Table/Tab-vmem-heavy.tex
\scalebox{0.77}{
\begin{tabular}{llrrrrrrrrr}
  Model &     & DJI & IXIC & SPX & FCHI & GDAXI & STOXX50E & HSI & KS11 & N225 \\ 
  \hline
  \multirow{2}{*}{\vMEM}   & statistic & 12.3104 & 25.1204 & 51.7463 & 61.3916 & 39.4824 & 31.1643 & 28.7689 & 23.0047 & 42.0453 \\ 
                           & p-value   & 0.0064 & 0.0000 & 0.0000 & 0.0000 & 0.0000 & 0.0000 & 0.0000 & 0.0000 & 0.0000 \\ 
  \multirow{2}{*}{\SpvMEM} & statistic & 16.1555 & 32.5538 & 83.4393 & 57.5994 & 76.6602 & 46.0709 & 56.6530 & 31.1979 & 87.8567 \\ 
                           & p-value   & 0.0011 & 0.0000 & 0.0000 & 0.0000 & 0.0000 & 0.0000 & 0.0000 & 0.0000 & 0.0000 \\ 
   \hline
\end{tabular}
}

%% file: Table/Tab-Inf-spvmem.tex
\scalebox{0.8}{
\begin{tabular}{lrrrrrrrrrrrrrrrrrr}
   & \multicolumn{2}{c}{DJI} & \multicolumn{2}{c}{IXIC} & \multicolumn{2}{c}{SPX} & \multicolumn{2}{c}{FCHI} & \multicolumn{2}{c}{GDAXI} & \multicolumn{2}{c}{STOXX50E} & \multicolumn{2}{c}{HSI} & \multicolumn{2}{c}{KS11} & \multicolumn{2}{c}{N225}\\
 & est & zstat & est & zstat & est & zstat & est & zstat & est & zstat & est & zstat & est & zstat & est & zstat & est & zstat \\ 
  \hline
$\beta^{*}_{1,1,1}$ & 0.9324 & 96.43 & 0.8424 & 44.84 & 0.9368 & 167.80 & 0.8474 & 51.25 & 0.9106 & 103.13 & 0.9307 & 95.30 & 0.8734 & 41.86 & 0.8353 & 23.35 & 0.8717 & 58.33 \\ 
  $\alpha_{1,1,1}$ & -0.0065 & -0.49 & -0.0270 & -1.86 & -0.0214 & -2.99 & -0.0419 & -4.15 & -0.0350 & -4.62 & -0.0537 & -5.15 & -0.0238 & -2.59 & 0.0288 & 1.54 & -0.0082 & -0.77 \\ 
  $\alpha_{1,2,1}$ & 0.1795 & 5.51 & 0.2340 & 5.31 & 0.1102 & 6.24 & 0.1587 & 6.68 & 0.1527 & 7.48 & 0.0856 & 3.66 & 0.0432 & 2.15 & -0.0190 & -0.48 & 0.1277 & 5.35 \\ 
  $\alpha_{1,3,1}$ & 1.0742 & 9.22 & 0.7873 & 5.96 & 1.0448 & 13.70 & 0.5222 & 13.17 & 0.9163 & 9.91 & 0.6964 & 6.53 & 0.5428 & 6.38 & 0.9552 & 5.92 & 0.3938 & 5.55 \\ 
  $\gamma_{1,1,1}$ & 0.0369 & 2.93 & 0.0612 & 3.56 & 0.0465 & 6.54 & 0.0675 & 6.47 & 0.0419 & 5.39 & 0.0699 & 5.81 & 0.0324 & 3.75 & 0.0009 & 0.05 & 0.0549 & 5.54 \\ 
  $\beta^{*}_{2,2,1}$ & 0.9369 & 114.88 & 0.8315 & 55.57 & 0.9409 & 200.29 & 0.8652 & 84.52 & 0.9134 & 137.19 & 0.9239 & 99.50 & 0.8484 & 63.42 & 0.8711 & 48.37 & 0.8602 & 77.88 \\ 
  $\alpha_{2,1,1}$ & 0.0179 & 2.30 & 0.0249 & 3.36 & 0.0157 & 3.76 & 0.0102 & 2.06 & 0.0132 & 3.11 & -0.0061 & -0.92 & 0.0222 & 4.59 & 0.0412 & 4.96 & 0.0402 & 6.18 \\ 
  $\alpha_{2,2,1}$ & 0.1243 & 6.09 & 0.1798 & 7.52 & 0.0576 & 5.30 & 0.0907 & 7.21 & 0.1165 & 9.55 & 0.0437 & 2.62 & 0.0582 & 4.59 & 0.0072 & 0.36 & 0.0955 & 6.30 \\ 
  $\alpha_{2,3,1}$ & 0.8819 & 13.43 & 0.8018 & 12.20 & 0.9443 & 24.00 & 0.4381 & 13.61 & 0.8066 & 16.18 & 0.7747 & 11.80 & 0.5685 & 12.85 & 0.6636 & 9.53 & 0.3878 & 8.90 \\ 
  $\gamma_{2,2,1}$ & 0.0310 & 2.94 & 0.0155 & 1.38 & 0.0229 & 4.17 & 0.0653 & 10.04 & 0.0327 & 5.74 & 0.0546 & 5.60 & 0.0143 & 2.33 & 0.0031 & 0.34 & 0.0500 & 6.61 \\ 
  $\beta^{*}_{3,3,1}$ & 0.9662 & 161.30 & 0.9046 & 100.94 & 0.9754 & 308.19 & 0.8758 & 127.73 & 0.9457 & 194.48 & 0.9461 & 122.43 & 0.9190 & 147.93 & 0.9344 & 95.21 & 0.9111 & 145.11 \\ 
  $\alpha_{3,1,1}$ & 0.0048 & 1.38 & 0.0033 & 1.11 & 0.0016 & 0.79 & 0.0055 & 2.88 & 0.0036 & 2.43 & 0.0038 & 1.16 & 0.0033 & 1.77 & -0.0008 & -0.21 & -0.0008 & -0.32 \\ 
  $\alpha_{3,2,1}$ & 0.0115 & 1.32 & 0.0028 & 0.30 & 0.0007 & 0.14 & 0.0101 & 2.30 & 0.0019 & 0.48 & -0.0013 & -0.18 & -0.0124 & -2.63 & -0.0032 & -0.35 & 0.0160 & 2.99 \\ 
  $\alpha_{3,3,1}$ & 0.8783 & 29.88 & 0.8564 & 30.88 & 0.9173 & 49.83 & 0.8177 & 71.21 & 0.9613 & 54.36 & 0.9267 & 32.41 & 0.9298 & 56.29 & 0.9187 & 34.09 & 0.7496 & 45.20 \\ 
  $\gamma_{3,3,1}$ & -0.0046 & -1.37 & -0.0040 & -1.34 & -0.0067 & -3.12 & 0.0047 & 2.45 & -0.0048 & -2.76 & -0.0044 & -1.21 & -0.0037 & -2.39 & -0.0079 & -2.53 & 0.0002 & 0.13 \\ 
   \hline
  $\sigma_{1}$ & 0.8631 &  & 0.8511 &  & 0.8450 &  & 0.8304 &  & 0.8423 &  & 0.8699 &  & 0.8194 &  & 0.8304 &  & 0.8644 &  \\ 
  $\sigma_{2}$ & 0.3748 &  & 0.3306 &  & 0.3682 &  & 0.3745 &  & 0.3254 &  & 0.4139 &  & 0.3494 &  & 0.3873 &  & 0.4344 &  \\ 
  $\sigma_{3}$ & 0.0710 &  & 0.0684 &  & 0.0734 &  & 0.0683 &  & 0.0565 &  & 0.0726 &  & 0.0542 &  & 0.0682 &  & 0.0664 &  \\ 
$\rho_{1,2}$ & 0.5034 &  & 0.4499 &  & 0.5343 &  & 0.5238 &  & 0.4987 &  & 0.5707 &  & 0.5653 &  & 0.5844 &  & 0.6477 &  \\ 
  $\rho_{1,3}$ & 0.3123 &  & 0.3149 &  & 0.3048 &  & 0.1939 &  & 0.2316 &  & 0.2675 &  & 0.2398 &  & 0.2601 &  & 0.3196 &  \\ 
  $\rho_{2,3}$ & 0.4231 &  & 0.5311 &  & 0.3810 &  & 0.2678 &  & 0.3438 &  & 0.3341 &  & 0.2938 &  & 0.3205 &  & 0.3895 &  \\ 
   \hline
  $R^2(1)$ & 0.3610 &  & 0.2427 &  & 0.3457 &  & 0.2413 &  & 0.2676 &  & 0.2191 &  & 0.2269 &  & 0.2526 &  & 0.2437 &  \\ 
  $R^2(2)$ & 0.7135 &  & 0.6761 &  & 0.7019 &  & 0.6096 &  & 0.6986 &  & 0.5537 &  & 0.5758 &  & 0.5500 &  & 0.5099 &  \\ 
  $R^2(3)$ & 0.9608 &  & 0.9477 &  & 0.9599 &  & 0.9622 &  & 0.9728 &  & 0.9441 &  & 0.9733 &  & 0.9559 &  & 0.9570 &  \\ 
   \hline
$LB(5)$ & 0.0115 &  & 0.0003 &  & 0.0000 &  & 0.0000 &  & 0.0000 &  & 0.0033 &  & 0.0000 &  & 0.0105 &  & 0.0000 &  \\ 
  $LB(10)$ & 0.0051 &  & 0.0006 &  & 0.0000 &  & 0.0000 &  & 0.0000 &  & 0.1062 &  & 0.0000 &  & 0.0590 &  & 0.0000 &  \\ 
  $LB(15)$ & 0.0100 &  & 0.0002 &  & 0.0000 &  & 0.0000 &  & 0.0000 &  & 0.2393 &  & 0.0000 &  & 0.0736 &  & 0.0003 &  \\ 
$LB(20)$ & 0.0391 &  & 0.0013 &  & 0.0000 &  & 0.0000 &  & 0.0000 &  & 0.2361 &  & 0.0000 &  & 0.2317 &  & 0.0009 &  \\ 
   \hline
\end{tabular}
}

%% file: Table/Tab-spvmem-gof.tex
\scalebox{0.80}{
\begin{tabular}{llllrrrrrrrrrr}
                     Ind. &                 Test &          Distr. &    DJI &  IXIC & SPX & FCHI & GDAXI & STOXX50E & HSI & KS11 & N225 \\ 
  \hline
  \multirow{8}{*}{arVol}  & \multirow{4}{*}{AD}  & Gamma           & 0.0000 & 0.0835 & 0.0000 & 0.0000 & 0.0000 & 0.0117 & 0.0000 & 0.0000 & 0.0001 \\ 
                          &                      & Log-N           & 0.0000 & 0.0000 & 0.0000 & 0.0000 & 0.0000 & 0.0000 & 0.0000 & 0.0000 & 0.0000 \\ 
                          &                      & Beta$^{\prime}$ & 0.0000 & 0.0000 & 0.0000 & 0.0000 & 0.0000 & 0.0000 & 0.0000 & 0.0000 & 0.0000 \\ 
                          &                      & Log-L           & 0.0000 & 0.0000 & 0.0000 & 0.0000 & 0.0000 & 0.0000 & 0.0000 & 0.0000 & 0.0000 \\ 
                          & \multirow{4}{*}{CvM} & Gamma           & 0.0077 & 0.1368 & 0.0000 & 0.0000 & 0.0009 & 0.0120 & 0.0001 & 0.0356 & 0.0196 \\ 
                          &                      & Log-N           & 0.0000 & 0.0000 & 0.0000 & 0.0000 & 0.0000 & 0.0000 & 0.0000 & 0.0000 & 0.0000 \\ 
                          &                      & Beta$^{\prime}$ & 0.0000 & 0.0000 & 0.0000 & 0.0000 & 0.0000 & 0.0000 & 0.0000 & 0.0000 & 0.0000 \\ 
                          &                      & Log-L           & 0.0000 & 0.0000 & 0.0000 & 0.0000 & 0.0000 & 0.0000 & 0.0000 & 0.0000 & 0.0000 \\ \hline 
  \multirow{8}{*}{rkVol}  & \multirow{4}{*}{AD}  & Gamma           & 0.0003 & 0.0001 & 0.0000 & 0.0000 & 0.0000 & 0.0000 & 0.0000 & 0.0043 & 0.0000 \\ 
                          &                      & Log-N           & 0.3403 & 0.1577 & 0.4096 & 0.0113 & 0.0038 & 0.0001 & 0.0052 & 0.1816 & 0.0000 \\ 
                          &                      & Beta$^{\prime}$ & 0.4102 & 0.1964 & 0.6147 & 0.0248 & 0.0074 & 0.0003 & 0.0106 & 0.2603 & 0.0000 \\ 
                          &                      & Log-L           & 0.0123 & 0.0204 & 0.0000 & 0.0000 & 0.0132 & 0.4805 & 0.0066 & 0.0066 & 0.0240 \\ 
                          & \multirow{4}{*}{CvM} & Gamma           & 0.0033 & 0.0010 & 0.0005 & 0.0000 & 0.0000 & 0.0000 & 0.0000 & 0.0170 & 0.0000 \\ 
                          &                      & Log-N           & 0.4764 & 0.2425 & 0.3540 & 0.0315 & 0.0137 & 0.0010 & 0.0204 & 0.2419 & 0.0000 \\ 
                          &                      & Beta$^{\prime}$ & 0.5704 & 0.2878 & 0.5225 & 0.0583 & 0.0235 & 0.0022 & 0.0353 & 0.3227 & 0.0001 \\ 
                          &                      & Log-L           & 0.0260 & 0.0286 & 0.0000 & 0.0007 & 0.0243 & 0.6014 & 0.0099 & 0.0218 & 0.0267 \\ \hline
  \multirow{8}{*}{impVol} & \multirow{4}{*}{AD}  & Gamma           & 0.0000 & 0.0000 & 0.0000 & 0.0000 & 0.0000 & 0.0000 & 0.0000 & 0.0000 & 0.0000 \\ 
                          &                      & Log-N           & 0.0000 & 0.0000 & 0.0000 & 0.0000 & 0.0000 & 0.0000 & 0.0000 & 0.0000 & 0.0000 \\ 
                          &                      & Beta$^{\prime}$ & 0.0000 & 0.0000 & 0.0000 & 0.0000 & 0.0000 & 0.0000 & 0.0000 & 0.0000 & 0.0000 \\ 
                          &                      & Log-L           & 0.0000 & 0.0000 & 0.0000 & 0.0000 & 0.0000 & 0.0000 & 0.0000 & 0.0000 & 0.0000 \\ 
                          & \multirow{4}{*}{CvM} & Gamma           & 0.0000 & 0.0000 & 0.0000 & 0.0000 & 0.0000 & 0.0000 & 0.0000 & 0.0000 & 0.0000 \\ 
                          &                      & Log-N           & 0.0000 & 0.0000 & 0.0000 & 0.0000 & 0.0000 & 0.0000 & 0.0000 & 0.0000 & 0.0000 \\ 
                          &                      & Beta$^{\prime}$ & 0.0000 & 0.0000 & 0.0000 & 0.0000 & 0.0000 & 0.0000 & 0.0000 & 0.0000 & 0.0000 \\ 
                          &                      & Log-L           & 0.0000 & 0.0000 & 0.0000 & 0.0000 & 0.0000 & 0.0000 & 0.0000 & 0.0000 & 0.0000 \\ 
   \hline
\end{tabular}
}

%% file: Table/Tab-data.tex
\scalebox{0.7}{
\begin{tabular}{lllll}
  \textbf{Symbol} & \textbf{OMI Realized Library (arVol, rkVol)}      & \textbf{Investing.com (impVol)}            & \textbf{Start} & \textbf{End} \\
  \hline
  DJI      & Dow Jones Industrial Average              & VXD (DJIA Volatility)             & Aug 2, 2013  & Apr 13, 2021 \\
  IXIC     & Nasdaq 100                                & VXN (CBOE NASDAQ100 Volatility)   & Oct 6, 2011  & Apr 13, 2021 \\
  SPX      & S\&P 500                                  & VIX (CBOE Volatility index)       & Jan 3, 2000  & Apr 13, 2021 \\
  FCHI     & CAC 40                                    & VCAC (CAC40 VIX)                  & Jan 4, 2000  & Dec 31, 2020 \\
  GDAXI    & DAX                                       & V1XI (DAX New Volatility)         & May 11, 2001 & Apr 13, 2021 \\
  STOXX50E & EURO STOXX 50                             & V2TX (STOXX 50 Volatility VSTOXX) & Dec 28, 2012 & Apr 13, 2021 \\
  HSI      & HANG SENG                                 & VHSI (HSI Volatility)             & Jan 3, 2001  & Apr 13, 2021 \\
  KS11     & Korea Composite Stock Price Index (KOSPI) & KSVKOSPI (KOSPI Volatility)       & Aug 6, 2013  & Apr 13, 2021 \\
  N225     & Nikkei 225                                & JNIV (Nikkei Volatility)          & Nov 13, 2000 & Apr 13, 2021
\end{tabular}
}